\documentclass{aa}  

\usepackage{graphicx}
\usepackage{txfonts}
\usepackage{comment}
\usepackage{caption}
\usepackage{subfig}
\usepackage{lipsum}

\usepackage{tabularx}
\usepackage{booktabs}
\usepackage{arydshln}                   
\usepackage{orcidlink}

\newcommand{\abs}[1]{\lvert#1\rvert}

\newcommand{\rhill}{{\rm R}_{\rm Hill}}
\newcommand{\visc}{\alpha_{\rm visc}}
\newcommand{\rin}{r_{\rm in;\,p}}
\newcommand{\mcpd}{M_{\rm CPD}}
\newcommand{\mcsd}{M_{\rm CSD}}
\newcommand{\mj}{{\rm M}_{\rm J}}
\newcommand{\rj}{{\rm R}_{\rm J}}
\newcommand{\msun}{{\rm M}_\odot}
\newcommand{\rtranshat}{\hat{r}_{\rm trans}}
\newcommand{\rcpdhat}{\hat{r}_{\rm CPD}}
\newcommand{\ContrastStar}{\text{CPD}\,{:}\,{\rm Star}}
\newcommand{\ContrastGap}{\text{Gap}\,{:}\,\text{CPD}}
\newcommand{\mpdot}{\dot{M}_{\rm p}}
\newcommand{\lambdaS}{652\,{\rm nm}}
\newcommand{\lambdaM}{1.24\,\mu{\rm m}}
\newcommand{\lambdaL}{2.11\,\mu{\rm m}}
\newcommand{\dlim}{D_{\rm lim}}
\newcommand{\rpS}{5\,{\rm au}}
\newcommand{\rpM}{10\,{\rm au}}
\newcommand{\rpL}{50\,{\rm au}}
\newcommand{\lacc}{L_{\rm acc}}
\newcommand{\mcpdS}{0.1\cdot 10^{-3}\,\mj}

\newcommand{\rp}{r_{\rm p}}
\newcommand{\Mp}{M_{\rm p}}
\newcommand{\mpdotS}{0.4\cdot 10^{-6}\,\mj/{\rm yr}}
\newcommand{\mpdotL}{2.0\cdot 10^{-6}\,\mj/{\rm yr}}
\newcommand{\mcsdS}{0.001\,{\rm M}_\odot}
\newcommand{\mcsdL}{0.01\,{\rm M}_\odot}

\begin{document}

   \title{Feasibility of detecting and characterizing embedded low-mass giant planets in gaps in the VIS/NIR wavelength range}
        
       \titlerunning{Detecting and characterizing  embedded planets in gaps in the VIS/NIR}

   \author{A.~Krieger \orcidlink{0000-0002-3639-2435}
          \and
          S.~Wolf \orcidlink{0000-0001-7841-3452}
          }
  \authorrunning{A.~Krieger and S.~Wolf}

   \institute{University of Kiel, Institute of Theoretical Physics and Astrophysics, Leibnizstrasse 15, 24118 Kiel, Germany, \\ \email{akrieger@astrophysik.uni-kiel.de}
}

   \date{Received: November 12, 2021; Accepted: February 1, 2022 }

    \abstract
   {   
   High-contrast imaging in the visible and near-infrared (VIS/NIR) has revealed the presence of a plethora of substructures in circumstellar disks (CSDs). One of the most commonly observed substructures are concentric gaps that are often attributed to the presence of embedded forming planets. However, direct detections of these planets are extremely rare, and thus ambiguity  regarding the origin of most gap features remains. The aim of this study is to  investigate the capabilities of high-contrast VIS/NIR imaging of directly detecting and characterizing low-mass giant planets in gaps in a broad systematic parameter study. To this end, a grid of models of protoplanetary disks was generated. The models include a central T Tauri star   surrounded by a face-on CSD harboring an accreting planet, which itself is surrounded by a circumplanetary disk (CPD) and carves a gap. These gaps are modeled using empirically determined profiles, and the whole system is simulated fully self-consistently using the Monte Carlo radiative transfer code Mol3D in order to generate temperature distributions and synthetic observations assuming a generic dust composition consisting of astronomical silicate and graphite. Based on these simulations, we measured the impact the planet and its CPD have on contrast curves and quantified the impact of the observing wavelength and of five key parameters (planetary mass,  mass accretion rate,  distance to the star,   mass of the CPD, and   mass of the CSD) on the determined signal strength. Subsequently, we applied a detection criterion on our results and assess the capabilities of the instrument SPHERE/VLT of detecting the embedded planets. We find that a part of the investigated parameter space includes detectable planets, and we elaborate on the implication a non-detection has on the underlying parameters of a potential planet and its CPD. Furthermore, we analyze the potential loss of valuable information that would enable the detection of embedded planets by the use of a coronagraphic mask. However, we find this outcome to be extremely unlikely in the case of SPHERE. Finally, within the VIS/NIR wavelength range we  identify for each of the investigated basic properties of the planets and the disks the most promising observing wavelengths that enable us to distinguish between different underlying parameter values. In doing so, we find that the detectability and the characterization often benefit from different observing wavelengths, highlighting the complementarity and importance of multiwavelength observations.

   }


   \maketitle

\section{Introduction}

In recent years the quest to observe embedded accreting planets as they form has come a long way as a result of combined observations that are covering the ultraviolet   to the millimeter wavelength ranges. One instrument in particular  has been playing a major role in this, due to its diffraction-limited and high-contrast observations: the Spectro-Polarimetric High contrast imager for Exoplanets REsearch \citep[SPHERE;][]{2019A&A...631A.155B}, which covers the visible  and near-infrared (VIS/NIR) wavelength ranges. Together with other instruments, it has revealed a large abundance of substructures that are now known to be commonly present in protoplanetary disks (PPDs) ranging from gaps and rings to spiral arms, cavities, and various asymmetric features \citep[e.g.,][DSHARP]{2016A&A...588A...8G,2018A&A...620A..94G,2018ApJ...863...44A,2018ApJ...869L..41A}. These features are often interpreted as the result of embedded planets that interact with their natal circumstellar disk \citep[CSD; e.g.,][]{2005ApJ...619.1114W,2010A&A...518A..16F,2013A&A...549A..97R,Ober_2015,2016ApJ...826...75D}. However, other origins have been proposed as well \citep[e.g.,][]{2015A&A...574A..68F,2015ApJ...806L...7Z,2016A&A...590A..17R,2017MNRAS.467.1984G,2017MNRAS.468.3850S,2018A&A...609A..50D,2019MNRAS.484..107S}. Despite the ambiguity of their particular origins, the properties of the potential planets that may produce these features have been studied extensively, especially using hydrodynamics simulations \citep[e.g.,][]{2019ApJ...884L..41B,2020ApJ...888L...4T,2020MNRAS.498..639C}.
\cite{2017ApJ...835..146D}, for instance, analyzed observed gap features in order to determine the masses
of planets that may have caused them by using 2D and 3D hydrodynamics simulations with 3D radiative transfer simulations of five scattered light observations of CSDs around Herbig Ae/Be and T Tauri stars:  HD 97048 \citep{2016A&A...595A.112G}, TW Hya \citep{2017ApJ...837..132V}, HD 169142 \citep{2015PASJ...67...83M}, LkCa15 \citep{2016ApJ...828L..17T}, and RX J1615 \citep{2016A&A...595A.114D}. By assuming an $\alpha$-viscosity model \citep{Shakura_1973} with $\alpha_{\rm visc}=10^{-3}$ and single gap-opening planets as origins, they deduced that the corresponding planetary masses are all  of typical low-mass giant planets between about 0.1 and 1\,$\mj$.

While the kinematic detection of a planet  has proven to be very useful \citep{2018ApJ...860L..13P,2019NatAs...3.1109P}, in order to avoid much of the ambiguity that is present in the analysis of indirect features it is highly desirable to directly image the embedded planets. Unfortunately, this task is extremely challenging and observations are still rare. To date, the only two thus confirmed embedded planets,
 PDS 70 b and PDS 70 c, are both located around PDS~70  \citep{2018A&A...617A..44K,2018A&A...617L...2M,2019NatAs...3..749H,2019A&A...632A..25M}, in which case   it was even possible to  restrict the basic parameters of potential circumplanetary disks (CPDs) surrounding both of them \citep{2019ApJ...879L..25I,2021arXiv210807123B}.

In this study we focus on systems of low-mass giant planets carving gaps into the CSDs of T Tauri stars in which the planets themselves are surrounded by their own CPDs characterized by their accretion luminosity. Our goal is to assess the potential for direct detections and for the characterization of these planets and their CPDs in the VIS/NIR wavelength range in a systematic manner. Studying this particular class of planets is important as they are likely much more prevalent than their more massive counterparts;  the overall occurrence rate of 5 to 13\,$\mj$ companions located at orbital distances of 30 to 300\,au is only $0.6^{+0.5}_{-0.7}\,\%$ in systems with stellar masses of 0.1 to 3\,${\rm M}_\odot$ \citep{2016PASP..128j2001B}.
Therefore, we generated models of these systems and made use of the gap profiles that were empirically determined by \cite{2016PASJ...68...43K,2017PASJ...69...97K} and subsequently improved \citep{2019ApJ...884..142G}. We then conducted a broad parameter study by performing fully self-consistent Monte Carlo radiative transfer (MCRT) simulations based on these models using the code Mol3D \citep{Ober_2015}. The code was equipped with a method for dealing with the extremely high optical depths encountered in CPDs \citep{2020A&A...635A.148K} in order to significantly reduce noise in the determined temperature distributions and flux maps and, thus, improve the reliability of our simulations.

We then use the generated synthetic observations to assess the relevance of the different sources of flux observed in the VIS/NIR when trying to directly detect planets and their CPDs. Here we particularly elaborate on the relevance of proper simulations of thermal self-scattered flux of the CPDs as this turns out to be of high importance for this purpose. We generated convolved radial contrast profiles and quantified the impact of the planet and CPD on observations for each model at different observing wavelengths. The particular method that we used to measure the impact requires a conversion of the Cartesian detector grid to a polar detector grid, for which we present   for this purpose the specifically written python package CartToPolarDetector\footnote{\url{https://github.com/anton-krieger/CartToPolarDetector/}}. The obtained contrast values that measure the impact were then used as a basis for studying the detectability and ``characterizability'' of embedded planets and their CPDs with respect to the various underlying parameters of the model. With a particular focus on observing modes possible with SPHERE, we then discuss the possibility of losing valuable information due to the use of a coronagraph as a result of an inner working angle that is too small. And last, we deduce observing wavelengths that are best suited for detecting planets and CPDs and for characterizing  their various basic properties in the VIS/NIR wavelength range. 

The paper is structured as follows. In Sect. \ref{sec:setup_and_methods} the model setup and methods to reliably perform MCRT simulations are described. Instruments that are  analyzed later on are listed in Sect. \ref{sec:instruments}. Subsequently, Sect. \ref{sec:results_and_discussion} presents the method we use to measure the impact of embedded planets and their CPDs on observations, and discuss  the detectability and characterizability of these planets and their CPDs. Finally, Sect. \ref{sec:summary} presents a summary of our results and conclusions regarding observations with SPHERE and future instruments.

\section{Setup and methods}
\label{sec:setup_and_methods}
This study is performed on the basis of PPD models that are analyzed by the use of MCRT simulations and then evaluated with regard to the detectability and characterization of their embedded protoplanets. In this section we lay the foundation for the study by briefly summarizing the core principles of MCRT simulations and describing the components that constitute the models.

\subsection{MCRT}
In this study we perform MCRT simulations based on the code Mol3D \citep{Ober_2015}. Given a density distribution of dust and gas and a list of sources for radiation, Mol3D simulates the path and interactions of emitted photon packages individually and randomly according to their corresponding probability distributions. To this end, the model space is described by a grid whose cells have homogeneous and constant physical properties (e.g., density and temperature) in which the various sources are placed.  This eventually allows a precise temperature distribution to be
determined as well as wavelength-dependent flux maps (i.e., synthetic observations) to be produced. The high level of computational performance that is required to properly simulate complex systems is achieved by  utilizing a method of locally divergence-free continuous absorption of photon packages \citep{1999A&A...344..282L} together with  an immediate reemission scheme according to a temperature-corrected spectrum \citep{2001ApJ...554..615B} and   a method that utilizes a large database of precalculated photon paths in optically thick dusty media \citep{2020A&A...635A.148K}.

\subsection{Setup}
\label{sec:setup}
In the following we describe the simulated models in terms of their structural components as well as the instruments and corresponding wavelengths that we  studied. In general, the setup is composed of a star that is located in the center of a CSD, which harbors an accreting planet that itself is embedded in a CPD and carves a gap into the CSD. A list of the model parameters used can be found in Tab. \ref{tbl:parameters}.

\subsubsection{Stellar and CSD model}

The star and the planet are implemented as  point sources, described by a  blackbody spectrum corresponding to their assigned effective temperatures and radii, which are surrounded by their corresponding disks. The effective temperature of the star and its radius are given by $T_*$ and $R_*$, respectively, and the star is located at a distance $d$ from the observer. In this study we apply a disk model that is described by a three-dimensional parameterized  density distribution given by
\begin{equation}
\rho_n (r,z) = \frac{\Sigma_n(r)}{\sqrt{2\pi} h_n(r)} \exp\left\{ -\frac{1}{2}\left( \frac{z}{h_n(r)}\right)^2\right\},
\label{eq:general_density_distribution}
\end{equation}
where $r$ is the distance to its corresponding central object $n$ in the direction perpendicular to the $z$-axis, and  $\Sigma_n(r)$ and $h_n(r)$ are the surface density distribution and scale height of the disk, respectively. For the description of the CSD we use the subscript $n={\rm s}$ where s is referring to the star. In this case the surface density distribution and scale height are given by 
\begin{equation}
\Sigma_{\rm s}(r) = \Sigma_{\rm 0;\,s} \left( \frac{r}{r_{\rm 0;\,s}}\right)^{\beta_{\rm s} - \alpha_{\rm s}} \exp\left\{ -\left(\frac{r}{r_{\rm 0;\,s}} \right)^{2+\beta_{\rm s} - \alpha_{\rm s}}\right\} \hat{\Sigma}_{\rm gap}
\end{equation}
and
\begin{equation}
h_{\rm s}(r) = h_{\rm 0;\,s} \left(\frac{r}{r_{\rm 0;\,s}}  \right)^{\beta_{\rm s}},
\end{equation}
respectively, which is based on the description of \cite{1974MNRAS.168..603L} and \cite{1998ApJ...495..385H}; however, the CSD density distribution is additionally adapted for the purpose of modeling a gap that arises due to the presence of an embedded accreting planet. In general, the parameters $\alpha_n$ and $\beta_n$ describe the compactness and the flaring of the disk around the central object $n$, respectively; $\Sigma_{0;\,n}$ and $h_{0;\,n}$ describe the corresponding surface density and scale height evaluated at the reference radius $r_{0;\,n}$, respectively; and $r_{{\rm in};\,n}$ is the inner and $r_{{\rm out};\,n}$ the outer radius of the disk. For most model parameters we chose commonly used values constrained from observations of CSDs around T Tauri stars \citep[e.g., DSHARP;][]{2009ApJ...700.1502A,2015A&A...580A..26G}. Additionally, the surface density is perturbed by an accreting planet, which carves a gap into the disk, whose surface density profile is described by $\hat{\Sigma}_{\rm gap}$. 
The gap profile is based on an empirically determined \citep{2016PASJ...68...43K,2017PASJ...69...97K} and later improved \citep{2019ApJ...884..142G} model that is given by
 \begin{equation}
\hat{\Sigma}_{\rm gap} (r) = 
  \begin{cases} 
     \hat{\Sigma}_{\rm min}  &\text{if } \abs{r-\rp}<\Delta r_1\\
     \hat{\Sigma}_{\rm flank} (r)  &\text{if } \Delta r_1\leq \abs{r-\rp}<\Delta r_2\\
     1  &\text{if } \abs{r-\rp}\geq\Delta r_2,\\
   \end{cases}
   \label{eq:gap_profile_general}
 \end{equation}
with
\begin{equation}
\hat{\Sigma}_{\rm flank} (r) = \frac{2}{0.76} K'^{-1/4} \frac{\abs{r-\rp}}{\rp} - \frac{0.18}{0.76},
\label{eq:sigma_flank}
\end{equation}
\begin{equation}
\Delta r_1 = \left( 0.38\,\hat{\Sigma}_{\rm min} + 0.09 \right) K'^{1/4} \rp \quad {\rm and} \quad \Delta r_2 = 0.47K'^{1/4} \rp,
\label{eq:gap_widths}
\end{equation}
\begin{equation}
\hat{\Sigma}_{\rm min} = \frac{1}{1+0.046 K},
\end{equation}
\begin{equation}
K= \left( \frac{\Mp}{M_*}\right)^2\left( \frac{h_{\rm p}}{\rp} \right)^{-5} \visc^{-1} \quad {\rm and} \quad K' = \left( \frac{\Mp}{M_*}\right)^2\left( \frac{h_{\rm p}}{\rp} \right)^{-3}\visc^{-1},
\end{equation}
and $h_{\rm p} = h_{\rm s}(\rp)$. Here $\rp$, $\Mp$, and $h_{\rm p}$ are the radius of the planetary orbit, the mass of the planet, and the scale height of the CSD at the position of the planet, respectively. In addition, $M_*$ is the mass of the central star and $\alpha_{\rm visc}$ the parameter of the $\alpha$-viscosity model \citep{Shakura_1973}. The gap profile is azimuthally symmetric with the planet marking the center of a plateau ($\abs{r-\rp}<\Delta r_1$), where the profile reaches its minimal value $\hat{\Sigma}_{\rm min}$ as can be seen in Eq. \eqref{eq:gap_profile_general}. Outside of this minimum, the gap profile increases linearly in the radial direction until it reaches a value of $1$ at $\abs{r-\rp}=\Delta r_2$, as described by Eq. \eqref{eq:sigma_flank}. Outside the gap region (i.e., at $\abs{r-\rp}>\Delta r_2$) the CSD is unperturbed by the planetary gravitational impact. The quantities $\Delta r_1$ and $\Delta r_2$ from Eq. \ref{eq:gap_widths} are thus half of the radial width of the gap plateau and half of the total radial width of the gap, respectively.

\subsubsection{Planetary and CPD model}
\label{sec:planetary_and_cpd_model}
Similarly to the star, the planet is characterized by a point-source that is described by a blackbody spectrum. Since the planet is young and still accretes matter, its effective temperature is a combination of its intrinsic temperature and its accretion temperature.

The accretion temperature is linked via the Stefan-Boltzmann law to the accretion luminosity which itself is given by \citep[e.g.,][]{2017ApJ...836..221M}
\begin{equation}
L_{\rm acc} = \eta \,G \frac{\Mp \mpdot}{R_{\rm p}},
\label{eq:L_acc}
\end{equation}
where $G$ is the gravitational constant, $\mpdot$ the mass accretion rate onto the planet, $R_{\rm p}$ the radius of the planet, and $\eta$ the efficiency by which gravitational energy is converted into radiative energy as matter falls onto the planet. We note that the efficiency $\eta$ in general depends on various quantities; however, for simplicity we assume the case of cold accretion (i.e., $\eta=1$). In this study the parameter $\mpdot$ assumes two different values that  correspond to a relatively low and a relatively high mass accretion rate with respect to the chosen part of the parameter space (see Tab. \ref{tbl:parameters}). The chosen values were estimated using the empirical results of \cite{2016ApJ...823...48T}. The point-source approximation is justified since most of the accreted material is assumed to be accreted via the poles of the planet or in its radially close vicinity \citep{2006A&A...445..747K,2012ApJ...747...47T}.

Within the considered planetary mass range the intrinsic luminosity is expected to be given by roughly $L_{\rm int}=10^{-5}\,{\rm L}_\odot$ \citep[see, e.g., Fig. 7 in][]{2017A&A...608A..72M}. However, by comparing the corresponding expected intrinsic temperature of the simulated planets with their accretion temperature, we find that the latter is strongly dominating. In particular, the resulting effective temperature of our simulated planets changes only by a few percent, and thus we neglect the contribution from the intrinsic luminosity in this study.

The density distribution of the CPD is defined in Eq. \ref{eq:general_density_distribution} and uses the subscript $n={\rm p}$. Its corresponding surface density and scale height are then given by
\begin{equation}
\Sigma_{\rm p}(r) = \Sigma_{\rm 0;\,p} \left( \frac{r}{r_{\rm 0;\,p}}\right)^{\beta_{\rm p} - \alpha_{\rm p}} \exp\left\{ -\left(\frac{r}{r_{\rm 0;\,p}} \right)^{2+\beta_{\rm p} + \alpha_{\rm p}}\right\},
\end{equation}
and
\begin{equation}
h_{\rm p}(r) = h_{\rm 0;\,p} \left(\frac{r}{r_{\rm 0;\,p}}  \right)^{\beta_{\rm p}},
\end{equation}
respectively. Contrary to the observation-based choice of CSD parameters, the CPD parameters are based on results of hydrodynamic simulations. While these parameters may differ in different simulations and studies, we tried to choose generic values for our models. The chosen parameters are $\alpha_{\rm p}=2.5$ \citep[compare with][]{2012ApJ...747...47T,2017ApJ...842..103S}, $h_{\rm out;\,p}=h_{\rm p}(r_{\rm out}) = 0.5\,r_{\rm out}$ \citep[compare with][]{2009MNRAS.397..657A}, and 
\begin{equation}
r_{\rm out;\,p} = 0.6\,R_{\rm Hill} \quad {\rm with} \quad R_{\rm Hill} = \rp \left( \frac{\Mp}{3M_*}\right)^{1/3}
\end{equation}
for the outer radius of the CPD \citep[compare with][]{2017ApJ...842..103S}. The parameter $\beta_{\rm p}$ is obtained via the relation $\alpha = 3\,(\beta-0.5)$ \citep{Shakura_1973}. The parameter $M_{\rm CPD}$ typically reaches a value of ${\sim}10^{-3}\,{\rm M}_{\rm J}$ for a Jupiter-mass planet \citep{2017ApJ...842..103S} and is one of the parameters varied within this study. The reference radius for the CPD is chosen depending on the size of the Hill sphere in such a way that the density of the CPD reaches a value well below the ambient gap density to prevent abrupt density changes at the outer edge of the CPD. The exact values of these parameters can be found in Tab. \ref{tbl:parameters}.

The density distribution of the CPD is then added to the density distribution of the CSD and in a final step the region around the planet within its corresponding sublimation radius is cleared from all of its dust (i.e., the dust density is set to zero). However, the inner radius of the CPD $r_{\rm in;\,p}$, which is defined by the sublimation radius, has to be determined for each of the models individually before the actual MCRT simulations can be performed, which is   described in Sect. \ref{sec:determination_r_in_cpd}. In total, we simulate $72$ different models among which we vary $\rp$ (three values), $\Mp$ (three values), $\mpdot$ (two values), $\mcpd$ (two values), and $\mcsd$ (two values). Each model is then analyzed at three different wavelengths $\lambda$, which eventually results in a total of $216$ fully performed and subsequently analyzed simulations.

\begin{table}
\begin{center}
\begin{tabular}{lll}
\toprule
\textbf{Parameter} & \textbf{Variable} & \textbf{Value(s)}\\
\cmidrule{1-3}
\cmidrule{1-3}
\underline{Star:}  & &  \\
Luminosity & $L_*\,\left[ {\rm L}_\odot\right]$& $0.7$ \\
Temperature &  $T_*\,\left[{\rm K} \right]$& $3800$\\
Mass & $M_*\,\left[{\rm M}_\odot \right]$& $0.5$\\
\vspace{0.2em}Distance & $d\,\left[{\rm pc} \right]$& $140$\\
\underline{Circumstellar disk:}  & &  \\
Inner radius & $r_{\rm in;\,s}$ & sublimation\\
Outer radius & $r_{\rm out;\,s}\,\left[ {\rm au}\right]$ & $300$\\
Reference radius & $r_{\rm 0,\,s}\,\left[ {\rm au}\right]$ & $100$\\
Reference scale height & $h_{\rm 0;\,s}\,\left[ {\rm au}\right]$ & $7$ \\
Total mass$^*$ & $M_{\rm CSD}\,\left[ {\rm M}_\odot\right]$ & $0.01$ and $0.001$\\
Compactness parameter& $\alpha_{\rm s}$ & $1.8$\\
Flaring parameter & $\beta_{\rm s}$ & $1.1$\\
\vspace{0.2em}Inclination& $i_{\rm s}\,\left[ {}^\circ\right]$&$0$\\
\underline{Planet:}  & &  \\
Mass$^*$ & $M_{\rm p}\,\left[\mj \right]$ & $0.25$, $0.5$, and $1$\\
Mass accretion rate$^*$ & $\mpdot\,\left[\mj/{\rm yr} \right]$ & $4{\cdot} 10^{-7}$ and $2{\cdot} 10^{-6}$\\
Accretion efficiency & $\eta$ & $1$\\
Radius & $R_{\rm p}\,\left[\rj \right]$ & $1$\\
\vspace{0.2em}Radial position$^*$ & $r_{\rm p}\,\left[ {\rm au}\right]$ & $5$, $10$, and $50$\\
\underline{Circumplanetary disk:}  & &  \\
Inner radius$^{**}$ & $r_{\rm in;\,p}$ & sublimation\\
Outer radius$^{**}$ & $r_{\rm out;\,p}\,\left[ \rhill\right]$ & $0.6$\\
Reference scale height$^{**}$ & $h_{\rm 0;\,p}\,\left[r_{\rm out;\,p} \right]$ & $0.5$\\
Reference radius$^{**}$ & $r_{\rm 0;\,p}\,\left[ \rhill \right]$ & \hspace{-0.8em} $
        \begin{cases} 
     0.03  &\text{if } r_{\rm p}{=}5\,{\rm au}\\
     0.05  &\text{if } r_{\rm p}{=}10\,{\rm au}\\
     0.1  &\text{if } r_{\rm p}{=}50\,{\rm au}\\
   \end{cases}$\\
Total mass$^*$ & $M_{\rm CPD}\,\left[ \mj\right]$ & $0.1{\cdot}10^{-3}$ and $1{\cdot}10^{-3}$\\
Compactness parameter & $\alpha_{\rm p}$ & $2.5$\\
Flaring parameter & $\beta_{\rm p}$ & $4/3$\\
\vspace{0.2em}Inclination& $i_{\rm p}\,\left[ {}^\circ\right]$&$0$\\
\underline{Gas and dust:}  & &  \\
Viscosity parameter & $\visc$ & $0.001$ \\
Mass ratio & $M_{\rm gas}:M_{\rm dust}$ & $100:1$ \\
\bottomrule
\end{tabular}
\caption{Collection of selected model parameters. For details, see Sect. \ref{sec:setup}.\\${}^* $Varied parameter. ${}^{**}$Parameter model-dependent.}
\label{tbl:parameters}
\end{center}
\end{table}

 \subsubsection{Dust model}
While dust growth and settling of dust grains has a major impact on the grain size distribution close to the midplane of curcumstellar disks \cite[e.g.,][]{2007A&A...469..963P,2008A&A...489..633P,Madlener_2012,Grafe_2013,2021AJ....161..238W}, the disk layers traced in the VIS/NIR wavelength range are dominated by small interstellar medium-like dust grains \cite[e.g.,][]{2003ApJ...588..373W,2011A&A...527A..27S}.
 
For this reason we assume spherical dust grains with radii $a$ ranging from $5\,$nm to $250\,$nm that follow a grain size distribution $dn \sim a^{q_g} da$ with grain size distribution exponent $q_g=-3.5$ \citep{1977ApJ...217..425M}. With a bulk density of $\rho_\text{bulk}=2.5\,\text{g}\,\text{cm}^{-3}$, dust grains consist of $f_{\textrm Si}=62.5\,\%$ astronomical silicate and $f_\text{Gr}=37.5\,\%$ graphite,  for which we apply the $1/3\,$:$\,2/3$ approximation \cite[$f_{\parallel\text{-Gr}}\,$:$\,f_{\perp\text{-Gr}}$=1\,:\,2;][]{1993ApJ...414..632D}. The corresponding wavelength-dependent refractive indices of these components are obtained from \cite{1984ApJ...285...89D}, \cite{1993ApJ...402..441L}, and \cite{2001ApJ...548..296W} and used to calculate corresponding cross-sections under the assumption of Mie theory by using the code miex \citep{2004CoPhC.162..113W}. In a final step the mean optical properties of dust grains are computed \citep{Wolf_2003}. The ratio of gas to dust is set to the canonical value of $100{:}1$.

\subsection{Simulation grid}  
\label{sec:grid}
The model is embedded in a grid that is defined in spherical coordinates. It is centered around the star, has an inner radius $r_{\rm in;\,s}$   defined by the sublimation radius of dust using a sublimation temperature of $T_{\rm subl}=1500\,$K, and an outer radius $r_{\rm out;\,s}$, which is chosen large enough to ensure that the density drops sufficiently and the disk becomes optically thin at large radii. For the purpose of this paper, two grid regions have been defined:  a highly resolved region around the planet and  a region of low resolution far away from the planet. 

The high-resolution region centers around the planet and extends up to the Hill radius. Within this region, we use 61 cells in each of the directions  $r$, $\theta$, and $\phi$. Due to the high resolution, the cells in the vicinity of the planet strongly resemble a Cartesian grid with flat cell borders. The planetary cell (i.e., the cell whose center corresponds to the center of the planet) has a cell width that is defined by the diameter of the planet, thus barely fitting the planet inside it. Along the three orthogonal axes, cell widths  grow exponentially with a constant step-factor. Outside the high-resolution region, the cell widths are constant for both angular coordinates $\theta$ and $\phi$ and their respective widths are chosen such that they are larger than the widths of the adjacent cells in the high-resolution region. Combined, this results in 172 cells in the $\theta$-direction and 174 cells in the $\phi$-direction. Contrary to the linear sampling angular coordinates in the low-resolution region, we use a logarithmic sampling of cell borders in the  $r$-direction. To this end, we first define a step factor for the low-resolution region of $s_f=1.05$, and determine a radial width of the innermost cell that is as close as possible to but lower than $\Delta r_{\rm in;\,s} = 0.02\,r_{\rm in;\,s}$, which is done in such a way that ensures a smooth transition between the low- and high-resolution region in $r$-direction. We then increase the cell width in the $r$-direction exponentially, starting from the innermost cell close to the star and moving  outward with the chosen constant step factor until  the high-resolution region is reached. At large enough radii, even outside the highly resolved Hill sphere, the low-resolution region begins again. Here, cell widths also grow exponentially according to $s_f$, thus ensuring a smooth transition from the high- to the low-resolution region. As a consequence, the number of cells in the $r$-direction depends on the position and size of the Hill sphere of the planet, and ranges from $319$ to $344$ cells among all simulations. 

A consequence of setting up this specific grid structure is the abundance of irregularly shaped cells in regions where one coordinate has a fine sampling and another coordinate a coarse sampling, leading to very oblate and prolate cells. Unfortunately, such cells may pose a problem to MCRT simulations when placed in an optically thick region far away from any source of radiation. In these regions of the model space, photon package counts are relatively low, which leads to a poor representation of potential photon paths in these irregularly shaped cells. Consequently, the determined temperature  suffers from a relatively high level of noise. While the noise does not significantly change the overall integrated flux of these regions, it unfortunately has an impact on their spectra.

Even though it is not possible to completely eliminate the noise, there are methods for reducing it significantly. It is possible to check the effectiveness of such a method by generating flux maps and simply verifying whether or not flux maps of pure thermal emission are noisy. The method we apply   to overcome this issue is to overlay these particular regions of higher resolution outside the Hill sphere, which are regions of irregularly shaped cells, with a coarse grid. Next, we combine multiple adjacent irregularly shaped cells into groups of cells, such that the shape of the group is more regularly shaped. During the temperature calculation, whenever a photon package traverses a cell that belongs to such a cell group, the temperature of the whole group is increased equally and simultaneously. In particular, the deposited energy is distributed equally between all the dust grains present in the corresponding group of cells. In order to achieve this, we assign a weight to every member of a group of cells, which equals the ratio of dust grains the cell contains   to the total number within the whole group, and distribute the deposited energy accordingly. To this end, we define the groups of cells as follows. In both angular directions, the coarse grid is linearly sampled with a cell width of ${\sim}1\,\deg$. In the radial direction we use the same number of linearly sampled coarse grid cells. This particular definition has proven to result in a substantially reduced level of noise.

 \subsection{Instruments}
 \label{sec:instruments}

Detecting and characterizing embedded planets requires state-of-the-art instrumentation. In this study we particularly focus on two promising instruments that are installed on SPHERE:  IRDIS and ZIMPOL. These instruments allow us to perform high-contrast observations in the  VIS/NIR wavelength range using broadband filters in the classical imaging mode. Additionally, the use of a coronagraph further boosts their performance at detecting even comparably dim objects in the vicinity of a bright stars. To benefit from high planetary temperatures, we primarily focus on short observational wavelengths. Furthermore, we study the impact of the inner working angle (IWA) of coronagraphs on observations of close-in embedded planets. In particular, we measure the signal strength of simulated embedded planets that are hidden inside the IWA or located close to the its rim to assess the potential loss of detectable planets due to the application of a too small IWA. For this purpose, we simulate coronagraphs with the smallest possible IWA that are offered for their corresponding instruments and filters. Chosen instruments, modes, broadband filters, wavelengths, coronagraphs, and their corresponding IWAs are summarized in Tab. \ref{tbl:instrument_sphere}.

\begin{table*}
\begin{center}
\begin{tabular}{lllrlr}
\toprule
\textbf{Instrument} & \textbf{Mode} & \textbf{Filter} & \textbf{Wavelength} & \textbf{Coronagraph} & \textbf{IWA}\\
\cmidrule{1-6}
ZIMPOL & Zimpol\_I   & R\_PRIM & $\lambdaS$ & V\_CLC\_S\_WF & 93\,mas  \\
IRDIS  & IRDIS alone & B\_J    & $\lambdaM$ & N\_ALC\_YJ\_S & 80\,mas  \\
IRDIS  & IRDIS alone & B\_Ks   & $\lambdaL$ & N\_ALC\_Ks    & 200\,mas  \\
\bottomrule
\end{tabular}
\caption[bla]{Instrument specifications. Data was taken from the User Manual of periods 107 and 108. \footnotemark}
\label{tbl:instrument_sphere}
\end{center}
\end{table*} \footnotetext{User Manual of period 107 \& 108: \url{https://www.eso.org/sci/facilities/paranal/instruments/sphere/doc/VLT-MAN-SPH-14690-0430_P107_jan_2021_zwa.pdf}}

\section{Results and discussion}
\label{sec:results_and_discussion}
In this section we present the results of simulating embedded accreting planets, as described in Sect. \ref{sec:setup}. First, we present the findings regarding the determined inner radius of the CPD as well as its resulting temperature distribution. Then, we discuss resulting flux maps of all simulated systems regarding the relevance of their different sources of radiation. Subsequently, we describe a procedure with which we analyze all flux maps with respect to the detectability and characterization of the embedded planets and their CPDs. This method is applied and its results are used as a basis for discussing the relevance of different model parameters and deciding which SPHERE instrument, in terms of its observational wavelength, is best suited for characterizing which model parameter.

\subsection{Determination of the CPD inner radius}
\label{sec:determination_r_in_cpd}
As mentioned before, the inner radius $\rin$ is defined by the sublimation radius of dust in the vicinity of the planet (see Sect. \ref{sec:planetary_and_cpd_model}). In general, its determination requires an iterative procedure that aims to narrow down the radius from both sides. On the one hand, the radius has to be large enough to eliminate the possibility of the onset of unexpected sublimation of dust during the run of a simulation. On the other hand, the radius has to be small enough to enable the dust to reach temperatures as close as possible to its sublimation temperature. 

The temperature of any cell of the grid emerges as a consequence of exposure to direct or indirect stellar and planetary radiation. However, as expected at the sublimation radius of the planet we find that the contribution from the star is negligible compared to that of the planet. Therefore, during the determination of the inner radius of the CPD $\rin$ we only consider planetary radiation. Since MCRT simulations of highly optically thick regions are very time consuming and require the calculation of a high number of individual interactions per photon package, which in general should not be limited, we use $N_\gamma=10^5$ photon packages per test run, which is the number of photon packages that were emitted from the planet to determine the resulting temperature distribution. For every test run we set a value for $\rin$, beginning with the value $\rin=R_{\rm p}$, and calculate the resulting temperature distribution. Doing so, we are able to narrow down the value for $\rin$ from both sides. A value for $\rin$ is accepted, if (i) it did not result in the onset of unexpected sublimation during the test run, (ii) it is narrowed down to its actual value with an accuracy of $<10\%$ or $\Delta r<1 R_{\rm p}$, and (iii) it is narrowed down to its actual value with an accuracy of $\Delta r<2 R_{\rm p}$.
The reasoning for these criteria is as follows. The first criterion is generally required in any MCRT simulation of this type which determines equilibrium temperature distributions. The second and third criteria define relative and absolute levels of accuracy, respectively, with which $\rin$ has to be determined. However, since the grid has a finite resolution, the second criterion can also be satisfied if $\rin$ is determined down to the accuracy of the width of the planetary cell, which  in terms of its volume is the smallest cell in the vicinity of the planet. It is worth mentioning that this method results in inner radii that may be rather slightly overestimated than underestimated, which is a direct consequence of the first mandatory criterion. 

In general, the value of $\rin$ depends on the model parameters and needs to be determined individually for each simulation. However, we find that determining $\rin$ for models with a high value for $\mcsd$ first and then using the same sublimation radius for simulations with a lower value for $\mcsd$, where all other underlying parameters coincide, is justified. In order to test the validity of this approach, we compare the maximum temperature of dust just outside the sublimation radius of all models, which only differ in their assumed parameter value of $\mcsd$, and find that it   differs by $<3.3\,\%$ for all simulations. This suggests only a weak dependence of $\rin$ on $\mcsd$ and affirms the viability of our approach. 
A list containing all determined inner radii $\rin$ can be found in the Appendix in Tab. \ref{tbl:r_in_cpd}.

\paragraph*{Dependence on $\rin$:}\quad \\
Results from Tab. \ref{tbl:r_in_cpd} suggest clear trends that describe the effect of different model parameters on $\rin$. When keeping the remaining parameters constant, we find that $\rin$ increases if $\rp$ decreases, $\Mp$ increases, $\mpdot$ increases, or $\mcpd$ increases. There are two underlying processes that can explain these trends. First,  an increase in $\lacc$ results in an extended sublimation radius. Second,  an increased density at the inner rim of the CPD causes an enhanced back-warming effect, which leads to a larger sublimation radius. The effect can be described as follows. By increasing the density in the region behind the illuminated surface of the inner rim of the CPD, photons that cross that surface now have an increased probability of leaving the CPD through that same surface, as the probability of traveling through the already optically thick CPD and leaving it on the other side decreases.  In a state of equilibrium this leads to an amplified radiation field at the illuminated surface of the inner rim of the CPD, and thus to an increased dust temperature. A more in-depth description of the back-warming effect including a one-dimensional derivation is presented in the Appendix in Sect. \ref{sec:back-warming}.

\subsection{Temperature distribution}

  \begin{figure*}
   \centering
   \includegraphics{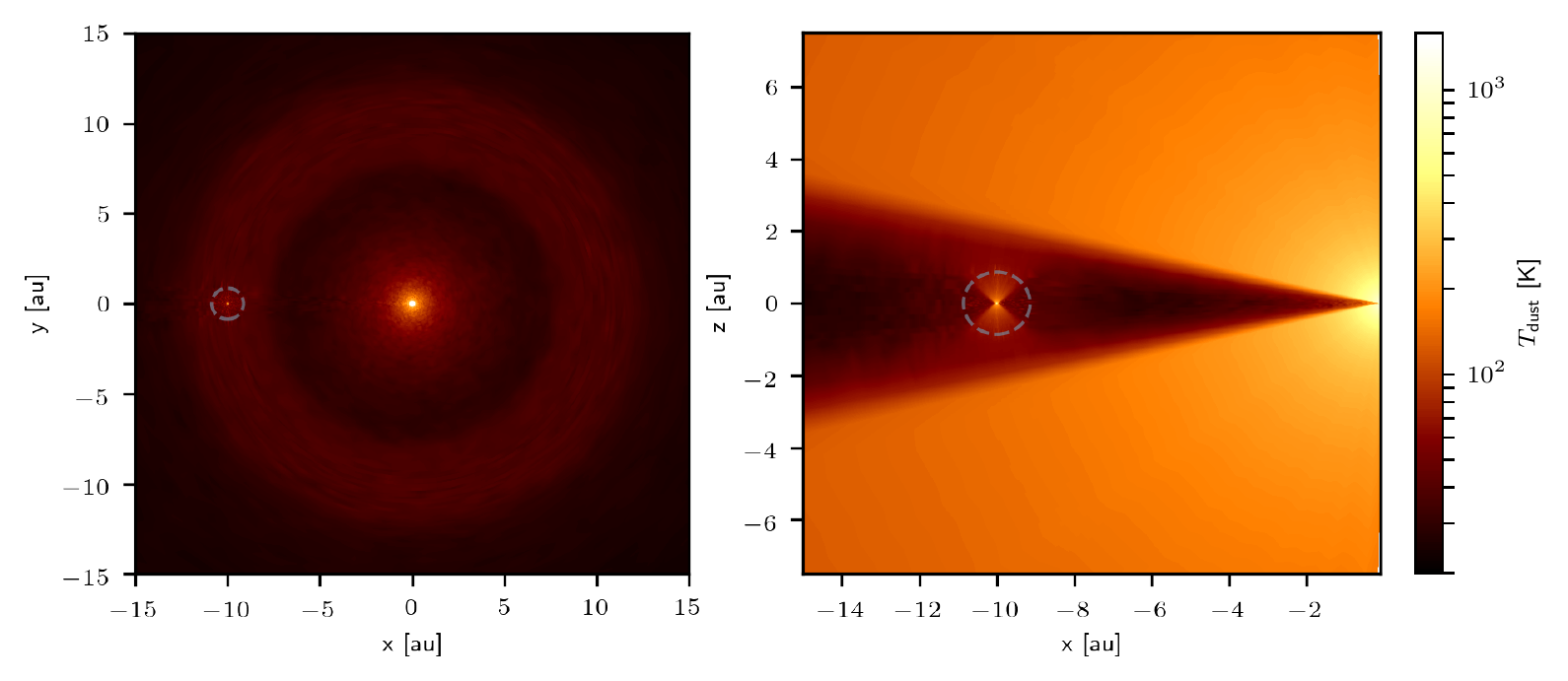}
   \caption{Temperature distribution in the midplane (left) and in a vertical cut through the midplane at the position of the planet (right) for a model with $\rp{=}10\,$au, $\Mp{=}1\,\mj$, $\mpdot{=}4\cdot10^{-7}\,\mj/$yr, $\mcpd{=}0.1\cdot 10^{-3}\,\mj$, and $\mcsd{=}0.001\,\msun$. The dashed gray circle with radius $r=\rhill$ indicates the size of the Hill sphere.}
              \label{fig:temperature_distribution}
   \end{figure*} 

Before the detectability of planets and their CPDs can be discussed, we determine the underlying temperature structure in a self-consistent manner. In order to do that for all models presented in Tab. \ref{tbl:parameters}, we set $\rin$ to the values deduced in the previous section and summarized in Tab. \ref{tbl:r_in_cpd}. Additionally, we use $r_{\rm in;\,s}=0.07\,$au for all simulations. The dust emission properties of cells were precalculated for 501 logarithmically sampled temperature values between $2.7$ and $3000\,$K and for 132 wavelenghts between $50\,$nm and $2\,$mm. For each model the temperature calculation is performed in two steps. In the first step, the star emits a total of $N_\gamma=10^8$ photon packages, which increase the temperature in the whole model space. In the second step, the  less luminous planet emits $N_\gamma=10^7$ photon packages, which mainly leads to a significant change in the temperature distribution in the immediate vicinity of the planet. However, there are two exceptions made in the case of models with the lowest accretion luminosity $\lacc$ and the greatest potential for high CPD densities (i.e., for the two models with the lowest values for $\Mp$, $\mpdot$, and $\rp$ and the highest value for $\mcpd$), where only $N_\gamma=10^6$ photon packages are used in order to avoid very long simulation times. It was verified that  these simulations, nonetheless, resulted in sufficiently smooth flux maps (see Sect. \ref{sec:grid}). Therefore, the reduced number of photon packages is not expected to reduce the quality of the results.

Figure  \ref{fig:temperature_distribution}  shows an example of the resulting temperature distribution for a model with $\rp{=}10\,$au, $\Mp{=}1\,\mj$, $\mpdot{=}4\cdot10^{-7}\,\mj/$yr, $\mcpd{=}0.1\cdot 10^{-3}\,\mj$, and $\mcsd{=}0.001\,\msun$ in the midplane (left) as well as in a vertical cut through the midplane at the azimuthal position of the planet (right). The temperature distribution reaches its highest values in the vicinity of the star and the planet, and rapidly falls off at greater distances from the two objects. This temperature decrease is particularly strong in the region of the planet (i.e., in the CPD) effectively leading to a spatially confined impact of the planet on the temperature distribution. Moreover, the vertical cut shows a strong shadowing effect in the sense that the immediate planetary light is blocked by the CPD, and thus casts a shadow inside the gap, which  keeps an otherwise hot region in the gap at a lower temperature. This effect is enhanced by the geometry of the CPD, which has an opening angle of ${\sim}90^\circ$ \citep[e.g.,][]{2017ApJ...842..103S}, which is much wider than that of the CSD.

\subsection{Ideal observations}
\label{sec:sphere_ideal}

In order to determine the detectability of planets and their CPDs we have to calculate and evaluate flux maps for all models. There are different sources for radiation that constitute the total flux map  detected by an observer: the star, the planet, and thermal radiation of dust. It is important to note that line emission may also play an important role in detecting embedded accreting planets; however, in this study we focus on the continuum emission of dust. In the case of thermal radiation, we additionally distinguish between unscattered radiation ($N_{\rm sca}=0$) and self-scattered radiation ($N_{\rm sca}\geq1$). The latter is often understimated and omitted; however, we  show that it is in general important in the case of embedded planets.

In order to determine the contribution of the different sources of radiation to the total flux map, all sources are calculated individually and their contributions are subsequently added up. For each model, wavelength, and radiation source the flux map is calculated individually, thus any declared number of photon packages refers to each  of these flux maps individually. For the simulation of stellar and planetary radiation we use $N_\gamma=10^7$ and $N_\gamma=10^6$ photon packages, respectively. The unscattered thermal dust radiation is calculated with a ray-tracing algorithm that   integrates and attenuates thermal dust radiation along the lines of sight in the direction of the detector. This method guarantees a smooth flux profile, but does not take into account self-scattering of thermal dust radiation. The process of self-scattering is simulated with $N_\gamma=10^8$ photon packages that are distributed among all dust containing grid cells proportionally to their respective total luminosity, meaning that more luminous cells emit more photon packages. In addition, our method assures that every cell  emits exactly its corresponding total luminosity.

  \begin{figure*}
   \centering
   \includegraphics{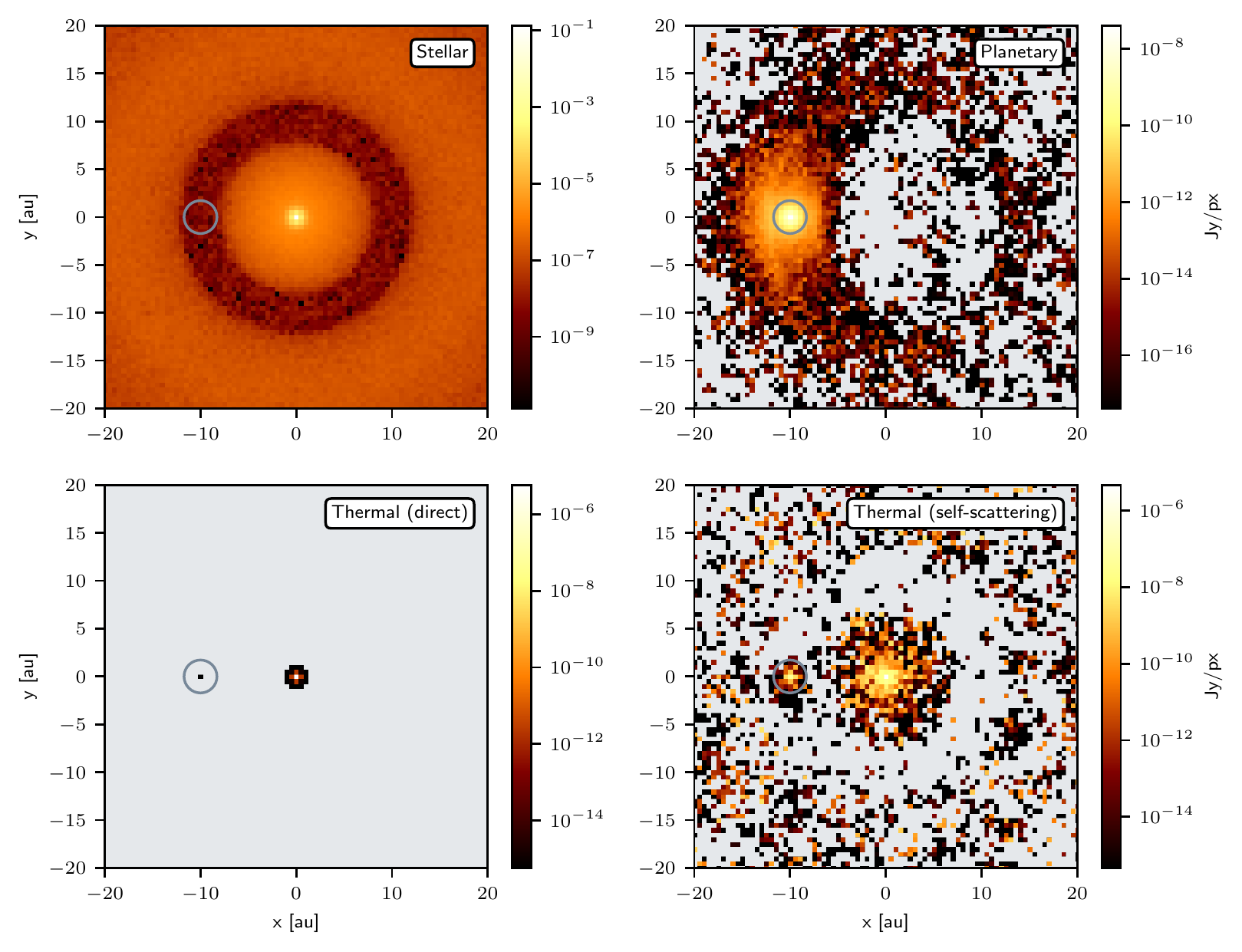}
   \caption{Flux maps at $\lambda=653\,$nm for a model with $\rp{=}10\,$au, $\Mp{=}1\,\mj$, $\mpdot{=}4\cdot10^{-7}\,\mj/$yr, $\mcpd{=}0.1\cdot 10^{-3}\,\mj$, and $\mcsd{=}0.001\,\msun$. The source for each of these plots is noted in its  upper right corner. The solid gray circle with radius $r=2\,\rhill$ indicates the position of the planet.}
              \label{fig:four_fluxes}
   \end{figure*} 

Figure \ref{fig:four_fluxes} shows examples of ideal flux maps for each of the four different sources at $\lambda=653\,$nm, which add up to the total flux of the model. Here, the model is described by $\rp{=}10\,$au, $\Mp{=}1\,\mj$, $\mpdot{=}4\cdot10^{-7}\,\mj/$yr, $\mcpd{=}0.1\cdot 10^{-3}\,\mj$, and $\mcsd{=}0.001\,\msun$, and thus corresponds to the temperature distribution  shown in Fig. \ref{fig:temperature_distribution}. Every map consists of $1201\times1201$ pixels, each with a width of ${\sim}0.5\,$au, with the star located in the central pixel. The resulting flux map due to direct and scattered stellar light (upper left plot) shows some striking features: first, a peak of flux at the position of the star; second, an overall decline of flux toward greater radii; and third, the gap carved by the planet into the CSD. The planetary flux map (upper right plot) shows a strong peak at the position of the planet and a decline in flux toward greater distances from the planet, meaning that the signal is broadened due to the optical depth between the planet and the observer. The direct (i.e., unscattered) thermal radiation of dust (lower left plot) has its peak at the central pixel and quickly falls of toward larger radii. A second local maximum is reached at the position of the planet; however, its contribution is generally the weakest. The self-scattered radiation (lower right plot), which is composed of thermal dust radiation that has scattered at least once, reaches its highest value at the central pixel, overall falls off toward larger radii, has a bright ring just outside a clear gap, and shows a stronger signal at the position of the planet and its CPD than anywhere else inside the gap. Comparing the resulting contributions of these four sources we find that in the vicinity of the star scattered stellar light dominates the flux making all other contributions negligible. In the vicinity of the planet, however, stellar radiation, planetary radiation, and self-scattering of thermal dust radiation all play a crucial role. The contribution due to direct thermal radiation is comparably low in the considered wavelength range. Finally, it is important to note that in later sections we take into account the level of noise present across all flux maps in order to avoid overinterpreting any results with weak planetary and CPD signals.

\begin{figure}
   \centering
   \includegraphics{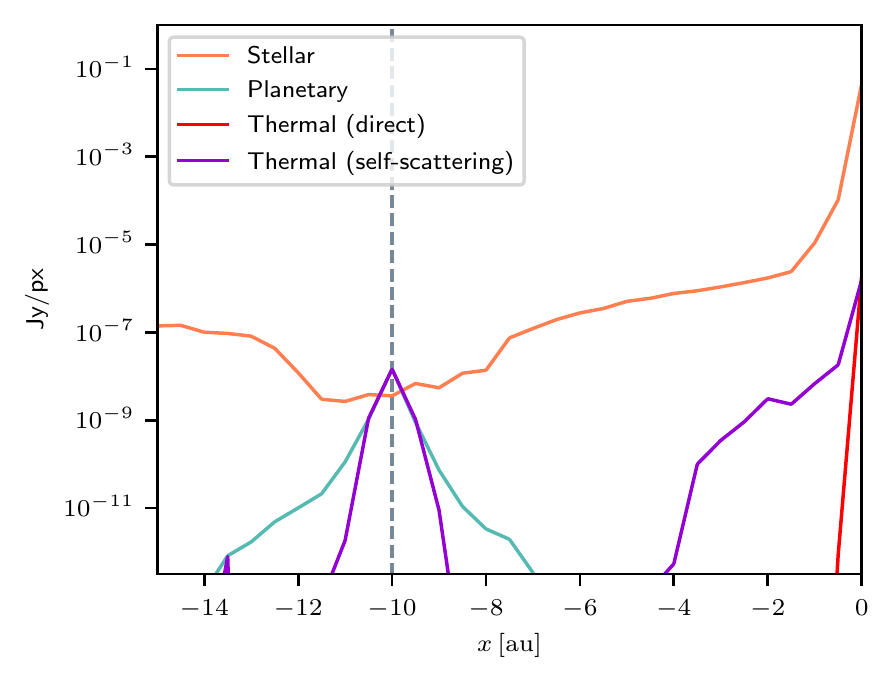}
   \caption{Vertical cut through various flux maps at $\lambda=653\,$nm along the x-axis for a model with $\rp{=}10\,$au, $\Mp{=}1\,\mj$, $\mpdot{=}4\cdot10^{-7}\,\mj/$yr, $\mcpd{=}0.1\cdot 10^{-3}\,\mj$, and $\mcsd{=}0.001\,\msun$. The gray vertical line gives the position of the planet. For details, see Sect. \ref{sec:sphere_ideal}.}
              \label{fig:flux_x_cut}
   \end{figure}

For comparison purposes, Fig. \ref{fig:flux_x_cut} shows a cut of these flux maps along the x-axis. In particular, a stripe centered at the x-axis with a width of $\Delta y = 2\rhill$ was averaged across the y-axis in order to reduce noise and better encompass the effect of the planet and the CPD. The plot clearly shows that despite the overall dominance of stellar flux, both the planetary radiation and the self-scattered thermal dust radiation may dominate the signal at the position of the planet. This emphasizes the need for a proper treatment of self-scattering in radiative transfer simulations in order to simulate embedded accreting planets.

\begin{figure}
   \centering
   \includegraphics{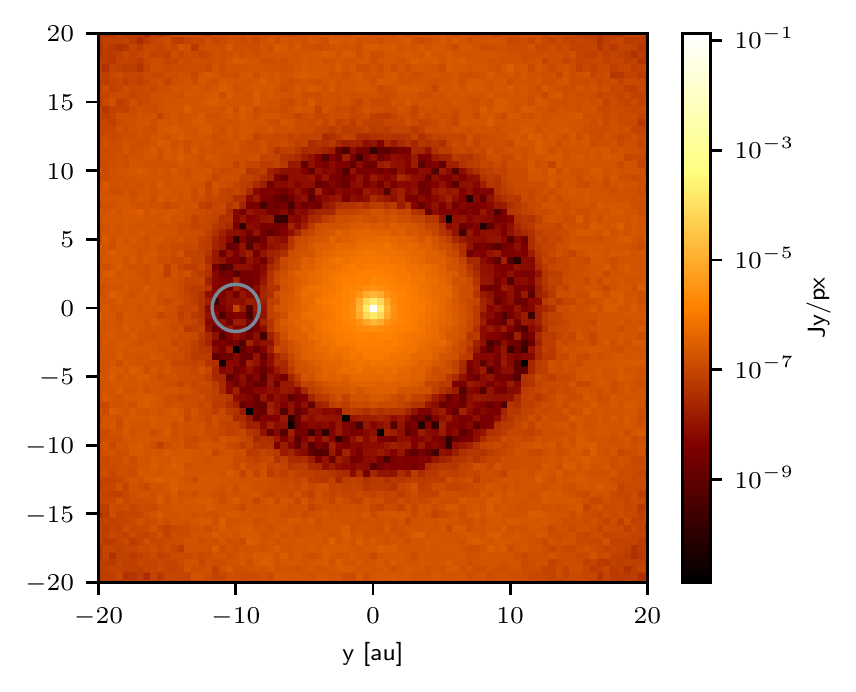}
   \caption{Total flux map at $\lambda=653\,$nm for a model with $\rp{=}10\,$au, $\Mp{=}1\,\mj$, $\mpdot{=}4\cdot10^{-7}\,\mj/$yr, $\mcpd{=}0.1\cdot 10^{-3}\,\mj$, and $\mcsd{=}0.001\,\msun$. The solid gray circle with radius $r=2\,\rhill$ indicates the position of the planet.}
              \label{fig:total_sum_four_fluxes}
   \end{figure}

Finally, Fig. \ref{fig:total_sum_four_fluxes} shows the total sum of all flux maps that are displayed individually in Fig. \ref{fig:four_fluxes}. It shows that for the most part the flux map is dominated by stellar radiation. There is also a very high contrast between the region around the star and the rest of the map, showing the need for    methods that suppress the stellar contribution in order to directly detect a planet.

\subsection{Convolved observations}
\label{sec:sphere_convolved}
In order to systematically assess the expected strength of the signal of the combined system of planet and CPD, we perform the following steps. First, the total flux maps are convolved with a point-spread-function (PSF) that is defined by a Gaussian with a full width at half maximum given by $FWHM=1.22 \lambda/D$, with $D=8.2\,$m as diameter\footnote{SPHERE is installed on the telescope UT3 of the VLT with a diameter of $D=8.2\,$m: \url{https://www.eso.org/sci/facilities/paranal/instruments.html}}. Second, the Cartesian detector grid is converted to a polar detector grid in order to make use of the high degree of azimuthal symmetry of the observed PPD. Third, three regions are defined, namely an azimuthal region where the combined planetary and CPD signal is the strongest, a clearly defined gap region that is effectively free from any planetary and CPD signal, and a transition region connecting the azimuthal CPD and the azimuthal gap region. Finally, by subtracting the mean azimuthal gap radial profile from the mean azimuthal CPD radial profile we deduce the excess signal strength of the planet and its CPD. \footnote{The planetary and CPD signal is derived from the total flux map. Therefore, it also includes any effect that the presence of the planet and the CPD has on observed scattered light of external sources, for example the star or dust outside the CPD.}

After an ideal (simulated) observation has been convolved using a wavelength-dependent beam size, a polar representation of the resulting map is constructed. The conversion of the Cartesian detector grid to a polar detector grid is performed with the python package CartToPolarDetector that was particularly written for this task. In order to find the polar representation of an originally Cartesian detector grid, the code overlays the Cartesian detector with a chosen polar grid and calculates the polar pixel values. Every Cartesian pixel is assumed to represent a homogeneous flux distribution across its full area. The associated value of a polar pixel is then the sum of all values of overlapping Cartesian pixels, each weighted with its geometrical intersection area with the polar pixel. Under the assumption of a homogeneous flux distribution within a Cartesian pixel, and since this approach does not rely on any method of interpolation, flux is conserved during the conversion. For the polar grid we choose $N_\phi=7200$ and $N_r=3000$ azimuthal and radial pixels, respectively, and place the center of the grid at the position of the star. This setup assures that the polar pixel area is smaller than the Cartesian pixel area in the vicinity of the planet, even in the case $\rp=50\,$au, where the polar planetary pixel covers about half the area of the Cartesian planetary pixel. For more details regarding the python package, see Sect. \ref{sec:package_cart_to_polar} in the Appendix. 

Next, we define three azimuthal regions. The azimuthal CPD region is defined as the smallest azimuthal interval that contains the full wavelength-dependent width of the beam in terms of its $FWHM$ centered at the position of the planet, meaning that the radius that defines this region is given by 
\begin{equation}
\rcpdhat=FWHM/2.
\end{equation}
Within this beam, the planetary pixel always has the highest absolute flux value within each    simulation. Similar to the previous definition, the azimuthal gap region as well as the transition region is defined by a wavelength-dependent radius $\rtranshat>\rcpdhat$ in such a way that the contamination due to a planetary or CPD signal becomes insignificant outside this radius. In general, there are three different effects that have to be considered in order to determine $\rtranshat$, each extending the apparent spread of the planetary and CPD signal. There is, first, the spread due to the extended beam size ($\tilde{r}_{\rm beam}$); second, the spread due to the optical depth between the planet and the observer ($\tilde{r}_{\tau}$); and third, the spread due to the geometrical extent of the CPD ($\tilde{r}_{\rm CPD}$). Apart from the third effect, which arises due to the geometrical extent of the area the signal originates from, the first two act on the signal solely by smearing it out radially. As a result, $\rtranshat$ can be approximated as the sum of these radii given by
\begin{equation}
\rtranshat = \tilde{r}_{\rm beam} + \tilde{r}_{\tau}+ \tilde{r}_{\rm CPD}.
\end{equation}
The last contributor (i.e., the geometrical extent of the CPD) can be estimated based on the position and mass of the star and its companion. To mimic the situation of a real observation, where both the position and mass of an accreting planet are initially not known, we simply use an upper bound for the planetary mass, which in our case is given by $1\,\mj$, and estimate $\tilde{r}_{\rm CPD}$ with the corresponding Hill radius as follows:
\begin{equation}
\tilde{r}_{\rm CPD} = \rp \left( \frac{1\,\mj}{3\,M_*}\right)^{1/3}.
\end{equation}
Both of the other contributions are generally more difficult to estimate as they are wavelength-dependent and in general unconfined. However, an empirical analysis of our data shows that the relation presented in Eq. \ref{eq:rtranshat_def} is reliable for all models and wavelengths, as we make clear in later sections. As a result, we find that 
\begin{equation}
\rtranshat = 2\,FWHM + \tilde{r}_{\rm CPD}
\label{eq:rtranshat_def}
\end{equation}
leads to a practically contamination-free azimuthal gap region in all simulations.
We note that according to this definition, for any planet that is too close to the star, no contamination-free azimuthal gap region can be defined. In particular, the method can only be properly applied to systems that satisfy the condition 
\begin{equation}
\rtranshat < 2\rp.
\label{eq:condition_azimuthal_gap_region}
\end{equation}
Simulated SPHERE observations for long wavelengths and $\rp{=}5\,$au, for instance, may not satisfy this condition (i.e., no clear gap signal can be deduced), which makes the determination of an excess planetary and CPD signal ambiguous. Nonetheless, even for these simulations we determine a signal, as   discussed below. 
Finally, the azimuthal transition region is defined by $\rcpdhat$ and $\rtranshat$ as the region that is neither part of the azimuthal CPD region nor of the azimuthal gap region. 

The resulting $\phi$ interval that covers the azimuthal CPD region has an extent of
\begin{equation}
\Delta \phi_{\rm CPD} = 4 \arcsin\left(\frac{\rcpdhat}{2 \rp} \right).
\end{equation}
In the case that the azimuthal gap region can be defined, it covers an angular extent of 
\begin{equation}
\Delta \phi_{\rm gap} = 2\pi - 4 \arcsin\left(\frac{\rtranshat}{2 \rp} \right).
\end{equation}

\begin{figure*}
   \centering
   \includegraphics{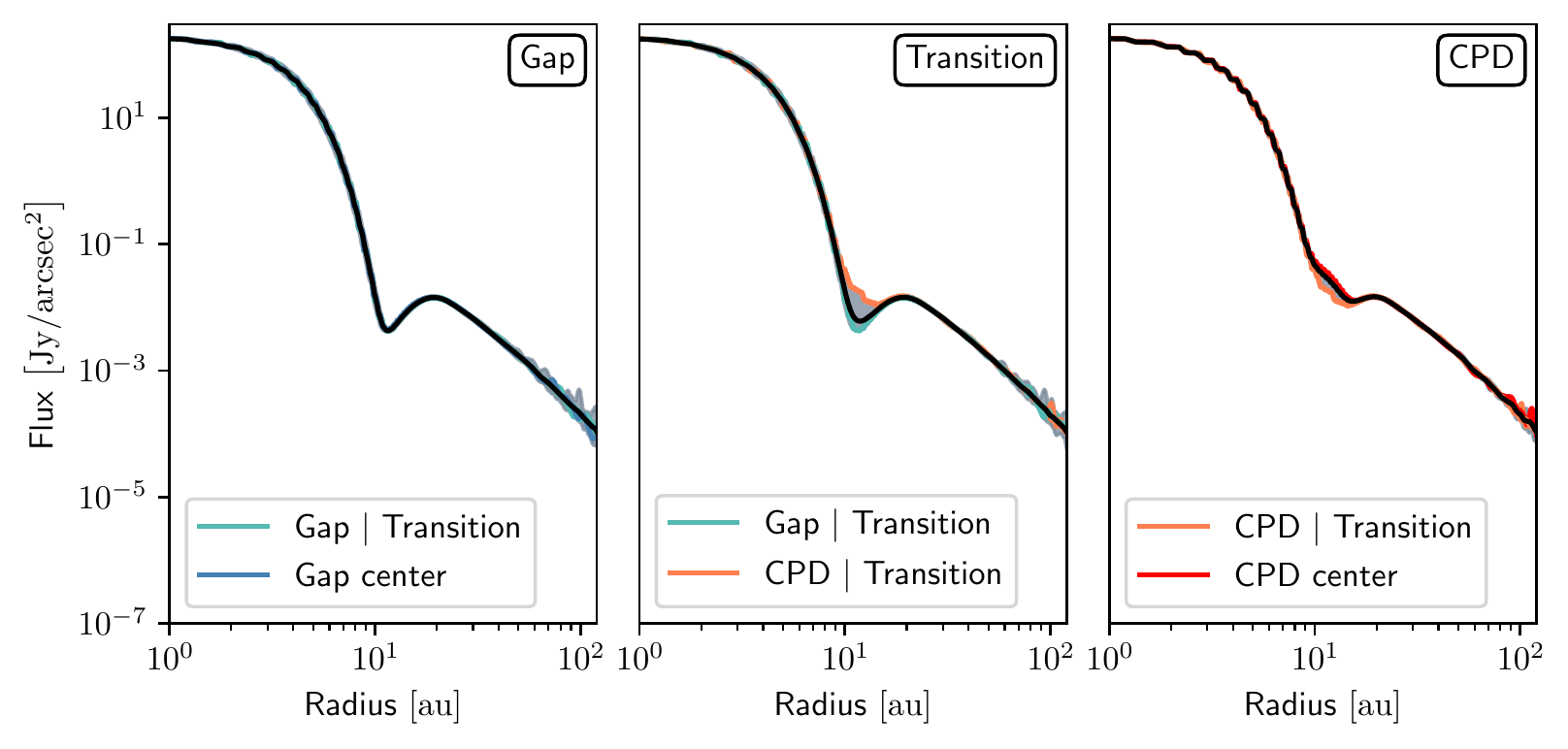}
   \caption{Radial profiles at $\lambda=1.24\,\mu$m for a model with $\rp{=}10\,$au, $\Mp{=}1\,\mj$, $\mpdot{=}4\cdot10^{-7}\,\mj/$yr, $\mcpd{=}0.1\cdot 10^{-3}\,\mj$, and $\mcsd{=}0.001\,\msun$. The gray shaded area highlights the region in the plot where individual radial profiles of their corresponding azimuthal regions (gap, transition, or CPD) lie. The black curve in each panel is the azimuthal mean of the individual radial profiles for their corresponding azimuthal regions. For details, see Sect. \ref{sec:sphere_convolved}.}
              \label{fig:three_regions_2_30}
   \end{figure*}

\begin{figure*}
   \centering
   \includegraphics{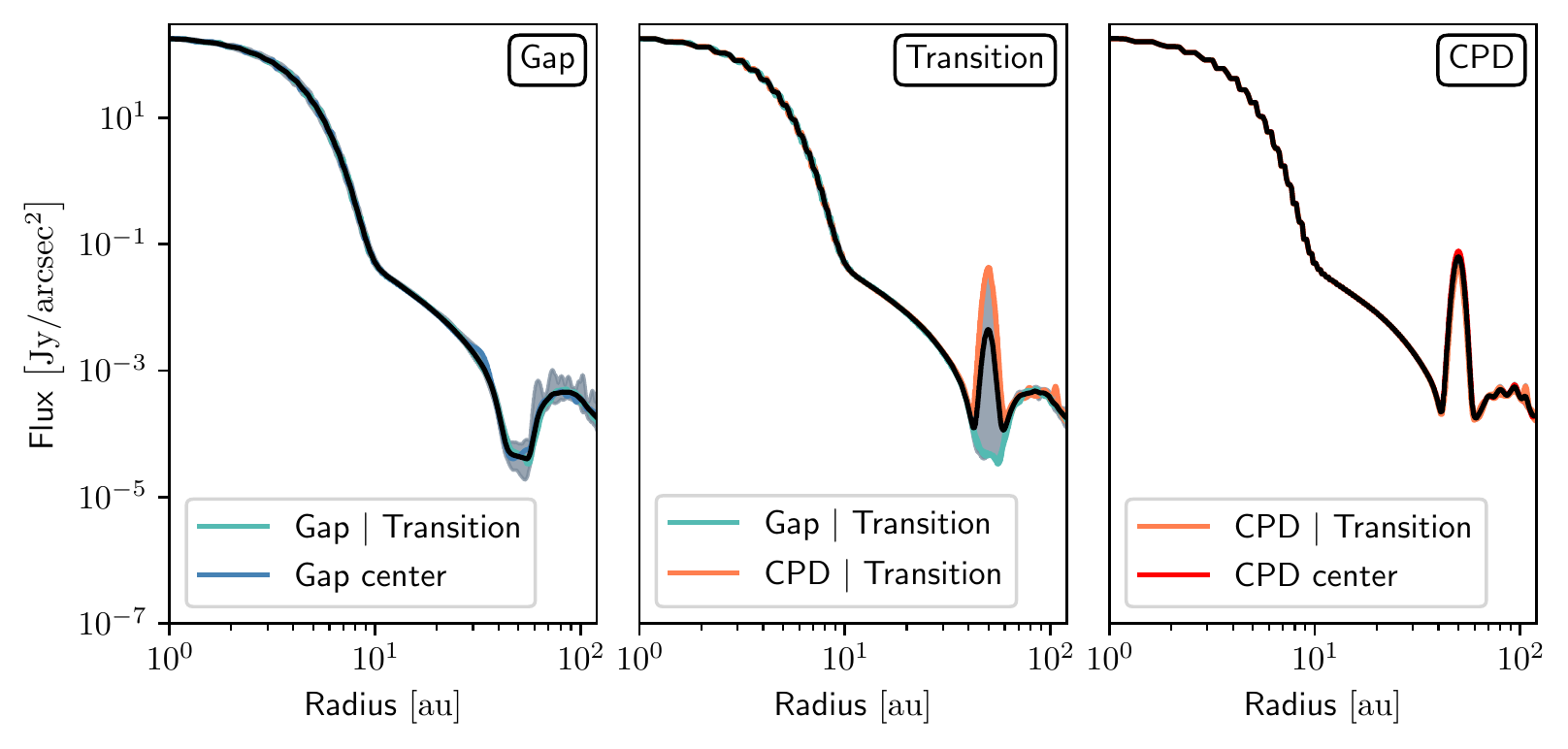}
   \caption{Radial profiles at $\lambda=1.24\,\mu$m for a model with $\rp{=}50\,$au, $\Mp{=}1\,\mj$, $\mpdot{=}4\cdot10^{-7}\,\mj/$yr, $\mcpd{=}0.1\cdot 10^{-3}\,\mj$, and $\mcsd{=}0.001\,\msun$. The gray shaded area highlights the region in the plot where individual radial profiles of their corresponding azimuthal regions (i.e., gap, transition, or CPD) lie. The black curve in each panel is the azimuthal mean of the individual radial profiles for their corresponding azimuthal regions. For details, see Sect. \ref{sec:sphere_convolved}.}
              \label{fig:three_regions_3_30}
   \end{figure*}

Figures \ref{fig:three_regions_2_30} and \ref{fig:three_regions_3_30} show radial profiles for two simulations at $\lambda=653\,$nm, whose parameters differ only in their values of $\rp$. The gray shaded area indicates the region in the plot where individual radial profiles that belong to  one of the three azimuthal regions lie. The black curve is the mean radial profile of its corresponding azimuthal region. Additionally, each subplot highlights radial profiles that are located at the edges of their corresponding azimuthal regions and the left and right subplot also shows the central radial profile for the azimuthal gap and the azimuthal CPD region, respectively. 

Both simulations show a distinct gap profile in their left plots. As expected, the gap depth is wider for the simulation with $\rp=50\,$au compared to that with $\rp=10\,$au. Even though they are  not as smooth as the
black mean curves, the light blue radial profiles (which indicate the border between the gap and the transition region)\  are not contaminated by planetary or CPD flux. Their level of statistical noise arises from   the use of the Monte Carlo method and the limited number of simulated photon packages. Due to the definition of $\rtranshat$, the gap is practically free from any planetary or CPD flux. The gap region adjoins the transition region, which is shown in the middle plots. They clearly show an overall increase in flux going from the azimuthal gap region border (light blue curve) to the border of the azimuthal CPD region (orange curve), which was expected. The azimuthal CPD region is shown in the right plots, which features the strongest signal at the azimuthal and radial position of the planet inside the gap among all individual radial profiles in that azimuthal CPD region. In Fig. \ref{fig:three_regions_2_30}, however, the effect of the planet and its CPD is not strong enough to result in a local maximum at the position of the planet in the radial profile. Nonetheless, by comparison with the mean azimuthal gap profile, it becomes clear that the planet is indeed contributing strongly to the observed local flux value. Another interesting feature can be seen in the azimuthal gap region plot in Fig. \ref{fig:three_regions_3_30}, where a clear kink appears in the radial profile at ${\sim}10\,$au. This kink marks the transition where the contribution through stellar light, which is dominating the radial profile at small radii, becomes secondary to the contribution from the CSD and its embedded objects. In the next step we analyze the obtained mean radial profiles regarding the planetary and CPD signal strength, which leads into a discussion of the detectability and characterizability of embedded planets.

\subsection{Planetary and CPD signal strength}
\label{sec:planetary_and_cpd_signal_strength}
\begin{figure*}
   \centering
   \includegraphics{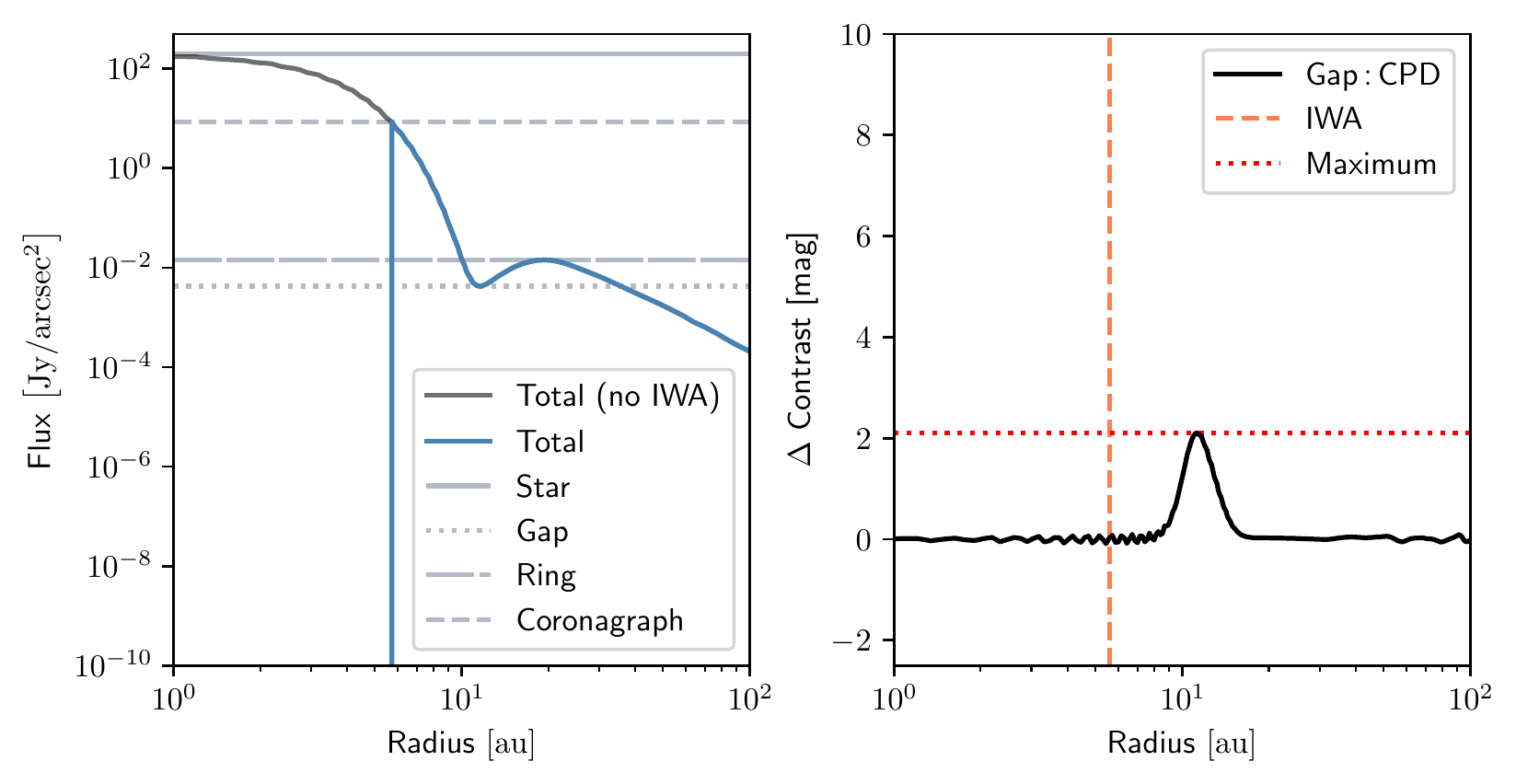}
   \caption{Flux thresholds used in determining contrast levels (left plot) and the contrast between the gap and CPD region (right plot). Results are shown at $\lambda=1.24\,\mu$m for a model with $\rp{=}10\,$au, $\Mp{=}1\,\mj$, $\mpdot{=}4\cdot10^{-7}\,\mj/$yr, $\mcpd{=}0.1\cdot 10^{-3}\,\mj$, and $\mcsd{=}0.001\,\msun$. A high contrast value means higher flux from the CPD region. For details, see Sect. \ref{sec:sphere_convolved}.}
              \label{fig:contrast_definitions_simu2_wl30}
\end{figure*}

The purpose of this section is to describe the method with which the planetary and CPD signal strength as well as other detectable features are quantified, which lays the groundwork for the subsequent discussion on the detectability of embedded planets and the characterization of their properties. To this end, the shape of the mean radial profile of the azimuthal gap region as well as that of the azimuthal CPD region exhibit some common features that can be analyzed. The left plot in Fig. \ref{fig:contrast_definitions_simu2_wl30} shows a typical mean radial profile for the azimuthal gap region (gray curve) and, in addition, various relevant flux levels. As expected, the highest level of flux always originates from the position of the star at $r=0\,$au (gray solid line). As discussed in Sect. \ref{sec:instruments}, we also simulate a coronagraph to study the impact of its inner working angle. In our simplified model of the coronagraph it acts on flux maps by reducing all incoming flux to zero within the full area of the IWA by granting unhindered transmission of flux outside of it. For the resulting exemplary observed radial profile (blue curve), we measure the flux level that is reached just outside the IWA of the coronagraph (short dashed line). If the radial profile features a gap, as can be seen in this plot, we also measure the minimum flux level inside the gap (dotted line) and the flux level of the adjacent ring (long dashed line), which is defined as the absolute maximum outside the gap region. In many simulations we find that no particular gap feature is present. This is usually the case, when the distance of the planet to the star is small and the observed wavelength relatively long. In this case, the aforementioned kink feature arises at a large radius and the stellar contribution, consequently, overlays any potential gap feature. However, we find that in this case the impact of an embedded planet shows mostly in a change of slope at the position of the kink, when comparing the mean radial profile of the azimuthal gap region with that of the azimuthal CPD region. Therefore, we identify those simulations and  measure the level of flux at the kink, which is defined as an inflection point with a locally maximum positive change in slope.

Furthermore, to measure the impact of the planet and its CPD, we compute the contrast between the mean radial profile of the azimuthal gap region and the azimuthal CPD region, which is  shown in the right plot in Fig. \ref{fig:contrast_definitions_simu2_wl30}. To infer the correct radial position and magnitude of the planetary and CPD signal, even in the case of a weak signal, it is crucial to search for it in the correct radial range. In the case of simulations with $\rp=50\,$au, this is done straightforwardly, since a clear plateau appears in the azimuthal gap profile between approximately $r_{\rm left}=45\,$au and $r_{\rm right}=55\,$au. Here, $r_{\rm left}$ and $r_{\rm right}$ denote the left and right border, respectively, of the interval within which the planetary and CPD signal is searched for, which itself is defined as the absolute maximum of the contrast profile in that interval. 

In the case of $\rp\leq 10\,$au we rely on a different method. If a gap feature is present, the gap depth is computed as the difference in flux between the gap and the ring. Next, the radial position of two points $r_{\rm left}^{50\%}$ and $r_{\rm right}^{50\%}$ are identified, where the gap reaches 50\,\% of its depth. By definition, one of these points ($r_{\rm left}^{50\%}$) is to the left of the radial location of the gap minimum and one point ($r_{\rm right}^{50\%}$) is to the right. Using these points to define the search range for the planetary and CPD signal can potentially lead to false results, however. This is   particularly the case if the gap is very shallow, which would often be accompanied by a very narrow search interval. To solve this issue, we introduce a minimum half interval length $\Delta r_{\rm min}^{50\%}$, which is the minimum distance the two interval borders individually need to be apart from the position of the gap minimum. We set $\Delta r_{\rm min}^{50\%} = 0.1\,r_{\rm feature}$, where $r_{\rm feature}$ is the radial position of the gap minimum. A review of all deduced planetary and CPD signals shows that this method results in a reliable identification of the planetary and CPD signal across all simulations and wavelengths. 
Furthermore, we also deduce a planetary and CPD signal in the absence of a gap feature. In this case we define the search interval such that the radial position of the kink feature is at its center and the two interval borders are placed at a distance of $\Delta r_{\rm min}^{50\%}$ both left and right of the position of the kink, with $r_{\rm feature}$ being the radial position of the kink feature. We note that in some cases the deduced signal strength of the planet and CPD in the contrast profile is at about the level of noise which is present in the contrast curves at very small or large radii, which can also be found in the right plot of Fig.~\ref{fig:contrast_definitions_simu2_wl30}. Consequently, we classify a signal strength of $<0.2\,$mag as noise. Choosing magnitude (mag) as the  units in our definition is suitable since SPHERE detection limits are typically defined in mag as well, which is   relevant in the following sections. 
In order not to lose potentially valuable simulation results, we also perform our analysis on simulations that are classified as contaminated. These simulations do not satisfy the condition in Eq. \eqref{eq:condition_azimuthal_gap_region}, and therefore do not have a defined azimuthal gap region that is free of contamination. In this case, we instead use the mean radial profile of the region on the opposite side of the planet in the CSD, which spans an angular range of $90^\circ$ as a replacement for the azimuthal gap region, and we perform all calculations using this mean radial profile.
After the contrast value between the gap and the planetary and CPD signal ($\ContrastGap$) and the radial position of that signal have been deduced, the mean radial profile of the azimuthal CPD region is evaluated at the position of the signal to obtain a corresponding planetary and CPD flux level. 

In a final step, contrast values are calculated between the stellar flux level and the determined flux levels of the coronagraph (C\,{:}\,Star), the gap (Gap\,{:}\,Star), the ring (R\,{:}\,Star), and the planet and CPD ($\ContrastStar$). The results of this analysis are listed in the Appendix in Tabs. \ref{tbl:results_wl24} to \ref{tbl:results_wl35}. These tables contain additional information. For every simulation, it is specified which type of feature was used in determining the signal strength of the planet and CPD, which can either be a kink or a gap feature. In the latter case the contrast values Gap\,{:}\,Star and R\,{:}\,Star are listed, since a kink based analysis is only performed in the absence of any gap and ring feature. The tables also specify whether the feature is hidden inside the IWA of the coronagraph, and whether or not a proper azimuthal gap region could be defined, meaning that Eq. \eqref{eq:condition_azimuthal_gap_region} is satisfied and the used mean azimuthal gap profile is free of planetary or CPD light (i.e., not contaminated).  
These results form the basis for the following study on the detectability of embedded planets, the impact of the coronograph, and the particular parameter dependences of crucial contrast values that inform  the detectability and characterization of these planets and their CPDs. 

\subsection{Detectability}
\label{sec:detectability}
\begin{figure}
   \centering
   \includegraphics{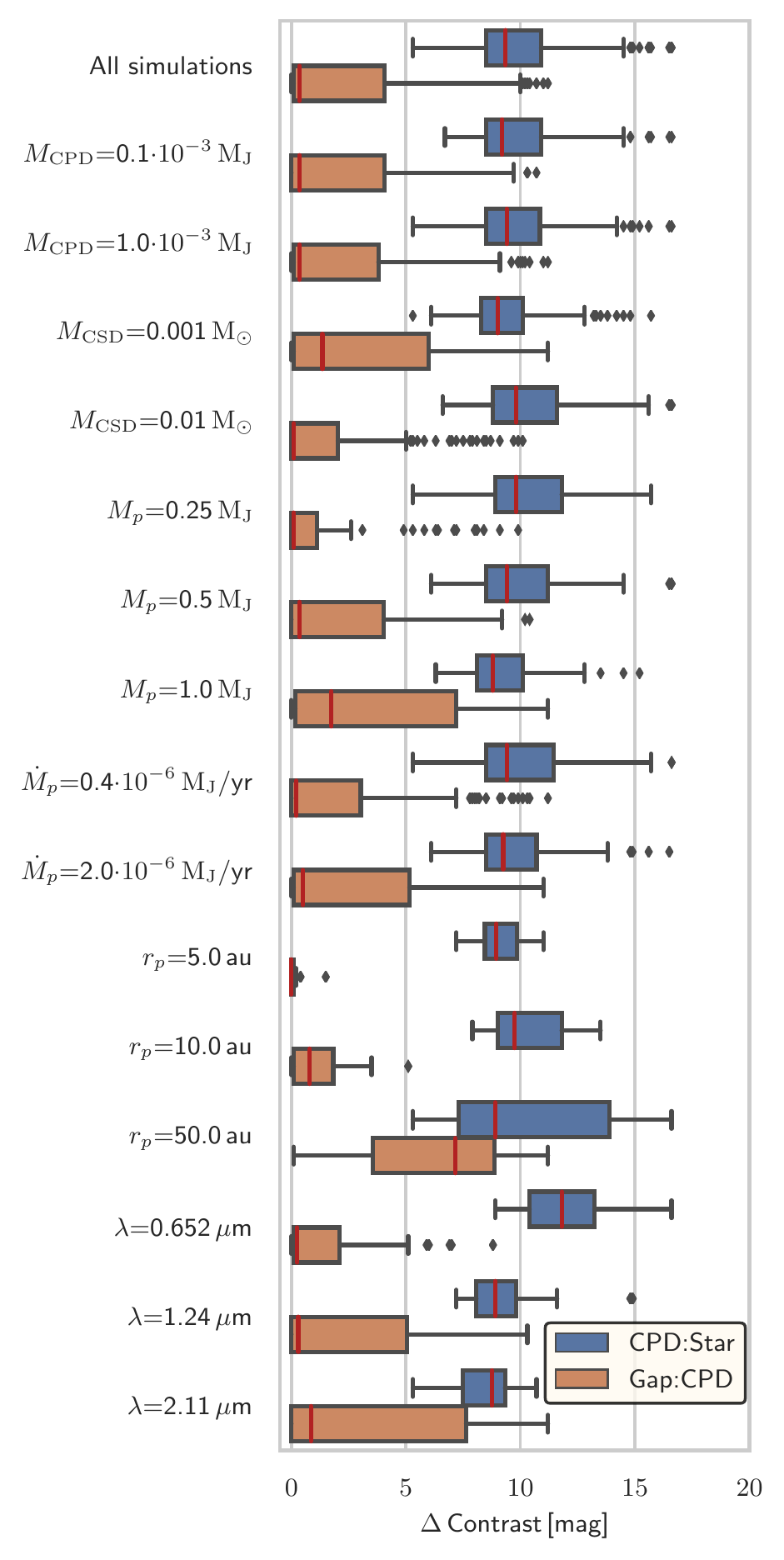}
   \caption{Distribution of contrast values $\ContrastStar$ and $\ContrastGap$ shown in Tabs. \ref{tbl:results_wl24} to \ref{tbl:results_wl35} using boxplots. Each median is highlighted by a red line, the middle $50\,\%$ of data are represented by a box, and the maximum whisker length equals $\Delta w = 1.5\,$IQR. The labels to the left characterize the data used for generating the corresponding boxplots to the right of the label. The label ``All Simulations'' in the first row refers to unrestricted data. In any other case the label refers to a shared parameter value. Outliers are shown as  black diamonds. For details, see Sect. \ref{sec:detectability}.}
              \label{fig:boxplot_spread}
   \end{figure}

In this section, we focus on two crucial contrast values,  $\ContrastStar$ and $\ContrastGap$, and analyze the detectability of planets within the parameter space as described in Tab. \ref{tbl:parameters}. 
Detectability strongly benefits from low contrast values of $\ContrastStar$ since that implies less difference in flux between the star and the dimmer planet and CPD. Moreover, a higher contrast value of $\ContrastGap$ is beneficial as well as it implies that the planet and CPD signal better stand out in the comparably dimmer gap.
To illustrate the results from Tabs. \ref{tbl:results_wl24} to \ref{tbl:results_wl35}, corresponding boxplots are displayed in Fig. \ref{fig:boxplot_spread} that concisely summarize the data. 
In these boxplots the red vertical line is the median, the box represents the middle $50\,\%$ of the data, and the left and right whiskers respectively spread to the minimum and maximum value within a range $\Delta w$ from the box. Here the maximum whisker range is chosen to be $\Delta w = 1.5\,$IQR, where IQR is the interquartile range (i.e., the width of the box). Data points outside the whisker range are classified as outliers, and are shown as  black diamonds. 

Every row in Fig. \ref{fig:boxplot_spread} contains a boxplot for the $\ContrastStar$ distribution and another boxplot for the $\ContrastGap$ distribution. The label to the left of each row describes the selected data for the corresponding boxplots to the right. In the case of the label ``All simulations,'' all of the simulations were used to generate the boxplots. Any other label refers to a parameter value that is shared by simulations whose data went into the corresponding boxplots. This illustration allows us to visually asses the relevance of different parameters and their particular values and to evaluate the overall state of the depicted parameter space. We note that since the whisker range depends on the IQR the number of outliers in general varies across different boxplots. Since all labels, except for  ``All simulations,'' display a subset of all results, the distributions differ and result in differently placed and sized boxes and whisker lengths and in different selected and displayed outliers. 

Considering all simulations, the median contrast value between the star and the planet and CPD is $\approx 9\,$mag and the contrast $\ContrastGap$ is roughly at the level of noise. This suggests that it is unlikely to detect a planet in this part of the parameter space, in particular a planet whose mass is $\Mp\leq1\,\mj$ since its detected flux is often too weak compared to the flux level of the gap. This is consistent with the fact that no such planet detection has yet been confirmed. 
However, a part of the parameter space seems to result in significantly better contrast values, and thus may represent candidates for future detections. By comparing the different median values we find that the best contrast values are achieved for a low value of $\mcsd$ and  for  high values of $\Mp$, $\rp$, and $\lambda$. This can be explained by considering the following effects. By reducing $\mcsd$, the optical depth (which dims the planetary and CPD light) is also reduced. This effect is relatively strong and affects both $\ContrastGap$ and $\ContrastStar$ positively. This drop in optical depth is also part of why for higher distances $\rp$ and higher masses $\Mp$ both $\ContrastGap$ and $\ContrastStar$ improve. Additionally, the $\Mp=1\,\mj$ simulations benefit from a relatively high accretion luminosity. 
Furthermore, simulations with a planet at $\rp=50\,$au form a gap whose detected flux level is relatively low, which is a consequence of the relatively small portion of direct and indirect stellar light that scatters far from the star, thus directly benefiting $\ContrastGap$. The majority of $\rp=50\,$au simulations even result in $\ContrastGap>3\,$mag, further highlighting the importance of the parameter $\rp$. However, if the optical depth between the observer and planet is too high, for instance due to a high value of $\mcsd$, even a planet at $\rp=50\,$au may not lead to a sufficiently strong planetary and CPD signal. On the contrary, the vast majority of $\rp=5\,$au planets do not generate a signal that can be spotted in the bright gap, since $\ContrastGap$ is mostly at the level of noise, and even though the median value of $\ContrastStar$ is lower for $\rp=5\,$au simulation than that of $\rp=10\,$au simulations, it is not actually a sign of a stronger planetary and CPD signal, but rather a consequence of the increased stellar flux that originates from all regions that are close to the star. The wavelength $\lambda$ also plays an intricate role. The spectral flux density originating from the planet and that from the comparably colder CPD both change with increasing $\lambda$. 

Additionally, the optical depth between the planet and the observer is directly linked to it; in particular, the optical depth decreases as $\lambda$ increases. Overall, the relatively low optical depth at $\lambda=2.11\,\mu$m allows for the most beneficial contrast values. 

Comparing different values of $\mpdot$, we find that the higher parameter value of $\mpdot=2\cdot 10^{-6}\,\mj/$yr  also slightly benefits the two contrast values. The effect of different values of $\mcpd$, however, seems to be overall rather weak. This is interesting, since its increase leads to an extended inner CPD radius due to the back-warming effect (see Sect. \ref{sec:determination_r_in_cpd}). This in turn changes the overall thermal and density structure, particularly in the hot regions of the CPD. Nonetheless, the impact of an altered CPD structure does not generally seem to    result in very different outcomes, making an observational determination of it solely based on SPHERE observations very challenging.

 \begin{figure*}
   \centering
   \includegraphics{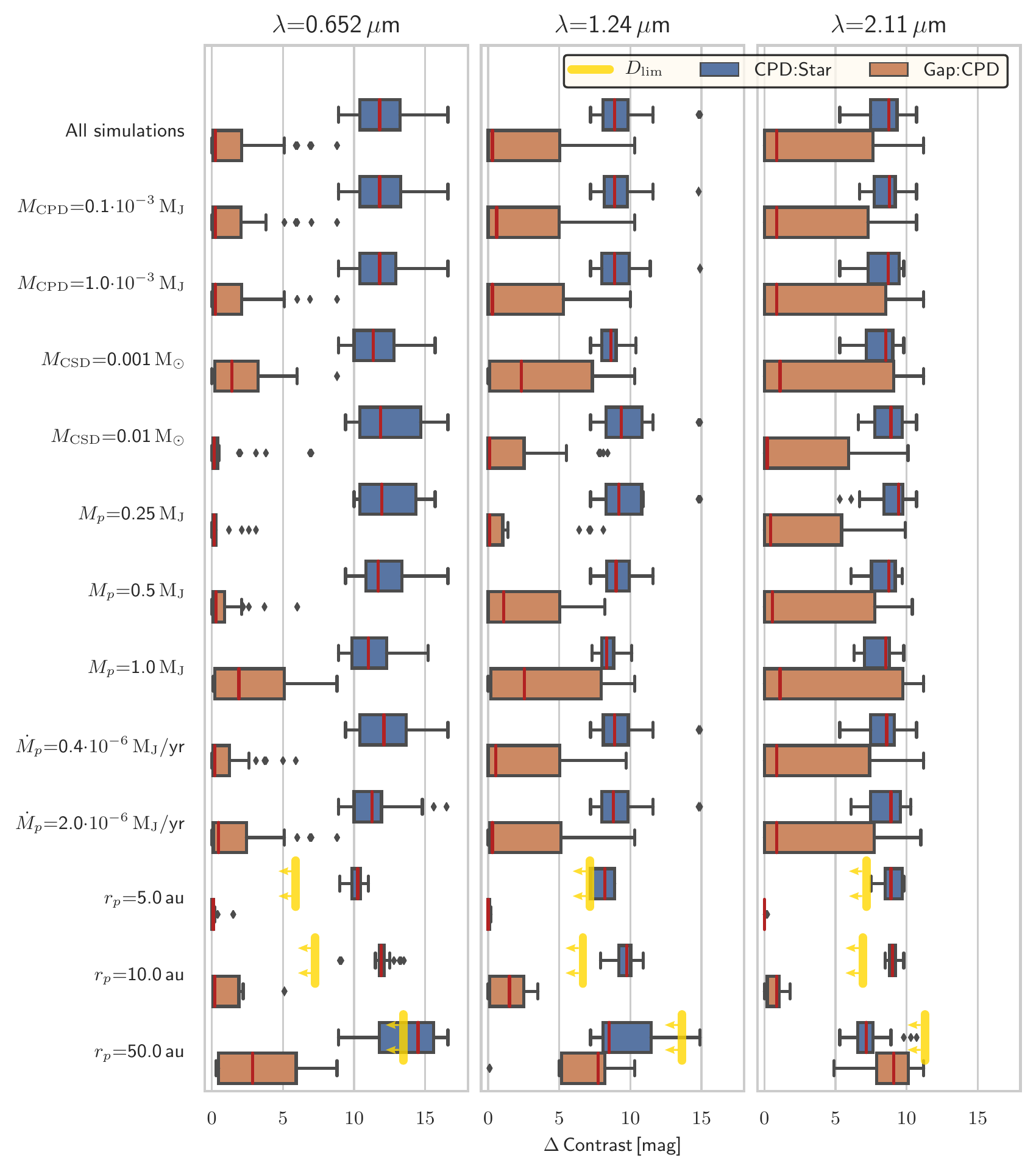}
   \caption{Wavelength-dependent distribution of contrast values $\ContrastStar$ and $\ContrastGap$ shown in Tabs. \ref{tbl:results_wl24} to \ref{tbl:results_wl35} using boxplots. Each median is highlighted by a red line, the middle $50\,\%$ of data are represented by a box, and the maximum whisker length equals $\Delta w = 1.5\,$IQR. The labels to the left characterize the data used in generating the corresponding boxplots to the right of the label. The label ``All Simulations'' in the first row refers to wavelength-dependent but otherwise unrestricted data. In any other case the label refers to a shared parameter value. The wavelength is shown above its corresponding column. The  5$\sigma$ detection limits $\dlim$ are shown as  vertical yellow lines, with arrows indicating the direction of detectable signals. Outliers are shown as  black diamonds. For details, see Sect. \ref{sec:detectability}.}
              \label{fig:boxplot_spread_wl_dep}
   \end{figure*}

The detection of embedded planets and their CPDs using SPHERE, though, is possible for a certain part of the parameter space. Thus, a detectable planetary and CPD signal restricts its underlying parameters, and can therefore already be used to restrict basic properties. We classify a planet and CPD as detectable if the contrast $\ContrastStar$ falls below a certain wavelength-dependent and $\rp$-dependent limit. The criterion we apply here is designed to be a weak criterion for detectability that solely considers the contrast $\ContrastStar$. However, as we   show in the following, this criterion alone already restricts the parameter space of detectable planets and their CPDs strongly. 

To estimate the detection limits, we make use of the SPHERE ESO exposure time calculator (ETC)\footnote{ESO exposure time calculator: \url{https://www.eso.org/observing/etc/}}. We use the properties of HL Tau as a proxy in order to generate generic estimates for T Tauri stars. Slight changes to the spectral type or provided J-band magnitude change these estimates only marginally. Additionally, we opt for a pupil-stabilized observation in this tool, choose filters as listed in Tab. \ref{tbl:instrument_sphere}, activate the use of a coronagraph, and set the exposure time to the standard 3600\,s with a DIT of 8\,s. As a result, the ETC generates $5\sigma$ performance curves that show the maximum contrast between the star and its companion that allows for a successful $5\sigma$ detection of the planetary signal. The obtained contrast values generally depend on the distance of the companion from the star. For the three considered observing wavelengths we generate these curves and read data points off of the performance curve, which are closest to 5, 10, and 50\,au, assuming a distance of $140\,$pc to the simulated star. The obtained contrast values are then used as detection limits $\dlim$ and are listed in Tab. \ref{tbl:detection_limit}.

\begin{table}
\begin{center}
\begin{tabular}{c|ccc}
$\dlim$ & $\rp=5\,$au & $\rp=10\,$au & $\rp=50\,$au \\
\cmidrule{1-4}
$\lambda=\lambdaS$ & 5.9\,mag & 7.3\,mag & 13.5\,mag \\
$\lambda=\lambdaM$ & 7.2\,mag & 6.7\,mag & 13.6\,mag \\
$\lambda=\lambdaL$ & 7.2\,mag & 6.9\,mag & 11.3\,mag \\
\end{tabular}
\caption[bla]{Detection limits $\dlim$ derived from ETC $5\sigma$ performance curves for three different values of $\lambda$ and $\rp$. For details, see Sect. \ref{sec:detectability}.}
\label{tbl:detection_limit}
\end{center}
\end{table}

The previous results shown in Fig. \ref{fig:boxplot_spread} already give a very general overview. To better study the significance of specific parameter values and assess the simulated parameter space with regard to the detectability of planetary and CPD signal, however, Fig. \ref{fig:boxplot_spread_wl_dep} explicitly shows the wavelength dependence of the results. Compared to Fig. \ref{fig:boxplot_spread}, it divides the data into sets for a single wavelength, which is shown above its corresponding column in Fig. \ref{fig:boxplot_spread_wl_dep}. Thus, the data is divided into groups that share a parameter $\lambda$ and one additional parameter, except for boxplots that belong to the ``All simulations'' label which only share a wavelength. Additionally, the detection limits $\dlim$ from Tab. \ref{tbl:detection_limit} are indicated in the three bottom rows of the plot by yellow vertical lines, with arrows that indicate the direction of detectable planetary and CPD signals. 

We find that all simulations with $\rp\leq 10\,$au produce planetary and CPD signals that satisfy $\ContrastStar\geq\dlim$, meaning that only in the best cases is the detection limit  barely reached, but most of these simulations exceed the limit and are thus not detectable. However, it is also clear that reducing the required significance level, for example  to $3\sigma$, would increase the number of planets that are classified as detectable. This means that some planetary and CPD signals might just fall short of being detectable, and  slightly improved performance curves or slightly increased detection limits would already allow   more detections to be performed. However, it is also important to note that in particular simulations with $\rp=5\,$au additionally suffer from an extremely low-contrast $\ContrastGap$, which would make improving performance curves a pointless endeavor as they would still miss their goal of detecting these close-in embedded planets. Instead, a detection of far-out planets at $\rp=\,50$au is much more likely, and in fact at $\lambda=\lambdaL$ all simulations produce a signal that is detectable, and  $\ContrastGap$ reaches values that often exceed $\approx7\,$mag. As a result, detections of far-out planets at $\lambda=\lambdaL$ have the best conditions for being detectable within the investigated parameter domain. 
It is also interesting to compare the position and IQR of the plotted boxes for fixed values of $\lambda$ and across different values of $\rp$. At $\lambda=\lambdaS$ the median of the $\ContrastStar$ values clearly increases for increasing $\rp$, while it first increases and then decreases for $\lambda=\lambdaM$, and mostly decreases for $\lambda=\lambdaL$. Apart from this change in sign of contrast value differences, it is also striking that the blue boxes that represent $\ContrastStar$ data points are covering very different intervals at $\lambda=\lambdaL$ across different values of $\rp$, which could be useful for identifying planetary properties. Even more remarkable is the fact that at this wavelength the contrast intervals that are covered by $\ContrastGap$ boxes are almost distinct and in the case of $\rp=50\,$au even reach values $>10\,$mag. For all three wavelengths, we find that the $\ContrastGap$ boxes are shifted increasingly toward higher contrast values for increasing $\rp$, which  benefits the detectability of these planets. At a fixed wavelength, $\rp$ is the only parameter that results in almost distinct contrast intervals for its different parameter values. 

While in many situations changes to $\rp$ can enable detections, changes in other parameters can be particularly detrimental for that purpose. At $\lambda=\lambdaS$, for instance, having $\Mp\leq 0.5\,\mj$ or $\mcsd=0.01\,$M$_\odot$ results in overall extremely low contrast values of $\ContrastGap$. In particular,  a planetary mass of $\Mp\,{=}\,0.25\,\mj$  struggles to produce significant contrast values, which is the case even at an increased wavelength of $\lambda=\lambdaM$. Overall, we find that observations at $\lambda=\lambdaL$ have the greatest potential for finding and identifying planetary signals in the studied part of the parameter space. However, in Sect. \ref{sec:parameter_impact} we   particularly focus on the potential to characterize embedded planets with regard to their underlying parameter values, where we   break down each individual parameter and come to more specific conclusions.

It is first worth mentioning the characteristics of simulations labeled ``Contaminated'' based on the results of Fig. \ref{fig:boxplot_spread_wl_dep}. These models are all observed at $\lambda\geq\lambdaS$ and share the parameter value $\rp=5\,$au. As mentioned before, these simulations do not satisfy the condition presented in Eq. \eqref{eq:condition_azimuthal_gap_region}, and thus they also do not allow  a proper azimuthal gap region to be defined that is contamination-free. As a consequence, the measured flux level of the gap is elevated and the $\ContrastGap$ value is reduced. This effect results in $\ContrastGap< 0.2\,$mag for all contaminated simulations (i.e., the signal is dominated by noise), which highlights the difficulty of detecting planets that fail to satisfy the condition in Eq. \eqref{eq:condition_azimuthal_gap_region}.

Finally, Fig. \ref{fig:boxplot_spread_wl_dep} also allows us to easily assess the role of the IWA and coronagraphs. In general, the importance of using a coronagraph is very obvious as it strongly suppresses stellar light, and therefore enables the detection of dimmer sources in the vicinity of the star. However, the IWA can be adjusted, and it is worth evaluating whether  currently available coronagraphs are blocking otherwise valuable planetary or CPD signals due to the extent of their mask. These planets are labeled  ``Hidden'' in Tabs. \ref{tbl:results_wl24} to \ref{tbl:results_wl35}, and are located at $\rp=5\,$au when observed at $\lambda=\lambdaS$. In Fig. \ref{fig:boxplot_spread_wl_dep} we find that these simulations show rather low $\ContrastStar$ values, which are well inside the undetectable range. Additionally, their $\ContrastGap$ values are extremely low as well. Only two of these simulations resulted in $\ContrastGap>1\,$mag, and both simulations are based on the highest possible values of $\Mp$ and $\mpdot$ and  on the lowest possible value for $\mcsd$ within the studied parameter ranges, only differing in their $\mcpd$ value. These results clearly suggest that decreasing the IWA would not result in any more detections of planets that are within the studied parameter domain when using SPHERE/ZIMPOL. Based on the presented data, we conclude that the currently available coronagraphs in terms of their IWAs are not limiting the ability of SPHERE to detect embedded planets within the studied part of the parameter space.

\subsection{Parameter impact and characterization}
\label{sec:parameter_impact}

 \begin{figure*}
   \centering
   \includegraphics{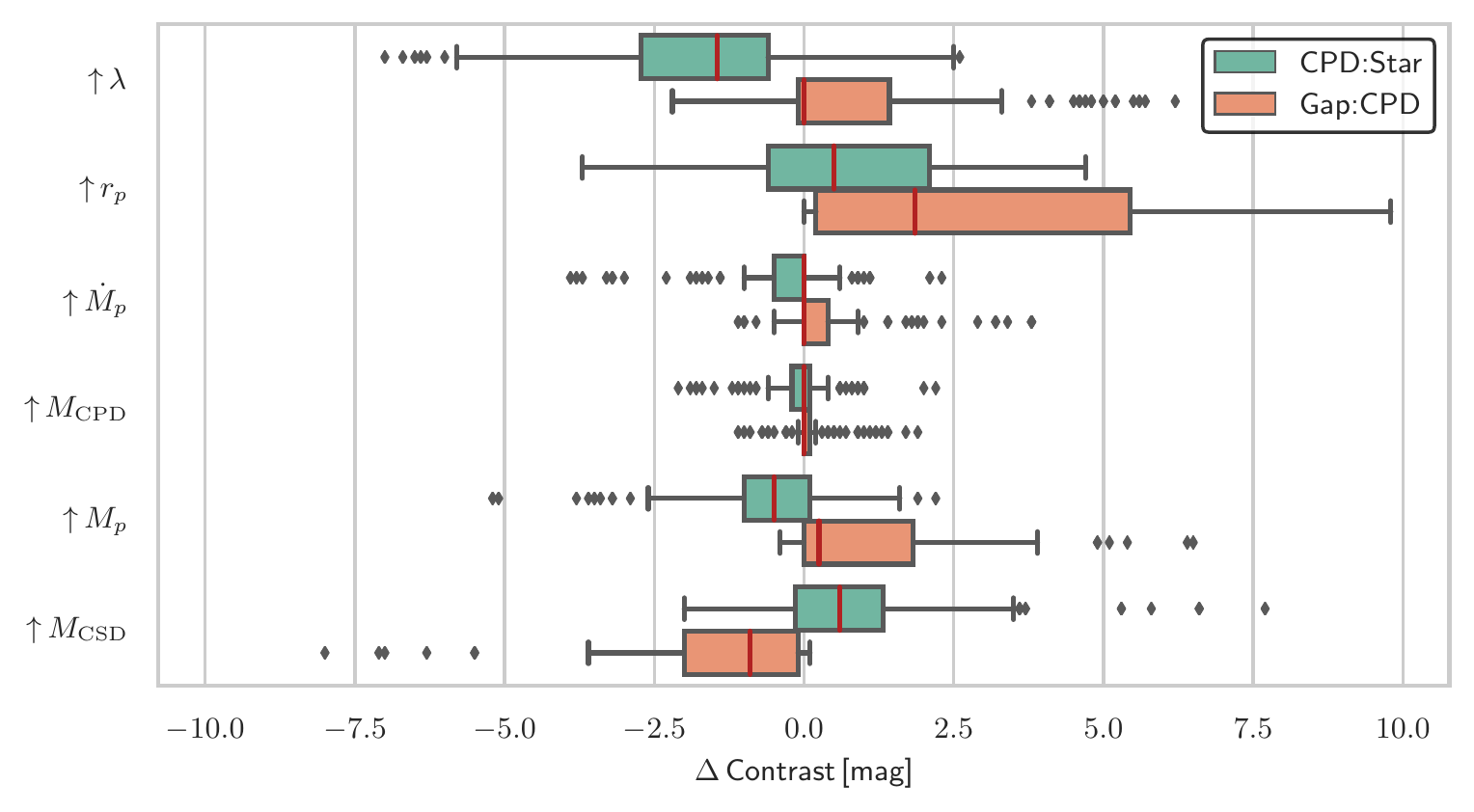}
   \caption{Distribution of contrast value changes of $\ContrastStar$ and $\ContrastGap$ with respect to the change in a single model parameter using boxplots. Each median is highlighted by a red line, the middle $50\,\%$ of data are represented by a box, and the maximum whisker length equals $\Delta w = 1.5\,$IQR. The labels to the left characterize the data used in generating the corresponding boxplots to the right of the label. A label refers to a parameter that has been increased to its next simulated value. Outliers are shown as  black diamonds. For details, see Sect. \ref{sec:parameter_impact}.}
              \label{fig:boxplot_Nconst0}
   \end{figure*} 

Thus far, we have presented an analysis of the potential for detecting embedded planets in the VIS/NIR with a focus on SPHERE. However, the gathered data can also be examined regarding the best approaches for characterizing these planets and their surrounding CPDs. As it turns out, these purposes do not benefit from the same observational strategies. To characterize single basic properties, it is crucial to understand the impact they have on observations, in particular, the sensitivity with which measured contrast values react to a change of a single parameter is important.

To better assess the impact of a parameter change on observations of different models and disentangle the information presented thus far, Fig. \ref{fig:boxplot_Nconst0} illustrates the effect of all varied parameters separately. 
In particular, we compute the difference of contrast values of two simulations whose underlying parameters differ only in the value of a single parameter and illustrate the distribution of contrast difference values with boxplots. 
To not restrict this analysis to a specific set of wavelength-dependent inner working angles, we consider all planetary and CPD signals, even if they are labeled  Hidden, which allows for a more general analysis.

Generally, an increase in any parameter value may lead to an improved or a worsened contrast. In this context, an improved contrast refers to a contrast change that is beneficial for a detection, for instance a reduced contrast $\ContrastStar$ or an increased contrast $\ContrastGap$. The presented results allow us to identify some clear trends that can be found for the majority of simulations. In order to identify these trends, we  mainly focus on the sign of the change in contrast values, particularly of the median, and on the distribution of the middle $50\,\%$ of data. We also note that even if the median is at about zero, there may still be a difference between the left and right sides of a distribution of data points, in that the data may be left-skewed or right-skewed. In the framework of detectability and characterizability, this translates to an advantageously skewed or disadvantageously skewed distribution of data points, depending on the type of contrast and direction of skewness. This means that an advantageously skewed distribution of data points is either a left-skewed $\ContrastStar$ distribution or a right-skewed $\ContrastGap$ distribution. A broad overview of the trends that can be found in Fig. \ref{fig:boxplot_Nconst0} is concisely summarized in Tab. \ref{tbl:upped_parameters}. Additionally, to highlight and discuss some of these findings, we   focus only on the key results. In the following we present an analysis of all studied varied parameters and how their changes affect measurements.

\begin{table}
\begin{center}
\begin{tabular}{lll}
\toprule
\textbf{Param.} & \textbf{Contrast} & \textbf{Effect}\\
\cmidrule{1-3}
${\uparrow}\lambda$ & $\ContrastStar$ & $\cdot$ Relatively strong effect\\ && $\cdot$ Improves very consistently\\ && $\cdot$ Advantageously skewed \vspace{0.2em}\\
                                                 & $\ContrastGap$ &  $\cdot$  Strongly advantageously skewed\\ 
                                                 \hdashline \vspace{-0.8em} \\ 
${\uparrow}\rp$         & $\ContrastStar$ &  $\cdot$ Worsens mostly \\ && $\cdot$ Disadvantageously skewed\vspace{0.2em}\\
                                                 & $\ContrastGap$ &  $\cdot$  Relatively strong effect\\ && $\cdot$ Improves strictly\\ && $\cdot$ Advantageously skewed\\
                                                 \hdashline \vspace{-0.8em} \\ 
${\uparrow}\mpdot$  & $\ContrastStar$ &  $\cdot$  Strongly advantageously skewed\\ && $\cdot$ Many outliers\vspace{0.2em}\\
                                                 & $\ContrastGap$ &  $\cdot$  Strongly advantageously skewed \\ && $\cdot$ Many outliers\\
                                                 \hdashline \vspace{-0.8em} \\ 
${\uparrow}\mcpd$    & $\ContrastStar$ &  $\cdot$  Relatively weak effect\\ && $\cdot$ Advantageously skewed\\ && $\cdot$ Many outliers\vspace{0.2em}\\
                                                 & $\ContrastGap$ &  $\cdot$  Relatively weak effect\\ && $\cdot$ Advantageously skewed\\ && $\cdot$ Many outliers\\
                                                 \hdashline \vspace{-0.8em} \\ 
${\uparrow}\Mp$       & $\ContrastStar$ &  $\cdot$  Improves mostly\\ && $\cdot$ Slightly disadvantageously skewed\vspace{0.2em}\\
                                                 & $\ContrastGap$ &  $\cdot$  Improves rather consistently\\ && $\cdot$ Strongly advantageously skewed\\
                                                 \hdashline \vspace{-0.8em} \\ 
${\uparrow}\mcsd$    & $\ContrastStar$ &  $\cdot$  Worsens mostly\\ && $\cdot$ Advantageously skewed\vspace{0.2em}\\
                                                 & $\ContrastGap$ &  $\cdot$  Worsens very consistently\\ && $\cdot$ Disadvantageously skewed\\
\bottomrule
\end{tabular}
\caption{Impact of changing a single parameter (Param.), which is illustrated in Fig. \ref{fig:boxplot_Nconst0}, on the contrast values $\ContrastStar$ and $\ContrastGap$. For details, see Sect. \ref{sec:parameter_impact}.}
\label{tbl:upped_parameters}
\end{center}
\end{table} 

\subsubsection{Wavelength dependence}

In this parameter study the observing wavelength $\lambda$ is a unique parameter in the sense that it is the only parameter that is in the control of the observer. Therefore, this section presents some general trends that can be found for $\lambda$ and we will have a dedicated in-depth discussion of the best choice for $\lambda$ for every other parameter separately. 

Overall, we find that increasing the observing wavelength $\lambda$ has the strongest beneficial effect on the contrast $\ContrastStar$, as can be seen in Fig. \ref{fig:boxplot_Nconst0}. Moreover, this effect is  consistent, meaning that all data points within the IQR are shifted toward a single side. This can be explained by the strong  change in optical depth between the observer and the planet. In particular, at $\lambda=\lambdaL$ the extinction cross-section, and thus the optical depth, is roughly 3 or 9 times smaller than that at $\lambda=\lambdaM$ or at $\lambda=\lambdaS$, respectively. It is also important to note that the temperature of dust in the CPD has a natural upper limit given by the sublimation temperature. Using Wien's displacememt law as a rough approximation, an upper temperature of ${\approx}1500\,$K yields a wavelength of $\lambda \gtrsim 3.4\,\mu$m at the peak of spectral flux per unit frequency. As a consequence, the optical depth decreases and the spectral flux of the CPD increases toward larger values of $\lambda$. Despite its higher temperature, the planet also increases its spectral flux and benefits from the decreased optical depth. 

To further elaborate on these finding, Figs. \ref{fig:boxplot_Nconst1a} and \ref{fig:boxplot_Nconst1b} give a more in-depth view into the impact of a single parameter on the contrast values, which is achieved by also  sorting the data into groups of shared parameter values. Similarly to Fig. \ref{fig:boxplot_spread_wl_dep}, the shared parameter value is displayed to the left. Here, the increased parameter is shown above its corresponding column. The results for the group of constant $\rp$ value in the left column (${\uparrow}\lambda$) of Fig. \ref{fig:boxplot_Nconst1a} is particularly interesting. Here we find that the benefit of an increasing wavelength seems to rise rather strongly with $\rp$, particularly in the case of $\ContrastStar$. However, an increase in $\lambda$ for $\rp=5\,$au simulations is leaving $\ContrastGap$ almost unchanged. At $\rp=10\,$au we find a  disadvantageously skewed distribution for $\ContrastGap$, meaning that the contrast tends to   worsen for increasing $\lambda$, before drastically improving toward $\rp=50\,$au. Moreover, for $\rp=50\,$au the beneficial effect is becoming very consistent. The fact that this effect is not strictly positive for increasing values of $\rp$ may be explained by considering the temperature of the star and its spectral flux, which reaches its highest value at $\lambda=\lambdaM$ and dominates the detected flux in the azimuthal gap region (see Sect. \ref{sec:sphere_ideal}). A negative impact on $\rp=10\,$au simulations suggests that despite the discussed benefits of an increased wavelength, the contrast $\ContrastGap$ may still decrease due to an elevated gap flux level induced by a change in spectral stellar flux. 

Overall we conclude that a detection of a planet at larger radii is more likely to benefit from a longer observational wavelength. At shorter distances $\rp$, though, it is rather unlikely to make any detection for different reasons. On the one hand, at shorter wavelengths the optical depth   leads to an overall low-contrast $\ContrastStar$ that we find to be deep inside the undetectable range. On the other hand, at longer wavelengths we find that the gap region becomes increasingly contaminated, as discussed in Sect. \ref{sec:detectability}, diminishing $\ContrastGap$ values to the level of noise. Nonetheless, in this case the $\ContrastStar$ values  come close to the detectable range. Finally, when comparing the impact of an increase in $\lambda$ for simulations with shared values of $\lambda$, we identify a diminishing return for higher wavelengths. Overall, the explored wavelength dependence suggests that observations at $\lambda=\lambdaL$ carry the greatest potential for future detections. A study of the best suited wavelength choice for characterizing basic planetary and CPD properties is  presented in each of the following sections for every parameter separately.

\subsubsection{$\rp$-dependence}
Among all investigated parameters, the radial position of the planet ($\rp$) is the only parameter that when varied usually improves one contrast while worsening the other (see Fig. \ref{fig:boxplot_Nconst0}). While an increase in $\rp$ improves $\ContrastGap$ strongly and consistently, its effect on $\ContrastStar$ varies strongly depending on the underlying model parameters. This can be clearly seen in the middle column (${\uparrow}\rp$) of Fig. \ref{fig:boxplot_Nconst1a}. We find that its increase  affects $\rp=\rpS$ and $\rp=\rpM$ simulations in a noticeably different way; specifically, it adversely affects the former and benefits the latter. The difference of median values from the displayed $\ContrastStar$ and $\ContrastGap$ distributions, which can be viewed as a measure for the overall impact on the observability of the models, shows a striking increase for increasing $\rp$ values. The same is true when increasing the observing wavelength $\lambda$, as can be seen in the bottom three rows of Fig. \ref{fig:boxplot_Nconst1a}. In particular the median of the $\ContrastStar$ simulations shifts from a clearly positive value at $\lambda=\lambdaS$ to a negative value at $\lambda=\lambdaL$, yet again highlighting the combined benefit of models with high $\rp$ value observed at longer wavelengths. This trend can also be observed for the shared parameter $\Mp$. Increasing the planetary mass $\Mp$ shifts the median of the $\ContrastStar$ distribution from positive values toward roughly zero, while clearly benefiting $\ContrastGap$ by shifting it consistently toward higher positive values. Therefore, we find that $\rp$ has the most diverse effect on all simulations. This behavior can be explained by two effects. First, $\rp$ strongly affects the stellar flux level at the position of the gap and, second, it also strongly impacts the optical depth between the observer and the planet. Therefore, an increase in $\rp$ on the one hand decreases the stellar flux level; on the other hand, it exposes the planetary and CPD signal. Our results suggest that both effects  compete at a similar level and either one can dominate. For instance, low $\Mp$ simulations suffer from a high optical depth, even at higher values of $\rp$, and as a consequence, these planets still appear  very dim, which in turn leads to an overall decrease in flux at the position of the planet and an increase in $\ContrastStar$. The higher the  $\Mp$, the more planetary and CPD flux is revealed, and the gap further deepens, which leads to a  better $\ContrastStar$ value change. The same effect as described for fixed increasing $\Mp$ values holds for increasing $\lambda$ values as well, but it is even stronger. 

Finally, the parameter $\rp$ is unique in that its determination on the basis of directly imaged can be estimated by simply measuring the position of the peak in flux. In that regard, its determination can be best achieved at a wavelength that shows the strongest signal, which we find to be $\lambda=\lambdaL$.

\subsubsection{$\Mp$-dependence}
For the majority of simulations, higher masses $\Mp$ result in better contrast values (see Fig. \ref{fig:boxplot_Nconst0}). In particular, the $\ContrastGap$ distribution is strongly advantageously skewed, meaning, more models benefit strongly from an increase in $\Mp$ than weakly. Moreover, the right column (${\uparrow}\Mp$) of Fig. \ref{fig:boxplot_Nconst1a} shows interesting  dependences on $\rp$ and $\lambda$. We find that an increase in $\Mp$ is particularly beneficial for far-out planets regarding both $\ContrastStar$ and $\ContrastGap$. Here the IQR of $\ContrastStar$ is especially noticeable, which is the narrowest for $\rp=\rpM$ simulations, where in addition  the median is close to but smaller than zero. This is interesting since an increase in $\Mp$   causes various effects. First, the optical depth between planet and observer is decreased. Second, the accretion luminosity is increased. Third, the inner radius of the CPD is widened, which changes the density and thermal structure of the CPD as a whole. Fourth, the optical depth in radial direction within the gap is decreased due to the its increased depth, which reduces the scattering probability of stellar photons. The small width of the distribution suggests that these effects are competing with each other and the overall result in similar outcomes when increasing $\Mp$. This is contrary to the scenarios of planets that are located either farther in or farther out. The latter especially tend to greatly benefit from an increase in $\Mp$. In order to identify one of the modeled planets it is crucial to observe at a wavelength where the observed contrast levels are the most diverse. Judging from the IQRs of both contrasts for different wavelengths we find that it is the widest for short wavelengths (i.e., at $\lambda=\lambdaS$). Combining these results with results from Sect. \ref{sec:detectability}, we find that the chances for detectability and the capability of characterizing the planetary mass directly counteract each other in the sense that a single wavelength observation can improve one only by diminishing the other.

\subsubsection{$\mpdot$-dependence}
An increase in the accretion rate onto the planet ($\mpdot$) leads to a proportional increase in $\lacc$, which leads to an advantageously skewed distribution of contrast changes according to Fig. \ref{fig:boxplot_Nconst0}. Compared to the other parameters, these differences are small as the median is at zero and the IQR is rather narrow. On the one hand, the fact that all of the medians of the boxes match their corresponding box edges, which are placed with contrast differences of zero, shows that a large part of models would not benefit from an increase in $\mpdot$. On the other hand, the large number of outliers also shows that a significant portion of simulations are actually  affected far beyond the level of noise. Moreover, we find that an increase often  results in a negative impact on both contrast values as well. The left column (${\uparrow}\mpdot$) in Fig. \ref{fig:boxplot_Nconst1b} also shows that negatively impacted outliers for the two contrast values are caused under different circumstances. In particular, negatively impacted outliers of $\ContrastStar$, which distinguish themselves due to high positive values, are typically based on simulations with higher planetary masses $\Mp$, while outliers of $\ContrastGap$, which have high negative values, usually originate in simulations with low values of $\Mp$. The most positively impacted outliers, though, are caused for both contrasts in simulations with higher planetary mass. Similarly, we find that the most negatively impacted outliers of $\ContrastStar$ and $\ContrastGap$ are caused for close-in and far-out planets, respectively, while most positively impacted outliers come from far-out $\rp=\rpL$ simulations for both contrast distributions. Moreover, we find that the simulated observations react most sensitively to changes of $\mpdot$ when observed at the shortest simulated wavelength $\lambda=\lambdaS$. At longer wavelengths, the impact of $\mpdot$ is much weaker, although it has the potential for the highest negative outliers regarding $\ContrastStar$. 

As mentioned before, the primary effect of increasing $\mpdot$ is an increase in the accretion luminosity $\lacc$, which in turn results, first, in an extended inner radius $\rin$ of the CPD that is changing its density and thermal structure and, second, in higher spectral planetary flux. Since both effects are confined to the planet and its vicinity and since this region is deeply embedded in the CSD, the impact of a change in $\mpdot$ at any observational wavelength best shows when the region is exposed the most, which means that  when the optical depth between the observer and the planet is the lowest. This explains the increase in the IQR of simulations with higher values of $\Mp$ or $\rp$. Since the increase in $\mpdot$ strictly increases the planetary flux, a negative impact is mostly the result of a reduction of the observed spectral CPD flux. The most negatively impacted outliers of $\ContrastStar$, for instance, come from simulations with high values for $\Mp$ and $\mcpd$ and with low values of $\rp$. This coincides exactly with the dependences found for increasing $\rin$ values in Sect. \ref{sec:determination_r_in_cpd}. Thus, more extended CPDs are more likely to result in worse $\ContrastStar$ values when being further extended due to an increase in $\mpdot$. For $\ContrastGap$, however, the situation is more complicated. Due to the locality of induced changes, one might expect that the measured contrast values $\ContrastStar$ and $\ContrastGap$ behave in exactly the same way except for their sign (i.e., that the distributions are mirrored). However, this is not the case, as can already be seen in Fig. \ref{fig:boxplot_Nconst0}. Figure \ref{fig:boxplot_Nconst1b} shows mirrored distributions only for the $\rp\geq\rpM$ and for  $\Mp=0.25\,\mj$ simulations. What these simulations have in common is either that they harbor relatively far-out planets or generally planets with low $\lacc$. On the other hand, it is those simulations with close-in planets and high accretion rates $\lacc$ that   typically cause the asymmetry between the two distributions. 
The reason  is that for these simulations the aforementioned argument of locality does not apply, meaning that  a change in $\mpdot$ affects the whole contaminated gap region, which is why these simulations are labeled Contaminated in Tabs. \ref{tbl:results_wl24} to \ref{tbl:results_wl35}. The effect is only present in the $\rp=\rpS$ simulations and enhanced by higher values of $\lacc$ particularly at longer observation wavelengths, which is in agreement with our results regarding $\rtranshat$ and generally simulations that do not satisfy the condition in Eq. \eqref{eq:condition_azimuthal_gap_region}. Overall, we conclude that observations  aimed at detecting mass accretion rates ought to take into account primarily short observational wavelengths given their higher level of sensitivity regarding $\mpdot$.

\subsubsection{$\mcpd$-dependence}
Of all the investigated parameters, the mass of the CPD ($\mcpd$) has the smallest overall impact on the two contrast values. While both contrast distributions in Fig. \ref{fig:boxplot_Nconst0} are advantageously skewed, a large portion of their data points fall into a very narrow IQR. Moreover, outliers are even more abundant in these distributions than they are in those corresponding to an increase in the planetary mass accretion rate $\mpdot$.
In the middle column (${\uparrow}\mcpd$) of Fig. \ref{fig:boxplot_Nconst1b} we find that the widest IQRs can be found for $\rp=\rpL$ and $\lambda=\lambdaL$ simulations. The overall narrowness of most IQRs, though, can be explained by considering the effect that a change in $\mcpd$ induces. Already at $\mcpd=\mcpdS$ the CPD is optically thick at all investigated wavelengths; therefore, increasing it  mostly causes a strengthening of the back-warming effect (see Sect. \ref{sec:determination_r_in_cpd}). As a consequence, the inner radius $\rin$ increases and the thermal and density structure of the CPD changes. Within the investigated parameter space $\mcpd$ has the greatest impact on $\rin$, as can be seen in Tab. \ref{tbl:r_in_cpd}. Despite its overall weak impact, we generally find that this redistribution of dust grains induced by an increase in dust mass in the CPD is beneficial for a detection; this means that  the back-warming effect is acting in favor of this goal, in particular at $\lambda=\lambdaL$, where the impact is the greatest. Similarly to $\mpdot$, we also find that some distributions of $\ContrastGap$ and $\ContrastStar$ are linked, in that they are symmetric and only differ in the sign of change. Here we observe this type of symmetry, yet again, for far-out planets at $\rp\geq\rpM,$ as well as a relatively strong symmetry for wavelengths $\lambda\leq\lambdaM$. At $\rp=\rpS$, however, this effect cannot be found, which is most likely  due to contaminated contrast values, analogously to the case of $\mpdot$. Contrary to the results for $\mpdot$, we conclude that the highest sensitivity for an observationally based measurement of $\mcpd$ can be achieved at longer wavelengths, particularly at $\lambda=\lambdaL$. These results are expected since the temperature of the CPD typically reaches values far below the planetary temperature, and thus has its emission maximum shifted toward longer wavelengths.

\subsubsection{$\mcsd$-dependence}
\label{sec:mcsd_dependence}
The CSD mass $\mcsd$ plays a crucial role in the detectability of embedded planets, which can be seen in the corresponding contrast distributions in Fig. \ref{fig:boxplot_Nconst0}. Judging from the medians, we find that the majority of simulations strongly benefit from a reduction in $\mcsd$, particularly the $\ContrastGap$ distribution, which additionally shows a very consistent beneficial result. In the right column (${\uparrow}\mcsd$) of Fig. \ref{fig:boxplot_Nconst1b} we find some clear trends regarding the interplay of $\mcsd$ and $\rp$. Here, the total range of data points of both contrast values strictly increases with $\rp$ and the medians are shifted increasingly toward an unfavorable direction. At $\rp=\rpL$ both contrast distributions are very consistent, meaning that all data points within each IQR share the same sign. In contrast, at $\rp=\rpS$ the $\ContrastStar$ distribution is strongly advantageously skewed and the IQR extends deeply into the beneficial range. Therefore, there are instances where an increase in $\mcsd$ actually is beneficial for the detectability of an embedded planet. However, this is mostly restricted to close-in planets, which are unfortunately difficult to detect. Regarding the wavelength dependence we find that the IQR of $\ContrastStar$ distributions is significantly wider at $\lambda=\lambdaS$ compared to the longer wavelengths.  At this wavelength, we also find the largest fraction of beneficial contrast differences. According to the different results from Fig. \ref{fig:boxplot_Nconst1b} we find that the fraction of models that benefit from an increase in $\mcsd$ mostly consist of simulated planets at $\rp=\rpS$ when observed at $\lambda=\lambdaS$. Moreover, these planets are rather low in mass and particularly $\Mp=0.5\,\mj$ planets seem to cause this effect. Furthermore, since the $\ContrastStar$ distributions for the parameters $\mcpd$ and       $\mpdot$ only show  weak differences among shared values, we conclude that this is for the most part an optical depth effect. That this occurs more often for farther-in planets suggests that the effect is most likely caused by scattered stellar flux, and is therefore less dependent on the optical depth between the observer and the planet than on the optical depth as observed from photons that are emitted from the star and scatter off dust grains located inside the gap. This hypothesis is further supported by the fact that at $\rp=\rpS$ the $\ContrastGap$ distribution is extremely narrow and even centered around zero. When trying to deduce the mass of the CSD solely using VIS/NIR data, these results suggest that different values of $\mcsd$ can be best  distinguished when observing at $\lambda=\lambdaS$.

\subsection{Caveats}
This study presents and applies a method of testing the abilities of SPHERE to detect and characterize embedded planets. In this section we assess potential caveats regarding the  method and underlying assumptions. In this context we discuss the robustness of our method, the impact of the chemical composition of the dust and of  the inclination angle, and finally the existence and potential structure of the CPD. Further elaborating on these topics, for instance in the framework of a broad parameter study, would certainly deepen our understanding of the capabilities of SPHERE. Based on the results of this study, though, we can only attempt to estimate the relevance these topics. 

The method we used to quantify the strength of the planetary and CPD signal involves the determination of three different regions:  the azimuthal CPD region, the gap region, and the transition region. The robustness of this method thus depends on the capability of properly identifying them. This process, however, may be performed inaccurately either due to the misidentification of the position of the potential planet or due to an incorrect estimation of the radius $\rtranshat$ from Eq. \eqref{eq:rtranshat_def}. The consequence of such an error certainly depends on its severity; however, small errors can be expected to only result in slight changes of the outcome. If, for instance, the azimuthal CPD region is slightly shifted but still covers the brightest local region at the position of the planet, the subsequently determined signal strength of the planet and CPD will be reduced. However, due to the averaging that is involved in the calculation of the radial CPD region only a fraction of the flux is lost, which stems from the rim of the azimuthal CPD region. Similarly, if the estimation of the  azimuthal gap region is  slightly off, the azimuthal averaging of the gap profile suppresses potentially considered contaminated flux that once more stems from the rim of the azimuthal gap region. Thus, it can be expected that both errors only slightly lower the determined planetary and CPD signal at any observed wavelength.

Next, we discuss different assumptions that were made in the course of this study. The chemical composition of dust  in general plays a crucial role for the observed properties of a PPD. Even though silicates and carbonaceous material  make up a significant portion of dust in the interstellar medium and consequently in CSDs \citep[e.g.,][]{2003ARA&A..41..241D,2010A&A...515A..77L}, it can be expected that these disks contain various other materials as well (e.g., in the form of water ice). To reduce the complexity of the modeled systems we focused on these key components, even though an extended study with a particular focus on the dust composition, size, and shape would certainly add to our findings.

Furthermore, we assumed the disk to be observed face-on (i.e.,  an inclination angle of $i=0^\circ$). This angle has an impact on the optical depth between the observer and the embedded planet, and thus influences the attenuation of light originating from the planet and its CPD before its detection. As a function of the orbital phase angle of the planet the determined excess signal strength of the planet and CPD thus varies. Overall, it can be expected that an inclined CSD would often lead to an increase in the optical depth, and as a result would lower the possibility of unveiling planets, making their direct detections in more inclined disks more challenging.

Moreover, we made a particular choice for the model of a CPD. However, to date, none of the modern instruments is capable of providing a high-resolution image of any such disk. To provide the most reliable results, this study assumes models of a CPD that are constructed based on results of state-of-the-art hydrodynamics simulations. Since the thermal and density structure of the CPD determines its detectability, a different model or the complete absence of the CPD may very well change the determined signal strength. Consequently, conducting a similar study but using a model of a circumplanetary envelope or even assuming the lack of any circumplanetary material instead would certainly be insightful. However, since the presence of CPDs seems to be a very likely scenario for the range of considered planetary masses, it can be expected that the results of this study apply to a wide range of PPDs harboring young accreting planets.

\section{Summary and conclusions}
\label{sec:summary}

We present a study on the detectability and characterizability of embedded low-mass giant planets in gaps of CSDs of typical T Tauri stars in the VIS/NIR wavelength range. To this end, we performed fully self-consistent MCRT simulations using the code Mol3D \citep{Ober_2015} for 72 models and three different observing wavelengths ($\lambdaS$, $\lambdaM$, and $\lambdaL$) each. Our model is composed of a central star that is surrounded by a CSD in which a planet is harbored that carves a gap into the CSD as it accretes material. The models make use of empirically determined gap profiles \citep{2016PASJ...68...43K,2017PASJ...69...97K,2019ApJ...884..142G} and the embedded planets themselves are surrounded by optically thick CPDs and emit accretion luminosity, thus heating up the corresponding environment. The parameter domain that is covered by the models is described in Tab. \ref{tbl:parameters} and contains, in addition to the wavelength $\lambda$, five other parameters: the planetary mass $\Mp$, the planet's distance to the star $r_p$, its mass accretion rate $\mpdot$, the CPD mass $\mcpd$, and the CSD mass $\mcsd$.
In order to properly treat the extremely high optical depths encountered in CPDs we made use of a method for optically thick dusty media as introduced by \cite{2020A&A...635A.148K}. Furthermore, a method to access the impact of a planet and its CPD was introduced that distinguishes between three azimuthally divided regions in the CSD: the azimuthal gap region, the azimuthal CPD region, and the transition region that separates  these two regions. This method utilizes a procedure to convert a Cartesian detector grid to a polar detector grid, for which  the python package CartToPolarDetector was specifically written, and which is described in Sect. \ref{sec:package_cart_to_polar}. The measured and quantified impact of a planet and its CPD on observations in the VIS/NIR is summarized in Tabs. \ref{tbl:results_wl24} to \ref{tbl:results_wl35} in the Appendix, and subsequent analyses of these data were presented throughout this paper. The results are summarized in the following:

\begin{itemize}
\item[1.] We apply a detection criterion (see Sect. \ref{sec:detectability}) based solely on the contrast between the stellar and the planetary and CPD region ($\ContrastStar$) which is $\lambda$- and $r_p$-dependent, and find that the majority of simulations in the investigated part of the parameter space would result in a non-detection of the planetary and CPD signal when using SPHERE. However, there are simulations that satisfy our condition for detectability even within the studied parameter domain. In particular, we find that all investigated simulations at $r_p\leq10\,$au are not detectable, and only  far-out planets have a chance to be detected. 

\item[2.] According to our results, a non-detection far-out at $r_p{=}50\,$au can still be used to restrict the underlying properties of a potential embedded planet which depend on the wavelength at which the observation was performed. Considering the trends found in Fig. \ref{fig:boxplot_Nconst0} and described in Tab. \ref{tbl:upped_parameters}, these are our findings.

        First, using ZIMPOL at $\lambda{=}\lambdaS$, a non-detection implies
        \begin{itemize}
                \item[(i)] $\Mp<1\,\mj$ or $\mpdot<\mpdotL$\\ for $\mcsd=\mcsdL$, and
                \item[(ii)] $\left(\Mp<1\,\mj \,\land\, \mpdot<\mpdotL\right)$\\ or $\Mp<0.5\,\mj$ for $\mcsd=\mcsdS$.
        \end{itemize}           
        Second, using IRDIS at $\lambda{=}\lambdaM$, a non-detection implies
        \begin{itemize}
                \item[(i)] $\Mp<0.5\,\mj$ or $\mpdot<\mpdotS$\\ for $\mcsd=\mcsdL$, and
                \item[(ii)] $\Mp<0.25\,\mj$ or $\mpdot<\mpdotS$\\ for $\mcsd=\mcsdS$.
        \end{itemize}
        Third, using IRDIS at $\lambda=\lambdaL$, a non-detection implies $\Mp<0.25\,\mj$ or $\mpdot<\mpdotS$ for both $\mcsd=\mcsdL$ and $\mcsd=\mcsdS$.
Overall, our results within the VIS/NIR wavelength range suggest that observations at $\lambda=\lambdaL$ provide the greatest potential for future detections, which in the case of SPHERE is possible using IRDIS. 

\item[3.] The characterization of the planets and their CPDs has been studied with regard to the observing wavelength and five different other parameters. We find that  characterizing and detecting low-mass giant planets in gaps in the VIS/NIR wavelength range often benefit from the contribution of various observing wavelengths (see Sect. \ref{sec:parameter_impact}). In the following we present a list of our key findings regarding the relevance and impact of each of those studied parameters on the characterizability of directly imaged planets. 

\begin{itemize}
        \item[a)] The distance between the star and the planet $\rp$ is decisive for the detectability. In general, its increase can both improve and worsen the contrast $\ContrastStar$. However, far-out planets strongly benefit from high values of $\rp$ due to the effect on the optical depth between the planet and the observer, which decreases as $r_p$ increases and thus exposes the planetary and CPD signal. Since the position of a directly imaged planet can be inferred straightforwardly from the position on the image, we find that its determination has the greatest potential at longer wavelengths when the chances for a detection are the highest (i.e., at $\lambda=\lambdaL)$.
        \item[b)] The planetary mass $\Mp$ is highly responsible for the observed planetary and CPD signal. Judging from the high sensitivity with which the observed planetary signal reacts to a change in $\Mp$ within the studied parameter domain, we find that observations at $\lambda=\lambdaS$ are best suited for its characterization.
        \item[c)] The mass accretion rate $\mpdot$ onto the planet has a comparably weak impact on the observed contrast values. We find a correlation between $\ContrastStar$ and the inner radius of the CPD $\rin$ when varying $\mpdot$. In particular, for increasing $\mpdot$ values the contrast $\ContrastStar$ often worsens as $\rin$ further increases. Overall, our results suggest that $\lambda=\lambdaS$ is the best suited observing wavelength for distinguishing between different values of $\mpdot$.
        \item[d)] Overall, the CPD mass $\mcpd$ has the smallest impact on the measured contrast values, making it the most difficult to determine  in the investigated parameter domain. We find that its increase shows the trend to improving the observed signal of the planet and CPD. Moreover, its characterizability benefits the most from longer observing wavelengths (i.e., $\lambda=\lambdaL$) since an increase (decrease) of $\mcpd$ mainly induces a strengthening (weakening) of the back-warming effect (see Sect. \ref{sec:determination_r_in_cpd}), which primarily affects the thermal radiation of the CPD. 
        \item[e)] The CSD mass $\mcsd$ plays a crucial role in the detectability of an embedded planet. Most systems benefit from a reduction of $\mcsd$; however, we also identified a potentially beneficial effect for $\ContrastStar$ from its increase for far-in planets and in particular planets of mass $\Mp = 0.5\,\mj$, which we attribute to an optical depth effect (for details see Sect. \ref{sec:mcsd_dependence}). Within the VIS/NIR wavelength range we find that the best suited observing wavelength for distinguishing between different values of $\mcsd$ is $\lambda=\lambdaS$. 
\end{itemize}

\item[4.] An analysis of planetary signals that are blocked by a coronagraphic mask clearly suggests that the current capabilities of SPHERE regarding the detection of embedded planets are not limited by the use of a coronagraph (see Sect. \ref{sec:detectability}). In particular, we find that the majority of simulated planets whose signal is blocked by the coronagraphic mask, which in our case are planets at $r_p=\rpS$ that are observed with ZIMPOL at $\lambda=\lambdaS$, result in very low $\ContrastStar$ values, which are well inside the undetectable range. In addition, we find that the majority of these systems generate contrast values between the gap and the planetary and CPD region of $\ContrastGap<1\,$mag. Consequently, this clearly suggest that decreasing the IWA would not result in any more detections of planets that are within the studied parameter domain when using SPHERE/ZIMPOL.

\item[5.] Our applied method for quantifying the planetary and CPD signal includes the definition of a contamination-free azimuthal gap region, which is defined via the wavelength-dependent radius $\rtranshat$ and the distance between the star and the planet $r_p$. In Sect. \ref{sec:detectability} we find that the contrast $\ContrastGap$ is strikingly reduced for all contaminated systems, which is likely a consequence of an elevated flux level of the contaminated gap. The contrast reaches in all simulated cases values of only $\ContrastGap< 0.2\,$mag. This shows, first of all, that our definition of the azimuthal gap region via $\rtranshat$ in Eq. \eqref{eq:rtranshat_def} together with the condition for a contamination-free  gap in Eq. \eqref{eq:condition_azimuthal_gap_region} is reliably categorizing systems with low contrast values and, second, that the detection of embedded planets that do not satisfy the condition in Eq. \eqref{eq:condition_azimuthal_gap_region} is indeed extremely challenging.

\item[6.] We find that the inner radius of the CPD $\rin$ is strongly affected by the back-warming effect  described in Sect. \ref{sec:determination_r_in_cpd}. As a consequence, we find that when keeping the remaining parameters constant the inner radius of the CPD ($\rin$) increases if $\rp$ decreases, $\Mp$ increases, $\mpdot$ increases, or $\mcpd$ increases. Its dependence on $\mcsd$ is shown to be comparably weak. A change in $\rin$ leads to a change in the thermal and density structure of the CPD, which consequently affects its spectral energy distribution. This is relevant since we find that the thermal radiation of the CPD may strongly contribute to the measured flux at the position of the planet (see Sect. \ref{sec:sphere_ideal}), in particular the self-scattered part of its thermal radiation since the direct part is highly attenuated by the optical depth between the planet and the observer.
\end{itemize}
Finally, it is possible to evaluate the fact that to this day only the planets PDS 70 b and PDS 70 c have been confirmed by direct imaging. Even though these two giant planets are outside the investigated parameter domain, the trends we find are very much in line with the observations, since we find great benefits for the contrast for high values of $r_p$ and $\Mp$. We find this to be particularly the case for SPHERE/IRDIS observations, which were also part of the multiwavelength analysis that was performed to confirm the presence of these planets. It is interesting, though, that our parameter study reveals the possibility for detecting significantly lower planets with a mass of $\Mp=1\,\mj$ using SPHERE. However, such a detection seems to be only feasible in the case of far-out planets with high mass accretion rates and especially low CSD masses leading to a lower optical depth between the observer and the planet. 

Nonetheless, to reliably deduce the presence of a low-mass giant planet embedded in a gap of a CSD it is crucial to perform multiwavelength observations. This is not only increasing the chances for detecting such a planet but is at the same time beneficial for its characterization, since we have seen that different wavelengths, even within the VIS/NIR wavelength range, possess different potentials for characterizing the various basic properties of both the planet and its CPD. Therefore, to assess the full potential for detecting and characterizing embedded planets, it would be important to conduct another study focusing primarily on the millimeter and submillimeter wavelength range. This would allow us to evaluate the potential for detecting and confirming additional planets in the future either by using state-of-the-art or future instrumentation. In addition, it may provide new insights that help guide the planning and design of future observatories in the quest for observing embedded accreting planets while they form.

\section*{ORCID iDs}
A. Krieger \orcidlink{0000-0002-3639-2435}
\href{https://orcid.org/0000-0002-3639-2435}
     {https://orcid.org/0000-0002-3639-2435}\\
S. Wolf \orcidlink{0000-0001-7841-3452}
\href{https://orcid.org/0000-0001-7841-3452}
     {https://orcid.org/0000-0001-7841-3452}

\begin{acknowledgements}
         We thank all the members of the Astrophysics Department Kiel for helpful discussions and remarks. 
         We acknowledge the support of the DFG priority program SPP 1992 "Exploring the Diversity of Extrasolar Planets (WO 857/17-1)".
                 This research has made use of seaborn \citep{Waskom2021}, a library for making statistical graphics in python.
\end{acknowledgements}

\bibliographystyle{aa} 
\bibliography{literature} 

\begin{appendix}

\section{Back-warming in 1D}
\label{sec:back-warming}
To explain the back-warming effect, we use a one-dimensional mono-chromatic model that is composed of a isotropically radiating light source in a vacuum that illuminates a one-dimensional connected density distribution (slab) of total optical thickness $\tau_{b}=2b$. In this section we  show how the temperature distribution inside the slab changes as its density changes. In particular, we  show that the temperature at the inner edge of the slab (i.e., the illuminated edge) increases and the temperature of the outer edge decreases when the density of the slab is increased. Since the effect is independent of the albedo $A$ of the medium, we chose $A=1$. 

The derivation of the resulting temperature distribution presented in this section is fully based on the  eigenstate-based description of this problem introduced by \cite{2021A&A...645A.143K} (hereafter   KW21). 

The initial distribution of photons $\nu_0(x)$ is described by
\begin{equation}
\nu_0(x) = 2 \delta\left( x+b\right),
\end{equation}
where $x\in \left[ -b,b \right] $ is the optical depth coordinate and $\delta(x)$ the delta distribution. A portion of the emitted photons enter the slab and are absorbed and reemitted, potentially multiple times, before all photons eventually (i.e., in the limit of $n\rightarrow\infty$ interactions)  leave the slab through either  of its edges. The distribution of photons that interacted $n$ times is described by
\begin{equation}
\nu_n(x)= A^n \sum_{k=1}^\infty {\lambda_k}^n a_k \cos\Bigl( \sigma_k x\Bigr) + {\lambda_k^*}^n b_k \sin\Bigl( \sigma_k^* x\Bigr),
\end{equation}
where $a_k$ and $b_k$ are expansion coefficients of $\nu_0(x)$; $\lambda_k$ and $\lambda_k^*$ are the eigenvalues of eigenstates $\cos(\sigma_k x)$ and $\sin(\sigma_k x)$, respectively; and $A$ is the albedo, which in our case is 1. 
Applying Eqs. A.13 to A.16 from KW21 gives
\begin{equation}
a_k = \frac{2}{b+\lambda_k} \cos(\sigma_k b) \quad \text{and} \quad b_k = -\frac{2}{b+\lambda_k^*} \sin(\sigma_k^* b).
\end{equation}
The temperature at any point of the slab is determined by the ambient radiation field at that point. The quantity $\nu_{\rm tot}(x)$ is a measure of the radiation field and is given by 
\begin{equation}
\nu^\times_{\rm tot}(x) = \sum_{n=1}^\infty \nu_n(x).
\end{equation}
Using the previous equations and applying Eqs. A.6 to A.8, A.39, and A.43 from KW21, we arrive at
\begin{equation}
\nu^\times_{\rm tot}(x) = 1 - \frac{x}{b+1}.
\label{eq:analytic_rad_field}
\end{equation}
Therefore, the radiation field reaches its maximum at the inner edge, where $\nu^\times_{\rm tot}(-b)=1+b/(b+1)$, and its minimum at the outer edge, where $\nu^\times_{\rm tot}(b)=1-b/(b+1)$. Here, we use the optical depth coordinate $x$ rather than a spatial coordinate. Thus, an increase in total optical depth is equivalent to an increase in density of the medium. According to Eq. \ref{eq:analytic_rad_field}, the radiation field consequently increases with increasing optical depth $b$. In other words, if a photon exits a slab at its outer edge, it will no longer increase the temperature of the slab, unless more optical depth is added the outer edge, which has the potential to interact with the photon and warm up the slab from the back. This effect is called back-warming. 

Interestingly, even though the slab warms up at the inner edge and cools down at the outer edge, when increasing the density the average temperature of the slab in this one-dimensional setup stays constant. This is the case since the average radiation field, according to Eq. \ref{eq:analytic_rad_field}, is constant and in particular independent of the optical depth $\tau_{b}$. Additionally, this can be shown by calculating the mean traversed path length of photons traveling through the slab, which is given by
\begin{equation}
\left\langle l \right\rangle = \int_{-b}^{b} \sum_{n=0}^\infty\, \nu_n(x) l(x)\,{\rm d}x,
\label{eq:avg_length_traversed}
\end{equation}
where $l(x)$ is the mean optical depth traversed by a neutral photon package (i.e., weight equals 1), originating from position $x$ within one step of interaction. The quantity $l(x)$ contains contributions from photons that leave the slab through its inner edge, from photons that are absorbed inside the slab, and from photons that leave the slab through the outer edge. It is thus given by
\begin{equation}
l(x) = \frac{b+x}{2} \int_{b+x}^\infty e^{-\tau}\,d\tau + \frac12 \int_{-b}^b \abs{\tau-x}e^{-\abs{\tau-x}}\,d\tau  +  \frac{b-x}{2} \int_{b-x}^\infty e^{-\tau}\,d\tau.
\end{equation}
Performing the integration in Eq. \ref{eq:avg_length_traversed} yields
\begin{equation}
\left\langle l \right\rangle  = 2b.
\end{equation}
The mean length traversed by a photon that enters the slab is therefore exactly given by the optical depth of the slab. In this one-dimensional case, the mean deposited energy per optical depth unit for photons that enter the slab thus equals $\left\langle l \right\rangle/\tau_{b}=1$ and does not depend on $\tau_{b}$.

\section{Converting Cartesian to polar detector}
\label{sec:package_cart_to_polar}

In this section we describe the basic principle of the python package CartToPolarDetector which is used to convert a Cartesian detector into its polar representation. The original Cartesian detector is described by $N_x$ and $N_y$ pixels of equal width in $x$-direction and $y$-direction, respectively. Each of these pixels has an assigned flux density value measured in Jy. Properties of the desired polar grid can be defined by the user. In particular, the grid is defined by the position of its center with regard to the original grid (shift), its radial extent, the number of cells in $\phi$-direction $N_\phi$, and the number of cells in $r$-direction $N_r$. By default, the shift is deactivated, meaning the two grid centers coincide, and the radial extent is chosen, such that the whole Cartesian detector is embedded in the polar detector. $N_r$ and $N_\phi$ can take any integer number greater than zero; however, $N_\phi$ must be divisible by four. 

During its execution, the code loops over all Cartesian pixels and classifies each pixel's respective position relative to the polar grid center with terms such as left, right, top, bottom,  middle, or combinations of them. Then, the $\phi$-range is determined in which polar pixels may intersect with the Cartesian pixel. Since any determined intersection area of two pixels does not change when the detectors rotate, the detectors are rotated by an integer multiple of $\pi/2$ such that the new Cartesian pixel position is either in the top or in the top right position. An exception to this is made only in the case that the position is initially classified as middle, which means that the polar center is inside the pixel, but not in any of the pixel's edges or corners, in which case no rotation is performed. Such a pixel is unique and is treated separately. After the intersection area calculation step, the pixels are rotated back and the corresponding contributions of Cartesian pixels to polar pixels are added to the polar grid. The assigned value that is added to a polar pixel is a product of the value assigned to the Cartesian pixel and the ratio of the corresponding intersection area to the total area of the Cartesian pixel. In a last step, after all Cartesian pixels have been looped over and the polar grid has been calculated, it is stored. Additionally, all input parameters as well as a list of pixel border values for $\phi$ and $r$ are stored, where $\phi$ and $r$ are measured in radians and in original Cartesian pixels, respectively.

 \begin{figure}
   \centering
   \includegraphics{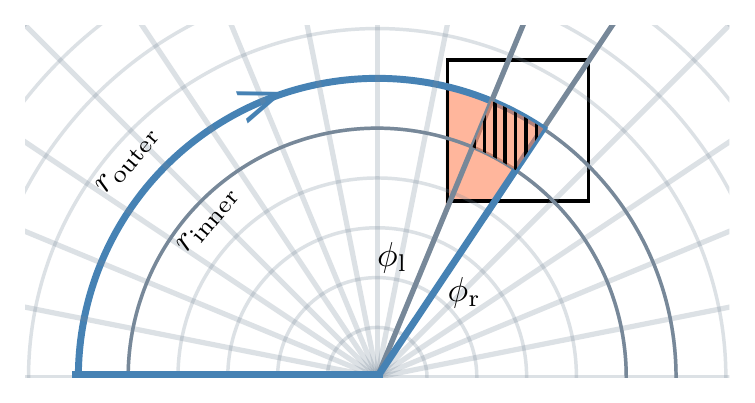}
   \caption{Integrated intersection area (orange). It is the intersection area of a Cartesian pixel (black) with a segment of a circle (blue) and it is assigned to a polar pixel (hatched region) that is defined by an inner radius $r_{\rm inner}$, an outer radius $r_{\rm outer}$, a left border $\phi_l$, and a right border $\phi_r$. The overlayed polar grid is shown in gray and the two radial and angular borders of a selected polar pixel are highlighted in dark gray.}
              \label{fig:integrated_intersection}
   \end{figure} 

The calculation of the intersection area is the most complicated step in terms of the number of possible ways an arbitrarily shaped polar pixel can intersect a Cartesian pixel. However, instead of showing the exact intersection formula for all cases, we briefly summarize the general approach by which they were obtained. However, to remove some of the complexity of this step, we divide it into two parts. In the first part we determine an integrated intersection area. Figure \ref{fig:integrated_intersection} illustrates the concept of the integrated intersection area. A polar pixel is defined by an inner and an outer radius, as well as two $\phi$ borders. Since we are dealing with pixels that, after a potential rotation, are located in the top region or top right region, we can define one $\phi$ border as the left and the other $\phi$ border as the right border, where the right border $\phi_r$ is pointing more toward the direction of the positive $x$-axis than the left border $\phi_l$. In the upper half of this plot a segment of a circle (blue) can be defined by the outer pixel radius and covering the range from $\phi=-\pi/2$ to the right border $\phi_r$, as can be seen in the plot, where $\phi=0$ corresponds to the direction of the positive $y$-axis. The integrated intersection area (orange) is then the intersection area of this segment of the circle with the Cartesian pixel (black). It is called the integrated intersection area as it contains the desired individual intersection area (hatched), but may also contain intersection areas that belong to other polar pixels.

In the second part of this calculation, the individual contributions can easily be obtained after determining the integrated intersection areas of all relevant polar pixels. Let $S_{i,j}$ the integrated intersection area of a polar pixel $(i,j)$, where $i$ and $j$ are non-negative integers with $i<N_\phi$ and $j<N_r$, and let $I_{i,j}$ the individual intersection area of that same pixel, then we find for all pixels in the upper half the relation
\begin{equation}
I_{i,j} = 
        \begin{cases}
                S_{i,j} - S_{(i-1)\,{\rm mod}\,N_\phi,j} - S_{i,j-1} + S_{(i-1)\,{\rm mod}\,N_\phi,j-1} & \text{if } j>0,\\
                S_{i,j} - S_{(i-1)\,{\rm mod}\,N_\phi,j} & \text{else,}\\
        \end{cases}
\end{equation}
where ${\rm mod}$ denotes  modulo operation. Due to the previously performed rotation, polar pixels in the lower half do not intersect with the Cartesian pixel. 

\section{Results}

\begin{table*}

\begin{center}
\resizebox*{!}{0.95\textheight}{%
\begin{tabular}{ccccc|c}
\toprule
$\rp\,\left[{\rm au}\right]$    &       $\Mp\,\left[\mj\right]$ &       $\mpdot\,\left[10^{-6}\,\mj/{\rm yr}\right]$     &       $\mcpd\,\left[10^{-3}\,\mj\right]$      &       $\mcsd\,\left[{\rm M}_\odot\right]$        &       $\rin\,\left[\rj\right]$        \\
\cmidrule{1-6}
5       &       1       &       0.4     &       0.1     &       0.01    &       10.2    \\
10      &       1       &       0.4     &       0.1     &       0.01    &       9.2     \\
50      &       1       &       0.4     &       0.1     &       0.01    &       8.3     \\
5       &       0.5     &       0.4     &       0.1     &       0.01    &       9.4     \\
10      &       0.5     &       0.4     &       0.1     &       0.01    &       8.6     \\
50      &       0.5     &       0.4     &       0.1     &       0.01    &       7.7     \\
5       &       0.25    &       0.4     &       0.1     &       0.01    &       7.4     \\
10      &       0.25    &       0.4     &       0.1     &       0.01    &       7       \\
50      &       0.25    &       0.4     &       0.1     &       0.01    &       7.2     \\
5       &       1       &       2       &       0.1     &       0.01    &       16.7    \\
10      &       1       &       2       &       0.1     &       0.01    &       15.8    \\
50      &       1       &       2       &       0.1     &       0.01    &       12.8    \\
5       &       0.5     &       2       &       0.1     &       0.01    &       13.8    \\
10      &       0.5     &       2       &       0.1     &       0.01    &       13.3    \\
50      &       0.5     &       2       &       0.1     &       0.01    &       11.7    \\
5       &       0.25    &       2       &       0.1     &       0.01    &       11.6    \\
10      &       0.25    &       2       &       0.1     &       0.01    &       10.5    \\
50      &       0.25    &       2       &       0.1     &       0.01    &       9.8     \\
5       &       1       &       0.4     &       1       &       0.01    &       22.6    \\
10      &       1       &       0.4     &       1       &       0.01    &       19.5    \\
50      &       1       &       0.4     &       1       &       0.01    &       17.6    \\
5       &       0.5     &       0.4     &       1       &       0.01    &       17.6    \\
10      &       0.5     &       0.4     &       1       &       0.01    &       16.7    \\
50      &       0.5     &       0.4     &       1       &       0.01    &       15.1    \\
5       &       0.25    &       0.4     &       1       &       0.01    &       9.8     \\
10      &       0.25    &       0.4     &       1       &       0.01    &       13.3    \\
50      &       0.25    &       0.4     &       1       &       0.01    &       12.2    \\
5       &       1       &       2       &       1       &       0.01    &       46      \\
10      &       1       &       2       &       1       &       0.01    &       35.4    \\
50      &       1       &       2       &       1       &       0.01    &       28      \\
5       &       0.5     &       2       &       1       &       0.01    &       38.1    \\
10      &       0.5     &       2       &       1       &       0.01    &       30.8    \\
50      &       0.5     &       2       &       1       &       0.01    &       23.3    \\
5       &       0.25    &       2       &       1       &       0.01    &       32.1    \\
10      &       0.25    &       2       &       1       &       0.01    &       25.4    \\
50      &       0.25    &       2       &       1       &       0.01    &       19.6    \\
5       &       1       &       0.4     &       0.1     &       0.001   &       10.2    \\
10      &       1       &       0.4     &       0.1     &       0.001   &       9.2     \\
50      &       1       &       0.4     &       0.1     &       0.001   &       8.3     \\
5       &       0.5     &       0.4     &       0.1     &       0.001   &       9.4     \\
10      &       0.5     &       0.4     &       0.1     &       0.001   &       8.6     \\
50      &       0.5     &       0.4     &       0.1     &       0.001   &       7.7     \\
5       &       0.25    &       0.4     &       0.1     &       0.001   &       7.4     \\
10      &       0.25    &       0.4     &       0.1     &       0.001   &       7       \\
50      &       0.25    &       0.4     &       0.1     &       0.001   &       7.2     \\
5       &       1       &       2       &       0.1     &       0.001   &       16.7    \\
10      &       1       &       2       &       0.1     &       0.001   &       15.8    \\
50      &       1       &       2       &       0.1     &       0.001   &       12.8    \\
5       &       0.5     &       2       &       0.1     &       0.001   &       13.8    \\
10      &       0.5     &       2       &       0.1     &       0.001   &       13.3    \\
50      &       0.5     &       2       &       0.1     &       0.001   &       11.7    \\
5       &       0.25    &       2       &       0.1     &       0.001   &       11.6    \\
10      &       0.25    &       2       &       0.1     &       0.001   &       10.5    \\
50      &       0.25    &       2       &       0.1     &       0.001   &       9.8     \\
5       &       1       &       0.4     &       1       &       0.001   &       22.6    \\
10      &       1       &       0.4     &       1       &       0.001   &       19.5    \\
50      &       1       &       0.4     &       1       &       0.001   &       17.6    \\
5       &       0.5     &       0.4     &       1       &       0.001   &       17.6    \\
10      &       0.5     &       0.4     &       1       &       0.001   &       16.7    \\
50      &       0.5     &       0.4     &       1       &       0.001   &       15.1    \\
5       &       0.25    &       0.4     &       1       &       0.001   &       9.8     \\
10      &       0.25    &       0.4     &       1       &       0.001   &       13.3    \\
50      &       0.25    &       0.4     &       1       &       0.001   &       12.2    \\
5       &       1       &       2       &       1       &       0.001   &       46      \\
10      &       1       &       2       &       1       &       0.001   &       35.4    \\
50      &       1       &       2       &       1       &       0.001   &       28      \\
5       &       0.5     &       2       &       1       &       0.001   &       38.1    \\
10      &       0.5     &       2       &       1       &       0.001   &       30.8    \\
50      &       0.5     &       2       &       1       &       0.001   &       23.3    \\
5       &       0.25    &       2       &       1       &       0.001   &       32.1    \\
10      &       0.25    &       2       &       1       &       0.001   &       25.4    \\
50      &       0.25    &       2       &       1       &       0.001   &       19.6    \\
\bottomrule
\end{tabular}
}
\caption{Determined value for $\rin$ shown for all simulated models, where the first five columns give the set of underlying parameter values. For details, see Sect. \ref{sec:determination_r_in_cpd}.}
\label{tbl:r_in_cpd}
\end{center}

\end{table*}

\begin{table*}

\begin{center}
\resizebox*{!}{0.88\textheight}{%
\begin{tabular}{ccccc|ccc|cccc|c}
\toprule
$\rp\,\left[{\rm au}\right]$    &       $\Mp\,\left[\mj\right]$ &       $\mpdot\,\left[10^{-6}\,\mj/{\rm yr}\right]$     &       $\mcpd\,\left[10^{-3}\,\mj\right]$      &       $\mcsd\,\left[{\rm M}_\odot\right]$        &       Type    &       Hidden  &       Cont.   &       C{:}Star        &       R{:}Star        &       Gap{:}Star &       $\ContrastStar$ &       $\ContrastGap$  \\
\cmidrule{1-13}
5       &       1       &       0.4     &       0.1     &       0.01    &       Gap     &       $\times$        &                &       10.7    &       9.7     &       11      &       10.1    &       0.1     \\ 
10      &       1       &       0.4     &       0.1     &       0.01    &       Gap     &                &                       &       10.7    &       10.7    &       13.9    &       13.5    &       0.2     \\ 
50      &       1       &       0.4     &       0.1     &       0.01    &       Gap     &                &                       &       9.1     &       15.4    &       18.4    &       14.5    &       3.8     \\ 
5       &       0.5     &       0.4     &       0.1     &       0.01    &       Gap     &       $\times$        &                &       9.6     &       9.2     &       9.8     &       9.4     &       0.0     \\ 
10      &       0.5     &       0.4     &       0.1     &       0.01    &       Gap     &                &                       &       10      &       11.3    &       13.1    &       11.9    &       0.1     \\ 
50      &       0.5     &       0.4     &       0.1     &       0.01    &       Gap     &                &                       &       9.7     &       14.7    &       17.1    &       16.6    &       0.4     \\ 
5       &       0.25    &       0.4     &       0.1     &       0.01    &       Gap     &       $\times$        &                &       10.6    &       10.5    &       10.7    &       10.4    &       0.1     \\ 
10      &       0.25    &       0.4     &       0.1     &       0.01    &       Gap     &                &                       &       10.8    &       11.3    &       13      &       11.8    &       0.1     \\ 
50      &       0.25    &       0.4     &       0.1     &       0.01    &       Gap     &                &                       &       10.2    &       14.7    &       16.1    &       15.6    &       0.3     \\ 
5       &       1       &       2       &       0.1     &       0.01    &       Gap     &       $\times$        &                &       10.7    &       9.7     &       11      &       10.7    &       0.2     \\ 
10      &       1       &       2       &       0.1     &       0.01    &       Gap     &                &                       &       10.7    &       10.7    &       13.9    &       11.9    &       2.0     \\ 
50      &       1       &       2       &       0.1     &       0.01    &       Gap     &                &                       &       9.1     &       15.4    &       18.4    &       11.3    &       7.0     \\ 
5       &       0.5     &       2       &       0.1     &       0.01    &       Gap     &       $\times$        &                &       9.6     &       9.2     &       9.8     &       9.4     &       0.0     \\ 
10      &       0.5     &       2       &       0.1     &       0.01    &       Gap     &                &                       &       10      &       11.3    &       13.1    &       11.9    &       0.1     \\ 
50      &       0.5     &       2       &       0.1     &       0.01    &       Gap     &                &                       &       9.7     &       14.7    &       17.1    &       16.5    &       0.5     \\ 
5       &       0.25    &       2       &       0.1     &       0.01    &       Gap     &       $\times$        &                &       10.6    &       10.5    &       10.7    &       10.4    &       0.1     \\ 
10      &       0.25    &       2       &       0.1     &       0.01    &       Gap     &                &                       &       10.8    &       11.3    &       13      &       11.8    &       0.1     \\ 
50      &       0.25    &       2       &       0.1     &       0.01    &       Gap     &                &                       &       10.2    &       14.7    &       16.1    &       15.6    &       0.3     \\ 
5       &       1       &       0.4     &       1       &       0.01    &       Gap     &       $\times$        &                &       10.7    &       9.7     &       11      &       10.1    &       0.1     \\ 
10      &       1       &       0.4     &       1       &       0.01    &       Gap     &                &                       &       10.7    &       10.7    &       13.9    &       12.5    &       0.2     \\ 
50      &       1       &       0.4     &       1       &       0.01    &       Gap     &                &                       &       9.1     &       15.4    &       18.4    &       15.2    &       3.1     \\ 
5       &       0.5     &       0.4     &       1       &       0.01    &       Gap     &       $\times$        &                &       9.6     &       9.2     &       9.8     &       9.4     &       0.0     \\ 
10      &       0.5     &       0.4     &       1       &       0.01    &       Gap     &                &                       &       10      &       11.3    &       13.1    &       11.9    &       0.1     \\ 
50      &       0.5     &       0.4     &       1       &       0.01    &       Gap     &                &                       &       9.7     &       14.7    &       17.1    &       16.6    &       0.4     \\ 
5       &       0.25    &       0.4     &       1       &       0.01    &       Gap     &       $\times$        &                &       10.6    &       10.5    &       10.7    &       10.4    &       0.1     \\ 
10      &       0.25    &       0.4     &       1       &       0.01    &       Gap     &                &                       &       10.8    &       11.3    &       13      &       11.8    &       0.1     \\ 
50      &       0.25    &       0.4     &       1       &       0.01    &       Gap     &                &                       &       10.2    &       14.7    &       16.1    &       15.6    &       0.3     \\ 
5       &       1       &       2       &       1       &       0.01    &       Gap     &       $\times$        &                &       10.7    &       9.7     &       11      &       10.7    &       0.2     \\ 
10      &       1       &       2       &       1       &       0.01    &       Gap     &                &                       &       10.7    &       10.7    &       13.9    &       12      &       1.9     \\ 
50      &       1       &       2       &       1       &       0.01    &       Gap     &                &                       &       9.1     &       15.4    &       18.4    &       11.4    &       6.9     \\ 
5       &       0.5     &       2       &       1       &       0.01    &       Gap     &       $\times$        &                &       9.6     &       9.2     &       9.8     &       9.4     &       0.0     \\ 
10      &       0.5     &       2       &       1       &       0.01    &       Gap     &                &                       &       10      &       11.3    &       13.1    &       11.9    &       0.1     \\ 
50      &       0.5     &       2       &       1       &       0.01    &       Gap     &                &                       &       9.7     &       14.7    &       17.1    &       16.5    &       0.5     \\ 
5       &       0.25    &       2       &       1       &       0.01    &       Gap     &       $\times$        &                &       10.6    &       10.5    &       10.7    &       10.4    &       0.1     \\ 
10      &       0.25    &       2       &       1       &       0.01    &       Gap     &                &                       &       10.8    &       11.3    &       13      &       11.8    &       0.1     \\ 
50      &       0.25    &       2       &       1       &       0.01    &       Gap     &                &                       &       10.2    &       14.7    &       16.1    &       15.6    &       0.3     \\ 
5       &       1       &       0.4     &       0.1     &       0.001   &       Gap     &       $\times$        &                &       10.4    &       9.7     &       10.6    &       10.4    &       0.1     \\ 
10      &       1       &       0.4     &       0.1     &       0.001   &       Gap     &                &                       &       10.7    &       11.7    &       14.1    &       12.2    &       1.9     \\ 
50      &       1       &       0.4     &       0.1     &       0.001   &       Gap     &                &                       &       10.4    &       15.6    &       18.1    &       11.9    &       5.9     \\ 
5       &       0.5     &       0.4     &       0.1     &       0.001   &       Gap     &       $\times$        &                &       11.1    &       10.5    &       11.2    &       10.9    &       0.1     \\ 
10      &       0.5     &       0.4     &       0.1     &       0.001   &       Gap     &                &                       &       10.9    &       11.5    &       13.6    &       13.2    &       0.2     \\ 
50      &       0.5     &       0.4     &       0.1     &       0.001   &       Gap     &                &                       &       9.8     &       15.4    &       17.3    &       13.5    &       3.7     \\ 
5       &       0.25    &       0.4     &       0.1     &       0.001   &       Gap     &       $\times$        &                &       10      &       9.7     &       10.1    &       10      &       0.0     \\ 
10      &       0.25    &       0.4     &       0.1     &       0.001   &       Gap     &                &                       &       9.8     &       11.1    &       12.4    &       12.1    &       0.2     \\ 
50      &       0.25    &       0.4     &       0.1     &       0.001   &       Gap     &                &                       &       9.4     &       15.7    &       17      &       15.7    &       1.2     \\ 
5       &       1       &       2       &       0.1     &       0.001   &       Gap     &       $\times$        &                &       10.4    &       9.7     &       10.6    &       9       &       1.5     \\ 
10      &       1       &       2       &       0.1     &       0.001   &       Gap     &                &                       &       10.7    &       11.7    &       14.1    &       9       &       5.1     \\ 
50      &       1       &       2       &       0.1     &       0.001   &       Gap     &                &                       &       10.4    &       15.6    &       18.1    &       8.9     &       8.8     \\ 
5       &       0.5     &       2       &       0.1     &       0.001   &       Gap     &       $\times$        &                &       11.1    &       10.5    &       11.2    &       10.6    &       0.4     \\ 
10      &       0.5     &       2       &       0.1     &       0.001   &       Gap     &                &                       &       10.9    &       11.5    &       13.6    &       11.5    &       2.2     \\ 
50      &       0.5     &       2       &       0.1     &       0.001   &       Gap     &                &                       &       9.8     &       15.4    &       17.3    &       11.2    &       6.0     \\ 
5       &       0.25    &       2       &       0.1     &       0.001   &       Gap     &       $\times$        &                &       10      &       9.7     &       10.1    &       10      &       0.0     \\ 
10      &       0.25    &       2       &       0.1     &       0.001   &       Gap     &                &                       &       9.8     &       11.1    &       12.4    &       12.1    &       0.2     \\ 
50      &       0.25    &       2       &       0.1     &       0.001   &       Gap     &                &                       &       9.4     &       15.7    &       17      &       13.8    &       3.1     \\ 
5       &       1       &       0.4     &       1       &       0.001   &       Gap     &       $\times$        &                &       10.4    &       9.7     &       10.6    &       10.4    &       0.1     \\ 
10      &       1       &       0.4     &       1       &       0.001   &       Gap     &                &                       &       10.7    &       11.7    &       14.1    &       12.8    &       1.3     \\ 
50      &       1       &       0.4     &       1       &       0.001   &       Gap     &                &                       &       10.4    &       15.6    &       18.1    &       12.8    &       5.0     \\ 
5       &       0.5     &       0.4     &       1       &       0.001   &       Gap     &       $\times$        &                &       11.1    &       10.5    &       11.2    &       11      &       0.1     \\ 
10      &       0.5     &       0.4     &       1       &       0.001   &       Gap     &                &                       &       10.9    &       11.5    &       13.6    &       13.3    &       0.2     \\ 
50      &       0.5     &       0.4     &       1       &       0.001   &       Gap     &                &                       &       9.8     &       15.4    &       17.3    &       14.5    &       2.6     \\ 
5       &       0.25    &       0.4     &       1       &       0.001   &       Gap     &       $\times$        &                &       10      &       9.7     &       10.1    &       10      &       0.0     \\ 
10      &       0.25    &       0.4     &       1       &       0.001   &       Gap     &                &                       &       9.8     &       11.1    &       12.4    &       12.1    &       0.2     \\ 
50      &       0.25    &       0.4     &       1       &       0.001   &       Gap     &                &                       &       9.4     &       15.7    &       17      &       14.2    &       2.6     \\ 
5       &       1       &       2       &       1       &       0.001   &       Gap     &       $\times$        &                &       10.4    &       9.7     &       10.6    &       9       &       1.5     \\ 
10      &       1       &       2       &       1       &       0.001   &       Gap     &                &                       &       10.7    &       11.7    &       14.1    &       9.1     &       5.1     \\ 
50      &       1       &       2       &       1       &       0.001   &       Gap     &                &                       &       10.4    &       15.6    &       18.1    &       8.9     &       8.8     \\ 
5       &       0.5     &       2       &       1       &       0.001   &       Gap     &       $\times$        &                &       11.1    &       10.5    &       11.2    &       10.6    &       0.4     \\ 
10      &       0.5     &       2       &       1       &       0.001   &       Gap     &                &                       &       10.9    &       11.5    &       13.6    &       11.5    &       2.1     \\ 
50      &       0.5     &       2       &       1       &       0.001   &       Gap     &                &                       &       9.8     &       15.4    &       17.3    &       11.2    &       6.0     \\ 
5       &       0.25    &       2       &       1       &       0.001   &       Gap     &       $\times$        &                &       10      &       9.7     &       10.1    &       10      &       0.0     \\ 
10      &       0.25    &       2       &       1       &       0.001   &       Gap     &                &                       &       9.8     &       11.1    &       12.4    &       12.1    &       0.2     \\ 
50      &       0.25    &       2       &       1       &       0.001   &       Gap     &                &                       &       9.4     &       15.7    &       17      &       14.8    &       2.1     \\ 

\bottomrule
\end{tabular}
}
\caption{Compilation of results for all simulated models at $\lambda=\lambdaS$. The first five columns give the set of underlying parameter values. Contrast values (in mag) are obtained based on either of two types of features (Gap or Kink). Hidden planets are flagged with a cross and refer to planets whose inferred position of the planetary and CPD signal is hidden inside the IWA of the corresponding coronagraph, which can be found in Tab. \ref{tbl:instrument_sphere}. Contaminated planetary and CPD signals are flagged with a cross in the ``Cont.'' column and correspond to simulations that do not satisfy the condition stated in Eq. \eqref{eq:condition_azimuthal_gap_region}. The last five columns refer to contrast values either with respect to the stellar (Star) flux or to the flux of the gap region (Gap), where the following abbreviations were used: Coronagraph (C), ring (R), planet and CPD (CPD). For details, see Sect. \ref{sec:planetary_and_cpd_signal_strength}.}
\label{tbl:results_wl24}
\end{center}

\end{table*}

\begin{table*}
\begin{center}
\resizebox*{!}{0.88\textheight}{%
\begin{tabular}{ccccc|ccc|cccc|c}
\toprule
$\rp\,\left[{\rm au}\right]$    &       $\Mp\,\left[\mj\right]$ &       $\mpdot\,\left[10^{-6}\,\mj/{\rm yr}\right]$     &       $\mcpd\,\left[10^{-3}\,\mj\right]$      &       $\mcsd\,\left[{\rm M}_\odot\right]$        &       Type    &       Hidden  &       Cont.   &       C{:}Star        &       R{:}Star        &       Gap{:}Star &       $\ContrastStar$ &       $\ContrastGap$  \\
\cmidrule{1-13}
5       &       1       &       0.4     &       0.1     &       0.01    &       Kink    &                &       $\times$        &       3.4     &       $-$     &       $-$     &       8.8     &       0.0     \\ 
10      &       1       &       0.4     &       0.1     &       0.01    &       Gap     &                &                       &       3.4     &       10.3    &       11.7    &       9.5     &       2.1     \\ 
50      &       1       &       0.4     &       0.1     &       0.01    &       Gap     &                &                       &       3.4     &       14.1    &       16.7    &       8.7     &       7.9     \\ 
5       &       0.5     &       0.4     &       0.1     &       0.01    &       Kink    &                &       $\times$        &       3.4     &       $-$     &       $-$     &       7.2     &       0.0     \\ 
10      &       0.5     &       0.4     &       0.1     &       0.01    &       Gap     &                &                       &       3.4     &       9.8     &       10.5    &       10      &       0.0     \\ 
50      &       0.5     &       0.4     &       0.1     &       0.01    &       Gap     &                &                       &       3.4     &       14.3    &       16.6    &       11.6    &       5.0     \\ 
5       &       0.25    &       0.4     &       0.1     &       0.01    &       Kink    &                &       $\times$        &       3.4     &       $-$     &       $-$     &       7.2     &       0.0     \\ 
10      &       0.25    &       0.4     &       0.1     &       0.01    &       Gap     &                &                       &       3.4     &       10.7    &       10.9    &       10.8    &       0.0     \\ 
50      &       0.25    &       0.4     &       0.1     &       0.01    &       Gap     &                &                       &       3.4     &       13.8    &       15.2    &       14.8    &       0.1     \\ 
5       &       1       &       2       &       0.1     &       0.01    &       Kink    &                &       $\times$        &       3.4     &       $-$     &       $-$     &       8.3     &       0.1     \\ 
10      &       1       &       2       &       0.1     &       0.01    &       Gap     &                &                       &       3.4     &       10.3    &       11.7    &       9       &       2.6     \\ 
50      &       1       &       2       &       0.1     &       0.01    &       Gap     &                &                       &       3.4     &       14.1    &       16.7    &       8.2     &       8.4     \\ 
5       &       0.5     &       2       &       0.1     &       0.01    &       Kink    &                &       $\times$        &       3.4     &       $-$     &       $-$     &       7.2     &       0.0     \\ 
10      &       0.5     &       2       &       0.1     &       0.01    &       Gap     &                &                       &       3.4     &       9.9     &       10.5    &       9.9     &       0.1     \\ 
50      &       0.5     &       2       &       0.1     &       0.01    &       Gap     &                &                       &       3.4     &       14.3    &       16.7    &       11.6    &       5.0     \\ 
5       &       0.25    &       2       &       0.1     &       0.01    &       Kink    &                &       $\times$        &       3.4     &       $-$     &       $-$     &       7.2     &       0.0     \\ 
10      &       0.25    &       2       &       0.1     &       0.01    &       Gap     &                &                       &       3.4     &       10.7    &       10.9    &       10.9    &       0.0     \\ 
50      &       0.25    &       2       &       0.1     &       0.01    &       Gap     &                &                       &       3.4     &       13.8    &       15.1    &       14.8    &       0.1     \\ 
5       &       1       &       0.4     &       1       &       0.01    &       Kink    &                &       $\times$        &       3.4     &       $-$     &       $-$     &       8.8     &       0.0     \\ 
10      &       1       &       0.4     &       1       &       0.01    &       Gap     &                &                       &       3.4     &       10.3    &       11.7    &       10.1    &       1.6     \\ 
50      &       1       &       0.4     &       1       &       0.01    &       Gap     &                &                       &       3.4     &       14      &       16.7    &       8.9     &       7.8     \\ 
5       &       0.5     &       0.4     &       1       &       0.01    &       Kink    &                &       $\times$        &       3.4     &       $-$     &       $-$     &       7.2     &       0.0     \\ 
10      &       0.5     &       0.4     &       1       &       0.01    &       Gap     &                &                       &       3.4     &       9.9     &       10.5    &       10      &       0.1     \\ 
50      &       0.5     &       0.4     &       1       &       0.01    &       Gap     &                &                       &       3.4     &       14.3    &       16.7    &       11.4    &       5.2     \\ 
5       &       0.25    &       0.4     &       1       &       0.01    &       Kink    &                &       $\times$        &       3.4     &       $-$     &       $-$     &       7.2     &       0.0     \\ 
10      &       0.25    &       0.4     &       1       &       0.01    &       Gap     &                &                       &       3.4     &       10.7    &       10.9    &       10.8    &       0.0     \\ 
50      &       0.25    &       0.4     &       1       &       0.01    &       Gap     &                &                       &       3.4     &       13.8    &       15.1    &       14.9    &       0.1     \\ 
5       &       1       &       2       &       1       &       0.01    &       Kink    &                &       $\times$        &       3.4     &       $-$     &       $-$     &       8.7     &       0.1     \\ 
10      &       1       &       2       &       1       &       0.01    &       Gap     &                &                       &       3.4     &       10.3    &       11.7    &       9.2     &       2.5     \\ 
50      &       1       &       2       &       1       &       0.01    &       Gap     &                &                       &       3.4     &       14.1    &       16.7    &       8.5     &       8.1     \\ 
5       &       0.5     &       2       &       1       &       0.01    &       Kink    &                &       $\times$        &       3.4     &       $-$     &       $-$     &       7.2     &       0.0     \\ 
10      &       0.5     &       2       &       1       &       0.01    &       Gap     &                &                       &       3.4     &       9.9     &       10.5    &       9.9     &       0.1     \\ 
50      &       0.5     &       2       &       1       &       0.01    &       Gap     &                &                       &       3.4     &       14.3    &       16.7    &       11.1    &       5.5     \\ 
5       &       0.25    &       2       &       1       &       0.01    &       Kink    &                &       $\times$        &       3.4     &       $-$     &       $-$     &       7.2     &       0.0     \\ 
10      &       0.25    &       2       &       1       &       0.01    &       Gap     &                &                       &       3.4     &       10.7    &       10.9    &       10.8    &       0.0     \\ 
50      &       0.25    &       2       &       1       &       0.01    &       Gap     &                &                       &       3.4     &       13.8    &       15.1    &       14.9    &       0.1     \\ 
5       &       1       &       0.4     &       0.1     &       0.001   &       Kink    &                &       $\times$        &       3.4     &       $-$     &       $-$     &       8.1     &       0.1     \\ 
10      &       1       &       0.4     &       0.1     &       0.001   &       Gap     &                &                       &       3.4     &       10.4    &       11.5    &       8.4     &       3.0     \\ 
50      &       1       &       0.4     &       0.1     &       0.001   &       Gap     &                &                       &       3.4     &       15.1    &       17.8    &       7.9     &       9.7     \\ 
5       &       0.5     &       0.4     &       0.1     &       0.001   &       Kink    &                &       $\times$        &       3.4     &       $-$     &       $-$     &       8.9     &       0.0     \\ 
10      &       0.5     &       0.4     &       0.1     &       0.001   &       Gap     &                &                       &       3.4     &       11.2    &       11.8    &       9.7     &       2.1     \\ 
50      &       0.5     &       0.4     &       0.1     &       0.001   &       Gap     &                &                       &       3.4     &       14.6    &       16.3    &       8.4     &       7.8     \\ 
5       &       0.25    &       0.4     &       0.1     &       0.001   &       Kink    &                &       $\times$        &       3.4     &       $-$     &       $-$     &       8.9     &       0.0     \\ 
10      &       0.25    &       0.4     &       0.1     &       0.001   &       Gap     &                &                       &       3.4     &       10.5    &       10.9    &       9.8     &       1.0     \\ 
50      &       0.25    &       0.4     &       0.1     &       0.001   &       Gap     &                &                       &       3.4     &       14.3    &       15.4    &       9       &       6.4     \\ 
5       &       1       &       2       &       0.1     &       0.001   &       Kink    &                &       $\times$        &       3.4     &       $-$     &       $-$     &       7.9     &       0.2     \\ 
10      &       1       &       2       &       0.1     &       0.001   &       Gap     &                &                       &       3.4     &       10.4    &       11.5    &       7.9     &       3.5     \\ 
50      &       1       &       2       &       0.1     &       0.001   &       Gap     &                &                       &       3.4     &       15.1    &       17.7    &       7.3     &       10.3    \\ 
5       &       0.5     &       2       &       0.1     &       0.001   &       Kink    &                &       $\times$        &       3.4     &       $-$     &       $-$     &       8.9     &       0.1     \\ 
10      &       0.5     &       2       &       0.1     &       0.001   &       Gap     &                &                       &       3.4     &       11.2    &       11.8    &       9.1     &       2.7     \\ 
50      &       0.5     &       2       &       0.1     &       0.001   &       Gap     &                &                       &       3.4     &       14.6    &       16.3    &       8.5     &       7.7     \\ 
5       &       0.25    &       2       &       0.1     &       0.001   &       Kink    &                &       $\times$        &       3.4     &       $-$     &       $-$     &       8.9     &       0.0     \\ 
10      &       0.25    &       2       &       0.1     &       0.001   &       Gap     &                &                       &       3.4     &       10.5    &       10.9    &       9.8     &       1.1     \\ 
50      &       0.25    &       2       &       0.1     &       0.001   &       Gap     &                &                       &       3.4     &       14.3    &       15.4    &       8.2     &       7.2     \\ 
5       &       1       &       0.4     &       1       &       0.001   &       Kink    &                &       $\times$        &       3.4     &       $-$     &       $-$     &       8.1     &       0.0     \\ 
10      &       1       &       0.4     &       1       &       0.001   &       Gap     &                &                       &       3.4     &       10.4    &       11.5    &       9       &       2.4     \\ 
50      &       1       &       0.4     &       1       &       0.001   &       Gap     &                &                       &       3.4     &       15.1    &       17.8    &       8       &       9.6     \\ 
5       &       0.5     &       0.4     &       1       &       0.001   &       Kink    &                &       $\times$        &       3.4     &       $-$     &       $-$     &       8.9     &       0.0     \\ 
10      &       0.5     &       0.4     &       1       &       0.001   &       Gap     &                &                       &       3.4     &       11.1    &       11.8    &       9.4     &       2.3     \\ 
50      &       0.5     &       0.4     &       1       &       0.001   &       Gap     &                &                       &       3.4     &       14.6    &       16.3    &       8.1     &       8.2     \\ 
5       &       0.25    &       0.4     &       1       &       0.001   &       Kink    &                &       $\times$        &       3.4     &       $-$     &       $-$     &       8.8     &       0.1     \\ 
10      &       0.25    &       0.4     &       1       &       0.001   &       Gap     &                &                       &       3.4     &       10.5    &       10.9    &       9.4     &       1.4     \\ 
50      &       0.25    &       0.4     &       1       &       0.001   &       Gap     &                &                       &       3.4     &       14.3    &       15.4    &       7.2     &       8.1     \\ 
5       &       1       &       2       &       1       &       0.001   &       Kink    &                &       $\times$        &       3.4     &       $-$     &       $-$     &       7.9     &       0.2     \\ 
10      &       1       &       2       &       1       &       0.001   &       Gap     &                &                       &       3.4     &       10.4    &       11.5    &       8       &       3.4     \\ 
50      &       1       &       2       &       1       &       0.001   &       Gap     &                &                       &       3.4     &       15.1    &       17.7    &       7.5     &       10.0    \\ 
5       &       0.5     &       2       &       1       &       0.001   &       Kink    &                &       $\times$        &       3.4     &       $-$     &       $-$     &       8.9     &       0.1     \\ 
10      &       0.5     &       2       &       1       &       0.001   &       Gap     &                &                       &       3.4     &       11.1    &       11.8    &       9.3     &       2.5     \\ 
50      &       0.5     &       2       &       1       &       0.001   &       Gap     &                &                       &       3.4     &       14.6    &       16.3    &       8       &       8.2     \\ 
5       &       0.25    &       2       &       1       &       0.001   &       Kink    &                &       $\times$        &       3.4     &       $-$     &       $-$     &       8.9     &       0.0     \\ 
10      &       0.25    &       2       &       1       &       0.001   &       Gap     &                &                       &       3.4     &       10.5    &       10.9    &       10.4    &       0.4     \\ 
50      &       0.25    &       2       &       1       &       0.001   &       Gap     &                &                       &       3.4     &       14.3    &       15.4    &       8.3     &       7.1     \\ 

\bottomrule
\end{tabular}
}
\caption{Compilation of results for all simulated models at $\lambda=\lambdaM$. The first five columns list  the set of underlying parameter values. Contrast values (in mag) are obtained based on either of the two types of features (Gap or Kink). Hidden planets are flagged with a cross and refer to planets whose inferred position of the planetary and CPD signal is hidden inside the IWA of the corresponding coronagraph, which can be found in Tab. \ref{tbl:instrument_sphere}. Contaminated planetary and CPD signals are flagged with a cross in the ``Cont.'' column and correspond to simulations that do not satisfy the condition stated in Eq. \eqref{eq:condition_azimuthal_gap_region}. The last five columns refer to contrast values either with respect to the stellar (Star) flux or to the flux of the gap region (Gap), where the following abbreviations were used: Coronagraph (C), ring (R), planet and CPD (CPD). For details, see Sect. \ref{sec:planetary_and_cpd_signal_strength}.}
\label{tbl:results_wl30}
\end{center}

\end{table*}

\begin{table*}
\begin{center}
\resizebox*{!}{0.88\textheight}{%
\begin{tabular}{ccccc|ccc|cccc|c}
\toprule
$\rp\,\left[{\rm au}\right]$    &       $\Mp\,\left[\mj\right]$ &       $\mpdot\,\left[10^{-6}\,\mj/{\rm yr}\right]$     &       $\mcpd\,\left[10^{-3}\,\mj\right]$      &       $\mcsd\,\left[{\rm M}_\odot\right]$        &       Type    &       Hidden  &       Cont.   &       C{:}Star        &       R{:}Star        &       Gap{:}Star &       $\ContrastStar$ &       $\ContrastGap$  \\
\cmidrule{1-13}
5       &       1       &       0.4     &       0.1     &       0.01    &       Kink    &                &       $\times$        &       7       &       $-$     &       $-$     &       9.2     &       0.0     \\ 
10      &       1       &       0.4     &       0.1     &       0.01    &       Kink    &                &                       &       7.1     &       $-$     &       $-$     &       8.9     &       0.8     \\ 
50      &       1       &       0.4     &       0.1     &       0.01    &       Gap     &                &                       &       7.1     &       13.9    &       16.5    &       7.4     &       9.1     \\ 
5       &       0.5     &       0.4     &       0.1     &       0.01    &       Kink    &                &       $\times$        &       7.1     &       $-$     &       $-$     &       9.7     &       0.0     \\ 
10      &       0.5     &       0.4     &       0.1     &       0.01    &       Kink    &                &                       &       7.1     &       $-$     &       $-$     &       9.2     &       0.2     \\ 
50      &       0.5     &       0.4     &       0.1     &       0.01    &       Gap     &                &                       &       7.1     &       13.9    &       15.5    &       8.4     &       7.2     \\ 
5       &       0.25    &       0.4     &       0.1     &       0.01    &       Kink    &                &       $\times$        &       7       &       $-$     &       $-$     &       7.8     &       0.0     \\ 
10      &       0.25    &       0.4     &       0.1     &       0.01    &       Kink    &                &                       &       7.1     &       $-$     &       $-$     &       8.9     &       0.0     \\ 
50      &       0.25    &       0.4     &       0.1     &       0.01    &       Gap     &                &                       &       7.1     &       14.1    &       15.6    &       10.7    &       4.9     \\ 
5       &       1       &       2       &       0.1     &       0.01    &       Kink    &                &       $\times$        &       7       &       $-$     &       $-$     &       8.8     &       0.0     \\ 
10      &       1       &       2       &       0.1     &       0.01    &       Kink    &                &                       &       7.1     &       $-$     &       $-$     &       8.9     &       0.9     \\ 
50      &       1       &       2       &       0.1     &       0.01    &       Gap     &                &                       &       7.1     &       13.9    &       16.6    &       6.9     &       9.7     \\ 
5       &       0.5     &       2       &       0.1     &       0.01    &       Kink    &                &       $\times$        &       7.1     &       $-$     &       $-$     &       9.7     &       0.0     \\ 
10      &       0.5     &       2       &       0.1     &       0.01    &       Kink    &                &                       &       7.1     &       $-$     &       $-$     &       9.3     &       0.2     \\ 
50      &       0.5     &       2       &       0.1     &       0.01    &       Gap     &                &                       &       7.1     &       13.8    &       15.9    &       8.4     &       7.5     \\ 
5       &       0.25    &       2       &       0.1     &       0.01    &       Kink    &                &       $\times$        &       7       &       $-$     &       $-$     &       7.5     &       0.0     \\ 
10      &       0.25    &       2       &       0.1     &       0.01    &       Kink    &                &                       &       7.1     &       $-$     &       $-$     &       8.9     &       0.0     \\ 
50      &       0.25    &       2       &       0.1     &       0.01    &       Gap     &                &                       &       7.1     &       14.1    &       15.6    &       10.3    &       5.3     \\ 
5       &       1       &       0.4     &       1       &       0.01    &       Kink    &                &       $\times$        &       7       &       $-$     &       $-$     &       7.5     &       0.0     \\ 
10      &       1       &       0.4     &       1       &       0.01    &       Kink    &                &                       &       7.1     &       $-$     &       $-$     &       8.7     &       1.0     \\ 
50      &       1       &       0.4     &       1       &       0.01    &       Gap     &                &                       &       7.1     &       14      &       16.7    &       6.6     &       10.1    \\ 
5       &       0.5     &       0.4     &       1       &       0.01    &       Kink    &                &       $\times$        &       7.1     &       $-$     &       $-$     &       7.6     &       0.0     \\ 
10      &       0.5     &       0.4     &       1       &       0.01    &       Kink    &                &                       &       7.1     &       $-$     &       $-$     &       9.2     &       0.3     \\ 
50      &       0.5     &       0.4     &       1       &       0.01    &       Gap     &                &                       &       7.1     &       13.9    &       15.8    &       7.2     &       8.5     \\ 
5       &       0.25    &       0.4     &       1       &       0.01    &       Kink    &                &       $\times$        &       7       &       $-$     &       $-$     &       9.8     &       0.1     \\ 
10      &       0.25    &       0.4     &       1       &       0.01    &       Kink    &                &                       &       7.1     &       $-$     &       $-$     &       9.8     &       0.1     \\ 
50      &       0.25    &       0.4     &       1       &       0.01    &       Gap     &                &                       &       7.1     &       14.1    &       15.2    &       8.9     &       6.3     \\ 
5       &       1       &       2       &       1       &       0.01    &       Kink    &                &       $\times$        &       7       &       $-$     &       $-$     &       9.8     &       0.0     \\ 
10      &       1       &       2       &       1       &       0.01    &       Kink    &                &                       &       7.1     &       $-$     &       $-$     &       9       &       0.8     \\ 
50      &       1       &       2       &       1       &       0.01    &       Gap     &                &                       &       7.1     &       13.9    &       16.6    &       6.7     &       9.9     \\ 
5       &       0.5     &       2       &       1       &       0.01    &       Kink    &                &       $\times$        &       7.1     &       $-$     &       $-$     &       9.7     &       0.0     \\ 
10      &       0.5     &       2       &       1       &       0.01    &       Kink    &                &                       &       7.1     &       $-$     &       $-$     &       9.2     &       0.2     \\ 
50      &       0.5     &       2       &       1       &       0.01    &       Gap     &                &                       &       7.1     &       14      &       16      &       7.3     &       8.7     \\ 
5       &       0.25    &       2       &       1       &       0.01    &       Kink    &                &       $\times$        &       7       &       $-$     &       $-$     &       9.7     &       0.0     \\ 
10      &       0.25    &       2       &       1       &       0.01    &       Kink    &                &                       &       7.1     &       $-$     &       $-$     &       9.7     &       0.1     \\ 
50      &       0.25    &       2       &       1       &       0.01    &       Gap     &                &                       &       7.1     &       14.2    &       15.5    &       9.8     &       5.8     \\ 
5       &       1       &       0.4     &       0.1     &       0.001   &       Kink    &                &       $\times$        &       7.1     &       $-$     &       $-$     &       8.5     &       0.0     \\ 
10      &       1       &       0.4     &       0.1     &       0.001   &       Kink    &                &                       &       7.2     &       $-$     &       $-$     &       8.8     &       1.2     \\ 
50      &       1       &       0.4     &       0.1     &       0.001   &       Gap     &                &                       &       7.1     &       14.8    &       17.4    &       7.1     &       10.3    \\ 
5       &       0.5     &       0.4     &       0.1     &       0.001   &       Kink    &                &       $\times$        &       7.1     &       $-$     &       $-$     &       9       &       0.0     \\ 
10      &       0.5     &       0.4     &       0.1     &       0.001   &       Kink    &                &                       &       7.1     &       $-$     &       $-$     &       9.1     &       0.8     \\ 
50      &       0.5     &       0.4     &       0.1     &       0.001   &       Gap     &                &                       &       7.1     &       14.9    &       16.4    &       7.2     &       9.2     \\ 
5       &       0.25    &       0.4     &       0.1     &       0.001   &       Kink    &                &       $\times$        &       7.1     &       $-$     &       $-$     &       8.6     &       0.0     \\ 
10      &       0.25    &       0.4     &       0.1     &       0.001   &       Kink    &                &                       &       7.2     &       $-$     &       $-$     &       9.6     &       0.9     \\ 
50      &       0.25    &       0.4     &       0.1     &       0.001   &       Gap     &                &                       &       7.1     &       14.4    &       15.2    &       7.2     &       8.0     \\ 
5       &       1       &       2       &       0.1     &       0.001   &       Kink    &                &       $\times$        &       7.1     &       $-$     &       $-$     &       8.5     &       0.1     \\ 
10      &       1       &       2       &       0.1     &       0.001   &       Kink    &                &                       &       7.2     &       $-$     &       $-$     &       8.7     &       1.3     \\ 
50      &       1       &       2       &       0.1     &       0.001   &       Gap     &                &                       &       7.1     &       14.7    &       17.4    &       6.7     &       10.7    \\ 
5       &       0.5     &       2       &       0.1     &       0.001   &       Kink    &                &       $\times$        &       7.1     &       $-$     &       $-$     &       8.5     &       0.0     \\ 
10      &       0.5     &       2       &       0.1     &       0.001   &       Kink    &                &                       &       7.1     &       $-$     &       $-$     &       9       &       1.0     \\ 
50      &       0.5     &       2       &       0.1     &       0.001   &       Gap     &                &                       &       7.1     &       14.9    &       16.4    &       7.3     &       9.1     \\ 
5       &       0.25    &       2       &       0.1     &       0.001   &       Kink    &                &       $\times$        &       7.1     &       $-$     &       $-$     &       9.5     &       0.0     \\ 
10      &       0.25    &       2       &       0.1     &       0.001   &       Kink    &                &                       &       7.2     &       $-$     &       $-$     &       9.5     &       1.0     \\ 
50      &       0.25    &       2       &       0.1     &       0.001   &       Gap     &                &                       &       7.1     &       14.4    &       15.2    &       6.7     &       8.4     \\ 
5       &       1       &       0.4     &       1       &       0.001   &       Kink    &                &       $\times$        &       7.1     &       $-$     &       $-$     &       8.1     &       0.0     \\ 
10      &       1       &       0.4     &       1       &       0.001   &       Kink    &                &                       &       7.2     &       $-$     &       $-$     &       8.6     &       1.4     \\ 
50      &       1       &       0.4     &       1       &       0.001   &       Gap     &                &                       &       7.1     &       14.7    &       17.5    &       6.3     &       11.2    \\ 
5       &       0.5     &       0.4     &       1       &       0.001   &       Kink    &                &       $\times$        &       7.1     &       $-$     &       $-$     &       8.5     &       0.0     \\ 
10      &       0.5     &       0.4     &       1       &       0.001   &       Kink    &                &                       &       7.1     &       $-$     &       $-$     &       8.5     &       1.4     \\ 
50      &       0.5     &       0.4     &       1       &       0.001   &       Gap     &                &                       &       7.1     &       14.9    &       16.5    &       6.1     &       10.4    \\ 
5       &       0.25    &       0.4     &       1       &       0.001   &       Kink    &                &       $\times$        &       7.1     &       $-$     &       $-$     &       9.4     &       0.2     \\ 
10      &       0.25    &       0.4     &       1       &       0.001   &       Kink    &                &                       &       7.2     &       $-$     &       $-$     &       8.7     &       1.8     \\ 
50      &       0.25    &       0.4     &       1       &       0.001   &       Gap     &                &                       &       7.1     &       14.4    &       15.3    &       5.3     &       9.9     \\ 
5       &       1       &       2       &       1       &       0.001   &       Kink    &                &       $\times$        &       7.1     &       $-$     &       $-$     &       8.6     &       0.0     \\ 
10      &       1       &       2       &       1       &       0.001   &       Kink    &                &                       &       7.2     &       $-$     &       $-$     &       8.8     &       1.2     \\ 
50      &       1       &       2       &       1       &       0.001   &       Gap     &                &                       &       7.1     &       14.7    &       17.5    &       6.4     &       11.0    \\ 
5       &       0.5     &       2       &       1       &       0.001   &       Kink    &                &       $\times$        &       7.1     &       $-$     &       $-$     &       9.4     &       0.0     \\ 
10      &       0.5     &       2       &       1       &       0.001   &       Kink    &                &                       &       7.1     &       $-$     &       $-$     &       9.1     &       0.9     \\ 
50      &       0.5     &       2       &       1       &       0.001   &       Gap     &                &                       &       7.1     &       14.9    &       16.4    &       6.2     &       10.2    \\ 
5       &       0.25    &       2       &       1       &       0.001   &       Kink    &                &       $\times$        &       7.1     &       $-$     &       $-$     &       9.7     &       0.0     \\ 
10      &       0.25    &       2       &       1       &       0.001   &       Kink    &                &                       &       7.2     &       $-$     &       $-$     &       9.8     &       0.7     \\ 
50      &       0.25    &       2       &       1       &       0.001   &       Gap     &                &                       &       7.1     &       14.4    &       15.2    &       6.1     &       9.1     \\ 

\bottomrule
\end{tabular}
}
\caption{Compilation of results for all simulated models at $\lambda=\lambdaL$. The first five columns give the set of  underlying parameter values. Contrast values (in mag) are obtained based on either of the  types of features (Gap or Kink). Hidden planets are flagged with a cross and refer to planets whose inferred position of the planetary and CPD signal is hidden inside the IWA of the corresponding coronagraph, which can be found in Tab. \ref{tbl:instrument_sphere}. Contaminated planetary and CPD signals are flagged with a cross in the ``Cont.'' column and correspond to simulations that do not satisfy the condition stated in Eq. \eqref{eq:condition_azimuthal_gap_region}. The last five columns refer to contrast values either with respect to the stellar (Star) flux or to the flux of the gap region (Gap), where the following abbreviations were used: Coronagraph (C), ring (R), planet and CPD (CPD). For details, see Sect. \ref{sec:planetary_and_cpd_signal_strength}.}
\label{tbl:results_wl35}
\end{center}
\end{table*}

 \begin{figure*}
   \centering
   \includegraphics{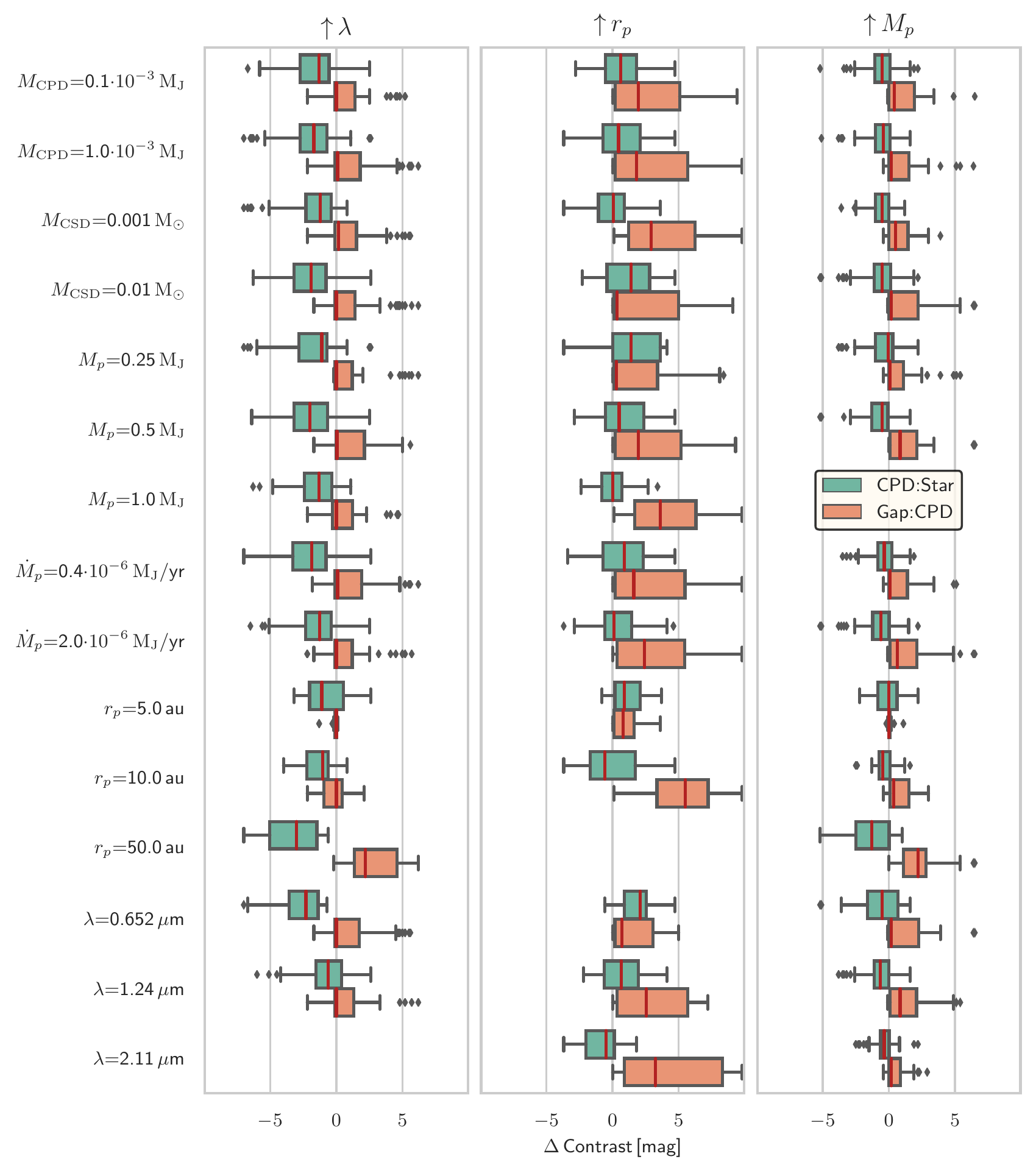}
   \caption{Distribution of contrast value changes in $\ContrastStar$ and $\ContrastGap$ with respect to the change in a single model parameter using boxplots. Each median is highlighted by a red line, the middle $50\,\%$ of data are represented by a box, and the maximum whisker length equals $\Delta w = 1.5\,$IQR. The labels to the left characterize the data used in generating the corresponding boxplots to the right of the label. A label refers to a shared parameter value. The label above any column refers to a parameter that has been increased to its next simulated value. Outliers are shown as  black diamonds. For details, see Sect. \ref{sec:parameter_impact}.}
              \label{fig:boxplot_Nconst1a}
   \end{figure*}

 \begin{figure*}
   \centering
   \includegraphics{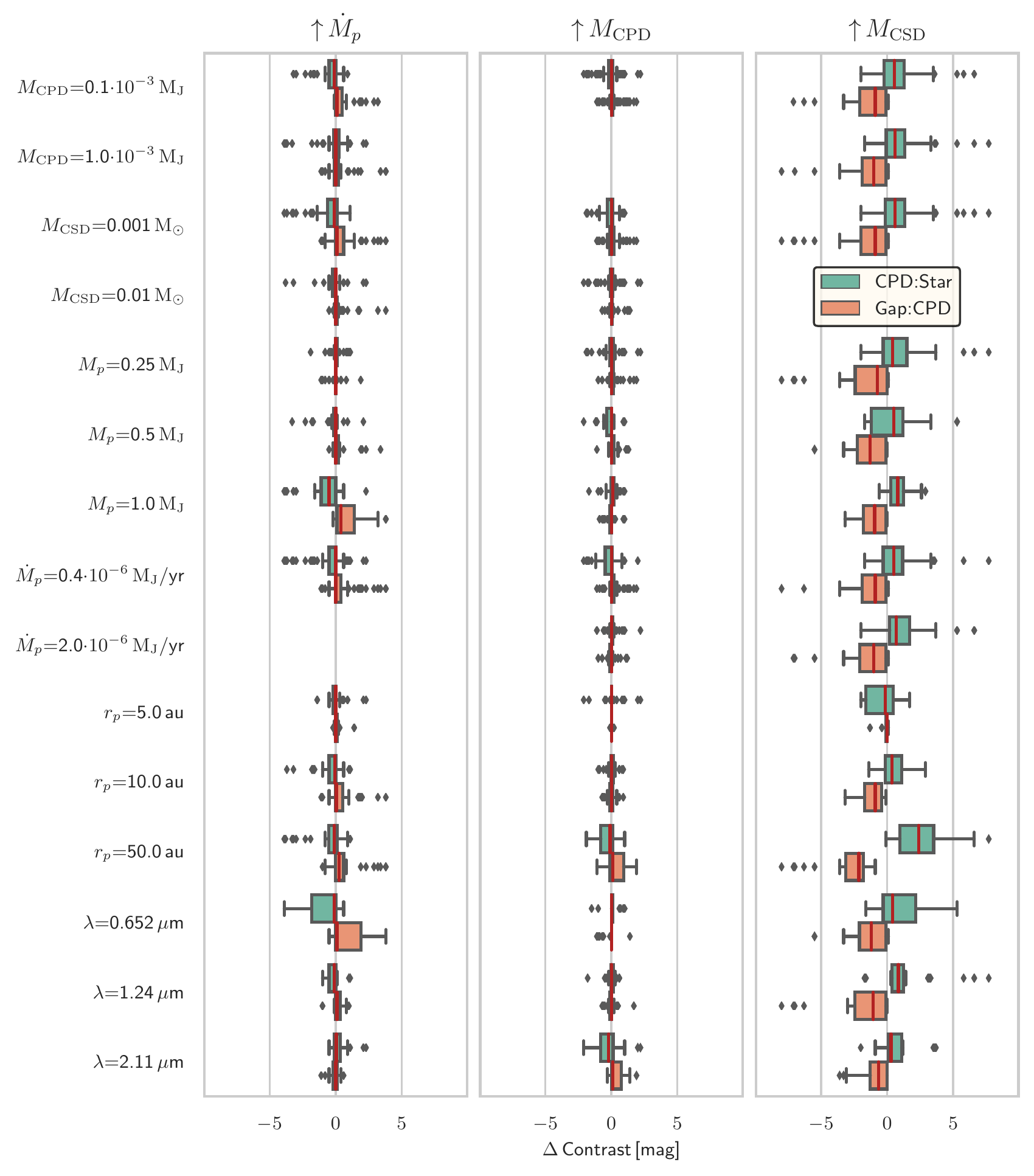}
   \caption{Distribution of contrast value changes in $\ContrastStar$ and $\ContrastGap$ with respect to the change in a single model parameter using boxplots. Each median is highlighted by a red line, the middle $50\,\%$ of data are represented by a box, and the maximum whisker length equals $\Delta w = 1.5\,$IQR. The labels to the left characterize the data used in generating the corresponding boxplots to the right of the label. A label refers to a shared parameter value. The label above any column refers to a parameter that has been increased to its next simulated value. Outliers are shown as   black diamonds. For details, see Sect. \ref{sec:parameter_impact}.}
              \label{fig:boxplot_Nconst1b}
   \end{figure*}

\end{appendix}

\end{document}